\documentclass[fleqn,usenatbib]{mnras}

\usepackage{newtxtext,newtxmath}

\usepackage[T1]{fontenc}

\DeclareRobustCommand{\VAN}[3]{#2}
\let\VANthebibliography\thebibliography
\def\thebibliography{\DeclareRobustCommand{\VAN}[3]{##3}\VANthebibliography}

\usepackage{graphicx}	
\usepackage{amsmath}	

\newcommand{\nustar}{{\it NuSTAR}}

\newcommand{\xmm}{{\it XMM-Newton}}
\newcommand{\chandra}{{\it Chandra}}
\newcommand{\eps}{erg s$^{-1}$}
\newcommand{\ecs}{erg cm$^{-2}$ s$^{-1}$}
\newcommand{\pcm}{cm$^{-2}$}
\newcommand{\source}{NGC~4507}
\newcommand{\M}{$M_{\odot}$}
\newcommand{\phc}{ph cm$^{-2}$ s$^{-1}$}
\newcommand{\kms}{km s$^{-1}$}
\newcommand{\nhl}{$N_{\rm H}^{\rm los}$}
\newcommand{\nht}{$N_{\rm H}^{\rm tor}$}
\newcommand{\nheq}{$N_{\rm H}^{\rm Eq}$}
\newcommand{\nh}{$N_{\rm H}$}
\newcommand{\mytc}{{\tt MYTC}}
\newcommand{\mytd}{{\tt MYTD}}
\newcommand{\mytz}{{\tt MYTZ}}
\newcommand{\myts}{{\tt MYTS}}
\newcommand{\mytl}{{\tt MYTL}}
\newcommand{\borus}{{\tt borus02}}
\newcommand{\xclumpy}{{\tt xclumpy}}
\newcommand{\rxtorus}{{\tt RXTorus}}

\title[The highly obscured AGN NGC~4507]{Absorption Variability of the Highly Obscured Active Galactic Nucleus NGC~4507}

\author[A. Jana et al.]{
Arghajit Jana,$^{1,2}$\thanks{E-mail: argha0004@gmail.com}
Claudio Ricci,$^{3,4}$
Sachindra Naik,$^{2}$
Atsushi Tanimoto,$^{5}$
Neeraj Kumari,$^{2,6}$
\newauthor
Hsiang-Kuang Chang,$^{1}$
Prantik Nandi,$^{2}$
Arka Chatterjee,$^{7}$
Samar Safi-Harb$^{7}$
\\
$^{1}$ Institute of Astronomy, National Tsing Hua University, Hsinchu 30013, Taiwan\\
$^{2}$ Astronomy \& Astrophysics Division, Physical Research Laboratory, Navrangpura, Ahmedabad, 380009, India\\
$^{3}$N{\'u}cleo de Astronom{\'i}a de la Facultad de Ingenier{\'i}a, Universidad Diego Portales, Av. Ej{\'e}rcito Libertador 441, Santiago, Chile\\
$^{4}$ Kavli Institute for Astronomy and Astrophysics, Peking University, Beijing 100871, People's Republic of China\\
$^{5}$ Department of Physics, The University of Tokyo, Tokyo 113-0033, Japan \\
$^{6}$ Department of Physics, Indian Institute of Technology, Gandhinagar- 382355, Gujarat, India\\
$^{7}$ Department of Physics and Astronomy, University of Manitoba, Winnipeg, MB R3T 2N2, Canada \\
}

\date{Accepted XXX. Received YYY; in original form ZZZ}

\pubyear{2021}

\begin{document}
\label{firstpage}
\pagerange{\pageref{firstpage}--\pageref{lastpage}}
\maketitle

\begin{abstract}
We present a detailed study of the highly obscured active galaxy NGC~4507, performed using four {\it Nuclear Spectroscopic Telescope Array} ({\it NuSTAR}) observations carried out between May and August in 2015 ($\sim 130$~ks in total). Using various phenomenological and physically motivated torus models, we explore the properties of the X-ray source and those of the obscuring material. The primary X-ray emission is found to be non-variable, indicating a stable accretion during the period of the observations. We find the equatorial column density of the obscuring materials to be $\sim 2 \times 10^{24}$ \pcm~ while the line of sight column density to be $\sim 7-8 \times 10^{23}$ \pcm. The source is found to be deeply buried with the torus covering factor $\sim 0.85$. We observe variability in the line-of-sight column density on a timescale of $<35$ days. The covering factor of the Compton-Thick material is found to be $\sim 0.35$, in agreement with the results of recent X-ray surveys. From the variability of the line-of-sight column density, we estimate that the variable absorbing material is likely located either in the BLR or in the torus. 
\end{abstract}


\label{sec:intro}
\section{Introduction}
Active galactic nuclei (AGNs) are classified as Type-1 and Type-2, based on the presence or absence of broad optical/UV emission lines. The simplified unification model of AGNs can explain different types of AGN based on different inclination angles with respect to an obscuring torus \citep{Antonucci1993}. In this framework, the Type-2 AGNs are seen edge-on (i.e. through the obscuring torus), while the Type-1 AGNs are observed face-on. Work carried out in the infrared has shown that the molecular torus is likely clumpy, rather than uniform \citep[e.g., ][]{Nenkova2008a,Nenkova2008b}. The level of obscuration toward the X-ray source is typically parametrized with the hydrogen column density (\nh).

Over the years, many AGNs are observed to show variable \nh~ in a timescale of hours to years \citep{Risaliti2002a}. The short-term variations (on timescales of $\sim$ days) are believed to be associated with the broad line emitting region (BLR), while the long-term variability (on timescales of months to years) are believed to be caused by the clumpy molecular torus \citep{Markowitz2014}. A growing number of AGNs, e.g., UGC~4203 \citep{Risaliti2010}, NGC~4151 \citep{Puccetti2007}, NGC~2992 \citep{Weaver1996,Murphy2007}, IC~751 \citep{Ricci2016}, NGC~6300 \citep{Guainazzi2002,AJ2020}, have shown variable $N_{\rm H}$ by repeated X-ray observation. In recent years, a new sub-class of AGNs, known as changing-look AGN has emerged. In these objects, the line of sight column density can go from a Compton-thin (\nh $<10^{24}$ \pcm) to a Compton-thick state (CT; \nh $>10^{24}$ \pcm) level, or vice-versa. These events can lead to a dramatic change in the observed X-ray spectrum, which can go from being reflection dominated (in the Compton-thick state) to transmission dominated (in the Compton-thin state), or vice versa \citep{Guainazzi2002,Matt2003}. These events are believed to be an important confirmation of the clumpiness of the BLR or torus \citep{Guainazzi2002,Elitzur2012,Yaqoob2015,AJ2020}.

\source~ is a nearby ($z$ = 0.0118) barred spiral galaxy, classified as SAB(s)ab \citep{Tueller2008,Winter2009}. \source~ is reported to be one of the brightest ($F_{\rm 2-10}^{\rm obs} \sim 10^{-11}$ \ecs) Seyfert 2 galaxies in the hard X-ray band ($>10$~keV; \citealp{Braito2013}), and was detected by INTEGRAL/ISGRI, Swift/BAT and CGRO/OSSE \citep{Bassani1995,Ricci2017apjs}. Over the years, several X-ray studies have revealed a variable \nh~ in the range of $\sim 1-9 \times 10^{23}$ \pcm~ based on the observations by {\it Ginga}, {\it ASCA}, {\it BeppoSAX}, \xmm~ and \chandra~ \citep{Awaki1991, Comastri1998,Risaliti2002b,Matt2004,Marinucci2013}. 

In this paper, we present a detailed X-ray spectral analysis of \source, obtained with four \nustar~ observations carried out between May and August 2015, with a total exposure time of $\sim 130$\,ks. We aim to probe the variability of the obscuration from these broad-band X-ray observations. The paper is structured as follows. In \S2, we present the data extraction procedure. The timing analysis is reported in \S3. In \S4, we report our detailed spectral analysis and our results. In \S5, we discuss our findings, and compare them to previous studies. Finally, in \S6, we summarize our findings.

\begin{table}
\caption{Log of the \nustar~ observations of NGC~4507 studied here.}
\label{tab:log}
\begin{tabular}{lcccccccccc}
\hline
ID & UT Date & Observation ID & Exp (s) & Count s$^{-1}$ \\
\hline
N1 & 2015-05-03 & 60102051002 & 30133 & $0.736\pm0.005$ \\
N2 & 2015-06-10 & 60102051004 & 34464 & $0.773\pm0.005$ \\
N3 & 2015-07-15 & 60102051006 & 32225 & $0.743\pm0.005$ \\
N4 & 2015-08-22 & 60102051008 & 30924 & $0.720\pm0.005$ \\
\hline
\end{tabular}
\end{table}

\label{sec:obs}
\section{Observation and Data Reduction}
\nustar~ observed \source~ four times between May and August 2015 with an interval of about five weeks between the different observations. In the present work, we studied all \nustar~ observations in a energy range of $3-60$~keV (see Table~\ref{tab:log}). \nustar~ is a hard X-ray focusing telescope, consisting of two identical modules: FPMA and FPMB \citep{Harrison2013}. Data were reprocessed with the \nustar~ Data Analysis Software ({\tt NuSTARDAS}, version 1.4.1). Cleaned event files were generated and calibrated by using the standard filtering criteria in the {\tt nupipeline} task, and the latest calibration data files available in the NuSTAR calibration database\footnote{\url{http://heasarc.gsfc.nasa.gov/FTP/caldb/data/nustar/fpm/}}. The source and background products were extracted by considering circular regions with 90\,arcsec radii, centred at the source coordinates and away from the source, respectively. The spectra and light curves were extracted using the {\tt nuproduct} task. We re-binned the spectra to ensure that they had at least 20 counts per bin by using the {\tt grppha} task. 

\begin{figure}
\centering
\includegraphics[width=8.5cm]{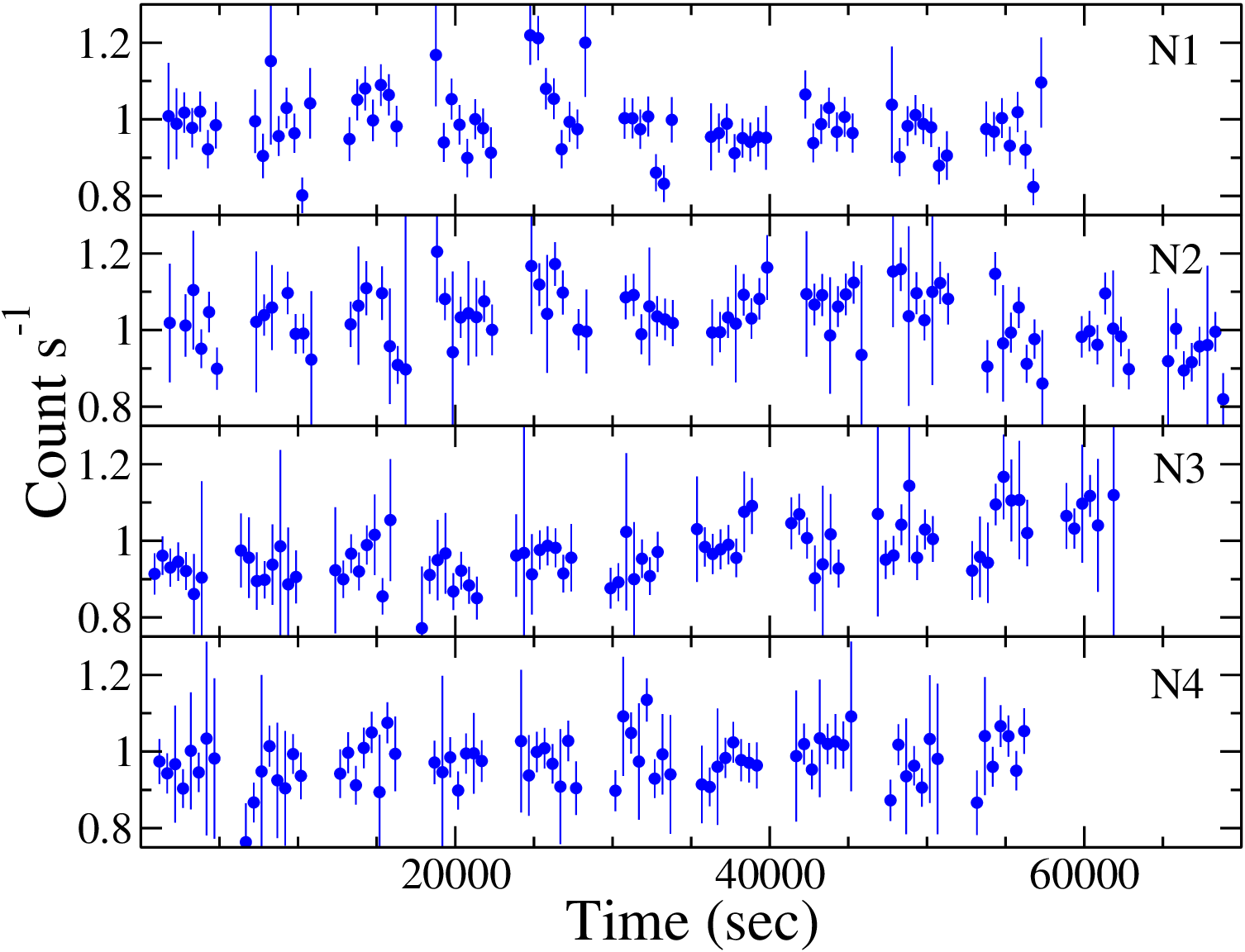}
\caption{Light curves of NGC 4507 in $3-60$~keV energy ranges for the observation N1, N2, N3 and N4. Each points represent 500~s.}
\label{fig:lc}
\end{figure}

\label{sec:timing}
\section{Timing Analysis}
We generated lightcurves in different energy ranges to study the variability in NGC~4507. Figure~\ref{fig:lc} shows the lightcurves in $3-60$~keV energy range for observations N1, N2, N3 and N4. We calculated the fractional rms variability ($F_{\rm var}$) to study the variability of the source \citep{Nandra1997, Edelson2002,Vaughan2003}. During the observation N1,  we obtained $F_{\rm var} < 5$\% in $3-10$~keV energy range. We did not observe any variability in $3-10$~keV energy range in other observations. No variability was observed in $10-60$~keV and $3-60$~keV energy ranges. We also calculated the fractional rms variability in $\sim 35$ days timescale. We found the fractional rms variability, $F_{\rm var} < 3\%$ in 35 days timescale.

\label{sec:spec}
\section{Spectral Analysis}
We carry out the spectral analysis in the $3-60$~keV energy range in {\tt XSPEC} v12.10\footnote{\url{https://heasarc.gsfc.nasa.gov/xanadu/xspec/}} \citep{Arnaud1996}. For the spectral analysis, we used various spectral models, based on slab and torus geometries. In all models, we included \textsc{phabs} component for the Galactic absorption with abundances set to those of \citet{Anders1989} and considered the photoelectric absorption cross-section of \citet{Verner1996}. We fixed the Galactic column density to $N_{\rm H} = 6.8 \times 10^{20}$ \pcm~ \citep{HI4PI2016}. We set the cosmological parameters to H$_{\rm 0} = 70$ \kms~ Mpc$^{-1}$, $\Lambda_{\rm 0}=70$, $\sigma_{\rm M} = 0.27$ \citep{Bennett2003}. We calculated uncertainties of all spectral parameter at the 90\% confidence level (1.6 $\sigma$).

\begin{figure*}
\centering
\includegraphics[width=8.4cm]{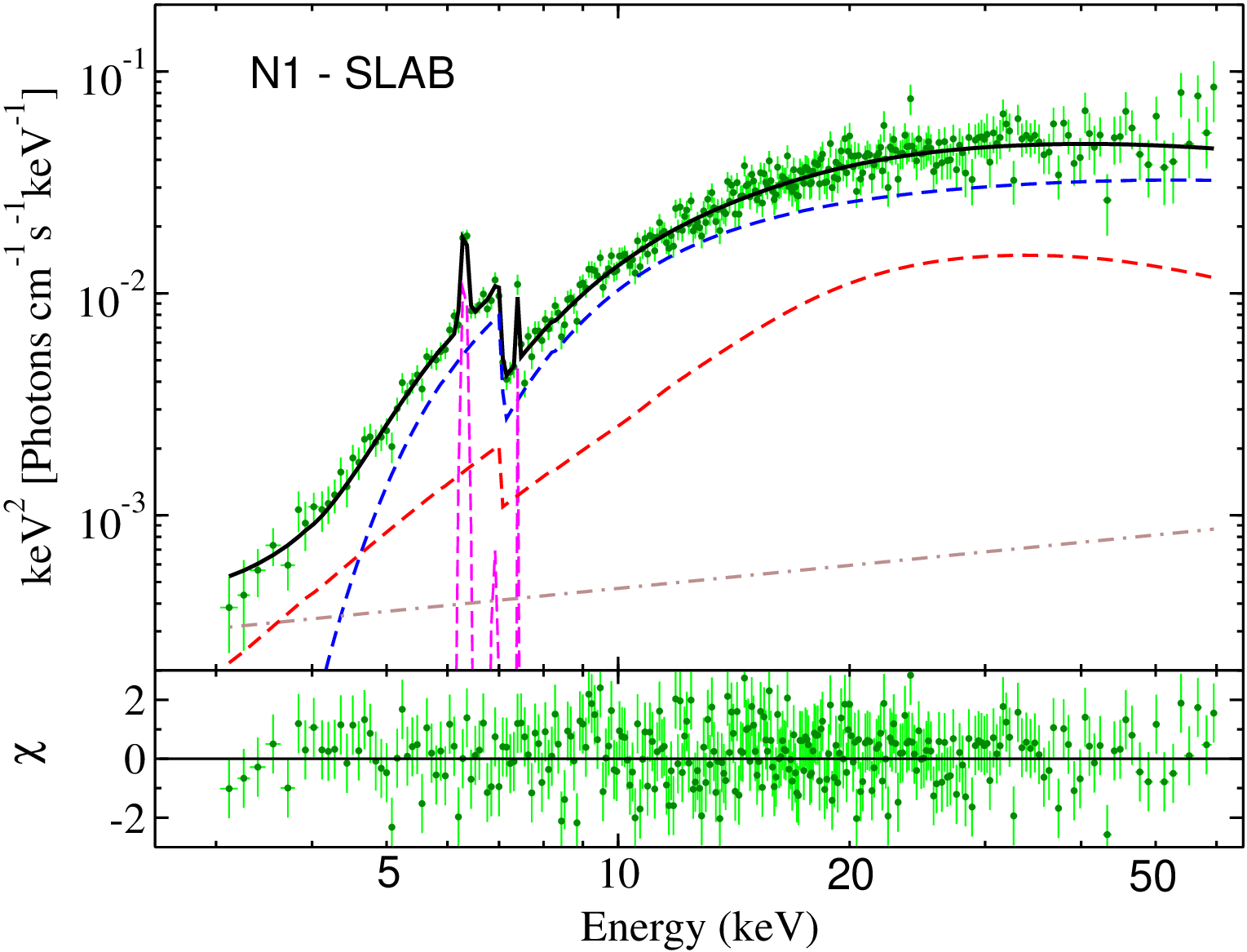}
\includegraphics[width=8.4cm]{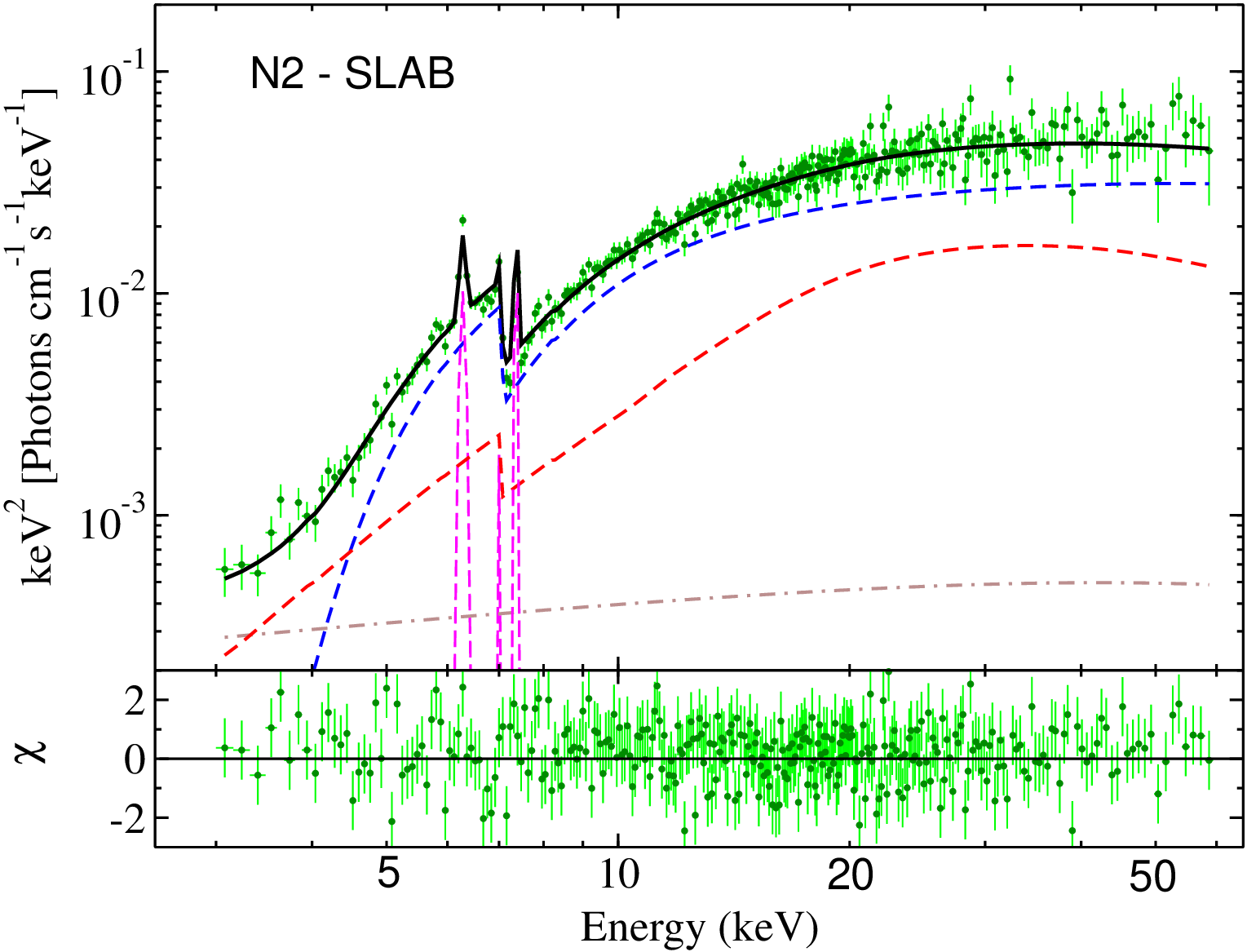}
\includegraphics[width=8.4cm]{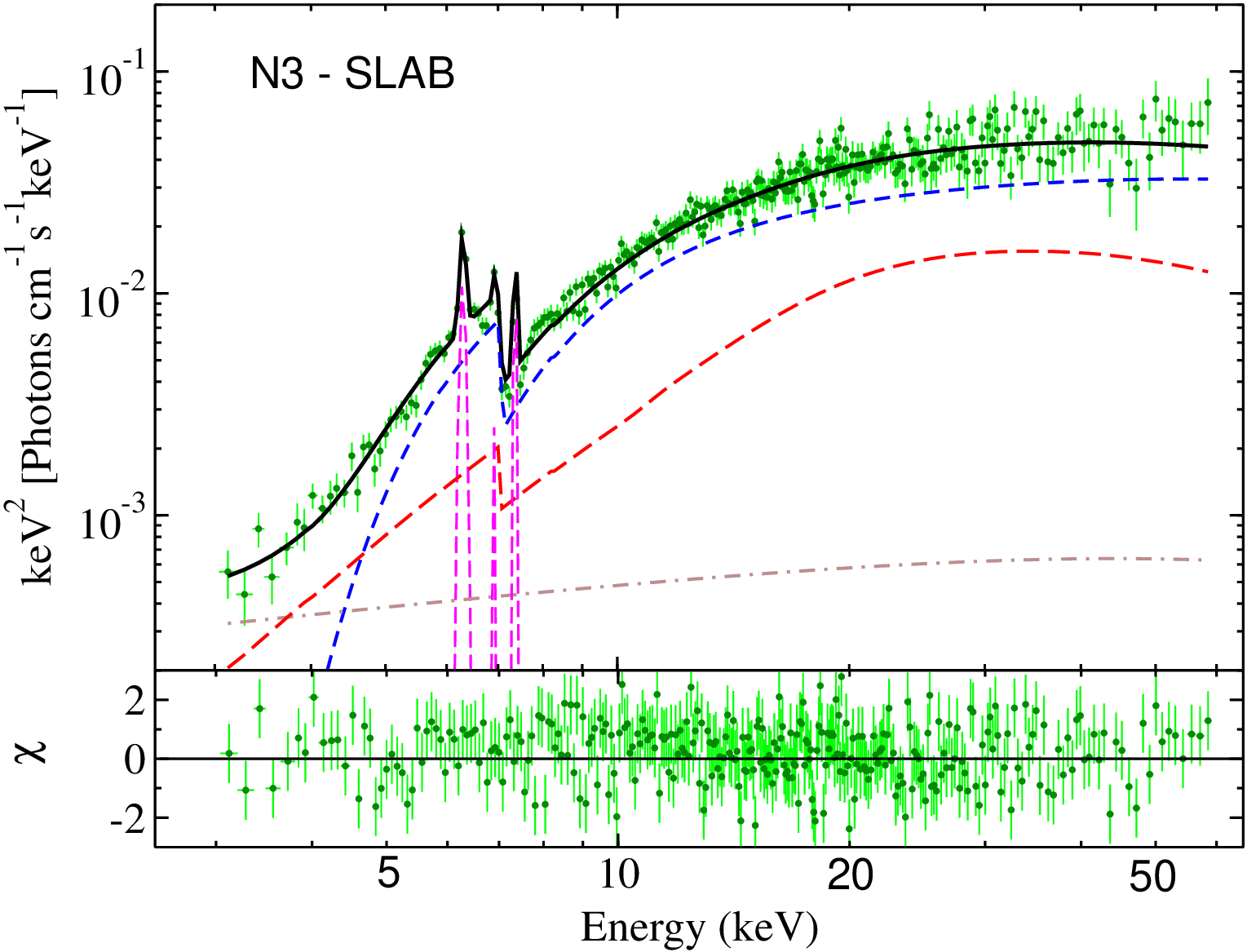}
\includegraphics[width=8.4cm]{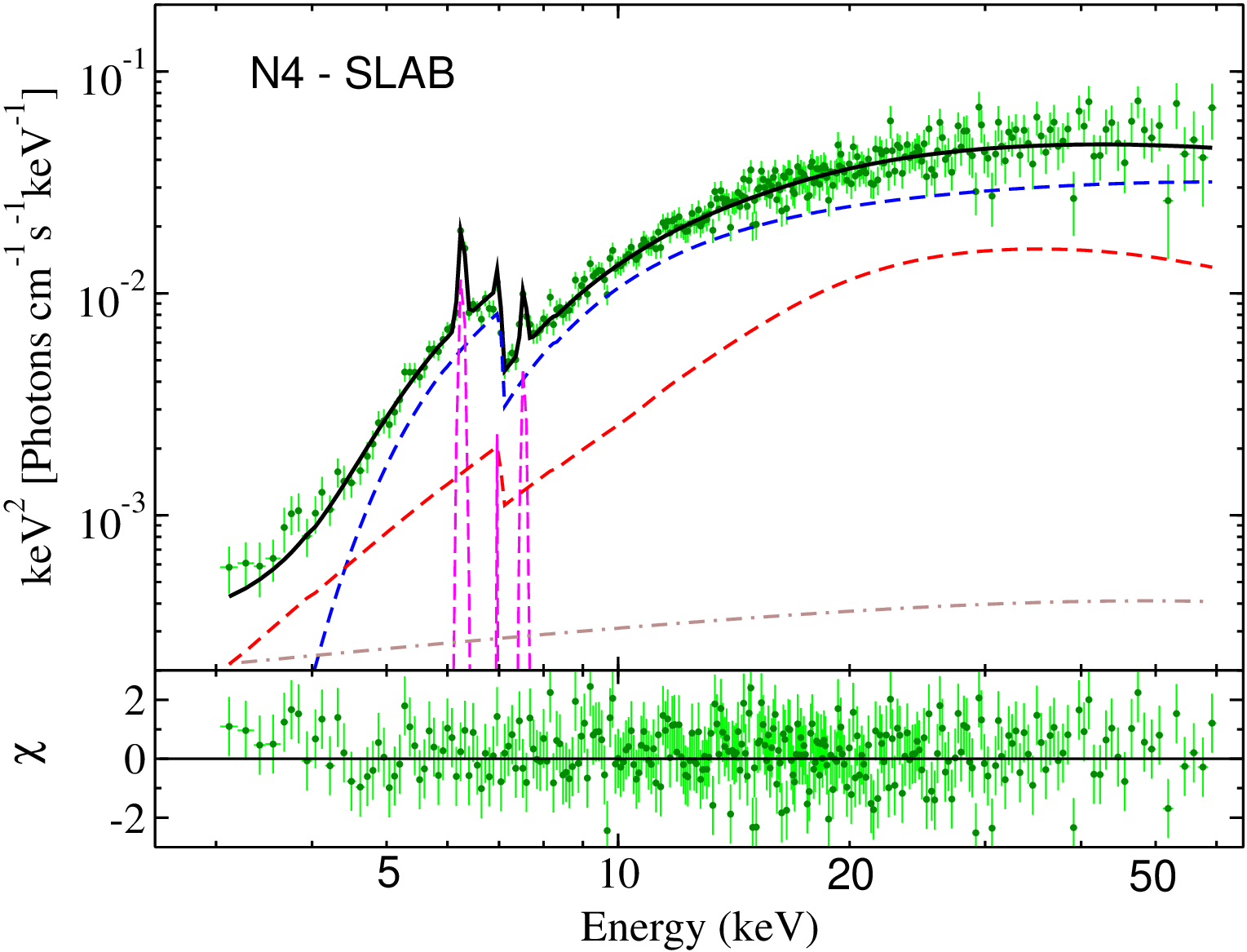}
\caption{Unfolded spectra fitted with the slab model for observations N1 (top left), N2 (top right), N3 (bottom left) and N4 (bottom right). Upper panel : Green points represent the data. The black, blue, red, magenta and brown lines represent the total emission, primary emission, reprocessed emission, line emission (Fe K$\alpha$, Fe K$\beta$ and Ni K $\alpha$), and scattered primary emission. Bottom panel: Corresponding residuals.}
\label{fig:pl-spec}
\end{figure*}

\label{sec:PL}
\subsection{Slab Model}
We started our spectral analysis with a simple absorbed power-law model with exponential cutoff \textsc{(zphabs1*cabs*zcutoffpl1)}. Here, \textsc{zphabs1} and \textsc{cabs} represent the line-of-sight absorption due to the photo-electric effect and Compton scattering, respectively. The hydrogen column density of \textsc{zphabs1} and \textsc{cabs} were tied to have the same value. The \textsc{zcutoffpl1} represents the primary continuum emission. This model did not provide an acceptable fit ($\chi^2=921$ for 598 degrees of freedom or dof). We, therefore, added four components, representing the reprocessed emission (modelled with \textsc{pexrav} \citealp{MZ98}) and three Gaussian lines, representing, Fe K$\alpha$, Fe K$\beta$, and Ni K$\alpha$ lines. The \textsc{pexrav} model describes the reprocessed X-ray emission from a cold, neutral, semi-infinite slab. We set the reflection fraction ($R_{\rm refl}$) to have a negative value, so that \textsc{pexrav} would only represent the reflection component. We linked the photon index, cutoff energy and normalization of \textsc{pexrav} to the values of the primary continuum. We set the iron abundance to one and the inclination angle to $i=60$\textdegree. Along with this, a scattered component is also observed in obscured AGNs \citep{Turner1997,Ueda2007,Gupta2021}. Hence, we included one additional power-law component multiplied by a constant (\textsc{const*zcutoffpl2}). We linked the photon index ($\Gamma$) and normalization of this power-law component (\textsc{zcutoffpl2}) to the values of the primary continuum (\textsc{zcutoffpl1}). The `\textsc{constant}' represents the scattering fraction ($f_{\rm Scat}$). The line width of the Gaussian components were fixed to 50~eV, 10~eV, and 10~eV for Fe K$\alpha$, Fe K$\beta$, and Ni K$\alpha$, respectively. The final model reads in {\tt XSPEC} as

\textsc{phabs1 * (zphabs2*cabs*zcutoffpl1 + 3*zgauss + pexrav + const*zcutoffpl2)}.

\begin{figure}
\centering
\includegraphics[angle=270,width=8.5cm]{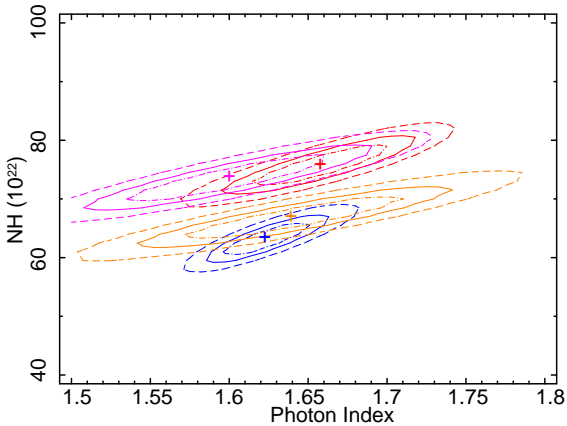}
\caption{Confidence contour between the photon index ($\Gamma$) and line of sight column density (\nh) in $10^{22}$ \pcm, fitted with the slab model. The red, blue, magenta and orange lines represent the the observation from N1, N2, N3 and N4, respectively. The dot-dashed, solid and dashed lines represent the contours at 1 $\sigma$, 2 $\sigma$ and 3 $\sigma$ level, respectively.}
\label{fig:cntr-pl}
\end{figure}

Here, \textsc{phabs1} represent the Galactic absorption in the direction of the source. This model (hereafter the slab model) provided a good fit for all observations. The results show that the X-ray source is highly obscured during all observations, with \nh~ varying in the range $6.3\pm 0.4-7.6\pm 0.6 \times 10^{23}$ \pcm. The photon index was found to be roughly constant ($\Gamma \sim 1.6-1.7$). The cutoff energy ($E_{\rm cut}$) was obtained to be constant ($E_{\rm cut} \sim 121^{+107}_{-41}-135^{+58}_{-73}$~keV) within the uncertainty. We detected strong Fe K$\alpha$ line emission in all four epochs, with a equivalent width (EW) of $237 \pm 7$~eV, $203 \pm 7$~eV, $235\pm 8$~eV, and $233\pm 8$~eV in observations N1, N2, N3, and N4, respectively (see Table\,\ref{tab:log}). The reflection was found be moderate ($R_{\rm refl} \sim 0.4$). The results obtained with this model are reported in Table~\ref{tab:pl}. To test if the variability of \nh~ is real and not associated to a degeneracy with the continuum parameters ($\Gamma$ and $E_{\rm cut}$), we fitted all the spectra simultaneously with $\Gamma$ and $E_{\rm cut}$ tied together. Using this approach, we found a similar variability of \nh~. We show the best-fitted unfolded spectra obtained with the slab model in the left panel of Figure~\ref{fig:pl-spec}, while the corresponding residuals are shown in the right panel of Figure~\ref{fig:pl-spec}. In the left panel of Fig~\ref{fig:pl-spec}, the black, red, green and blue solid lines represent the best-fitted slab model for N1, N2, N3 and N4, respectively, while the black, red, green and blue point represent the data for the observation N1, N2, N3 and N4, respectively. Figure~\ref{fig:cntr-pl} shows the confidence contour between the photon index ($\Gamma$) and line of sight column density (\nh).

\begin{table*}
\centering
\caption{The Slab model fitted spectral analysis result.}
\label{tab:pl}
\renewcommand{\arraystretch}{1.5} 
\setlength{\tabcolsep}{10pt} 
\begin{tabular}{lcccccccc}
\hline
 & &  N1 & N2 & N3 & N4 \\
\hline
(1) & \nhl ($10^{23}$ \pcm)&$7.6^{+0.5}_{-0.6}   $& $6.3^{+0.5}_{-0.5}  $ & $7.4^{+0.5}_{-0.4}    $  & $6.7^{+0.3}_{-0.5}  $ \\ 
(2) & $\Gamma$&$1.65^{+0.06}_{-0.05}  $& $1.62^{+0.05}_{-0.04}    $& $1.61^{+0.05}_{-0.05}    $  & $1.64^{+0.07}_{-0.06}     $ \\
(3) & $E_{\rm cut}$ (keV) & $121^{+107}_{-41}$&  $126^{+61}_{-37}$& $114^{+62}_{-21}$ & $135^{+58}_{-73}$ \\
(4) & $N_{\rm PL}$ ($10^{-2}$ \phc)&$2.20^{+0.06}_{-0.04}   $& $1.76^{+0.07}_{-0.08}    $& $1.88^{+0.10}_{-0.09}$  & $1.82^{+0.07}_{-0.06}  $ \\
(5) & $R_{\rm refl}$ & $0.38^{+0.08}_{-0.12}$ & $0.46^{+0.14}_{-0.23}$ & $0.29^{+0.10}_{-0.14}$ & $0.41^{+0.13}_{-0.24}$ \\
(6) & $f_{\rm Scat}$ (10$^{-2}$)& $0.97^{+0.04}_{-0.03}$  & $0.97^{+0.02}_{-0.02}$ & $1.21^{+0.02}_{-0.02}$ &$0.81^{+0.02}_{-0.02}$ \\
(7) & Fe K$\alpha$ ~~~LE (keV)&$6.38^{+0.04}_{-0.04}   $& $6.35^{+0.04}_{-0.04}    $& $6.37^{+0.04}_{-0.03}    $  & $6.33^{+0.06}_{-0.05}     $ \\
(8) & ~~~~~~~~~~~~~~~EW (eV)&$237^{+6}_{-7} $ & $203^{+6}_{-7}    $ & $235^{+6}_{-8} $  & $233^{+7}_{-8}   $ \\
(9) & ~~~~~~~~~~~~~~~Norm ($10^{-4}$ \phc)&$3.15^{+0.93}_{-0.68}$& $2.15^{+0.42}_{-0.61}$& $2.93^{+0.45}_{-0.62}$  & $2.61^{+0.45}_{-0.68} $ \\ 
(10)& Fe K$\beta$ ~~~LE (keV)&$6.99^{+0.09}_{-0.07}   $& $7.08^{+0.05}_{-0.06}    $& $6.98^{+0.09}_{-0.07}    $  & $7.04^{+0.07}_{-0.08}     $ \\
(11)& ~~~~~~~~~~~~~~~EW (eV)&$<24  $& $<26    $& $<31 $  & $<23    $ \\
(12)& ~~~~~~~~~~~~~~~Norm (10$^{-6}$ \phc)&$2.06^{+1.04}_{-1.24}$& $1.91^{+1.02}_{-0.88} $& $4.34^{+1.01}_{-0.97}$  & $3.87^{+1.53}_{-2.13}     $ \\
(13)& Ni K$\alpha$ ~~~LE (keV)&$7.47^{+0.10}_{-0.07}   $& $7.45^{+0.17}_{-0.13}    $& $7.45^{+0.11}_{-0.16}    $  & $7.62^{+0.07}_{-0.10}     $ \\
(14)& ~~~~~~~~~~~~~~~EW (eV)&$<48 $& $<35  $& $<32 $  & $<105    $ \\
(15)& ~~~~~~~~~~~~~~~Norm (10$^{-5}$ \phc)&$8.77^{+1.34}_{-1.53}$& $2.38^{+1.35}_{-1.43} $& $1.83^{+0.71}_{-0.88} $  & $10.01^{+1.45}_{-1.85} $ \\
(16)& $\chi^2$/dof&$632/590$    & $620/625$& $615/613$  & $567/586$ \\
\hline
(17)& $F_{\rm 2-10}^{\rm obs}$ (10$^{-11}$ \ecs) &$0.96_{-0.03}^{+0.03}$ & $1.06_{-0.03}^{+0.02}$&  $0.94_{-0.03}^{+0.04}$& $0.99_{+0.02}^{-0.01}$ \\
(18)& $L_{\rm 2-10}^{\rm intr}$ ($10^{43}$ \eps) &$3.70_{-0.15}^{+0.18}$ & $3.23_{-0.15}^{+0.17}$&  $3.62^{+0.14}_{-0.16}$ & $3.30^{+0.12}_{-0.14}$ \\
(19)& $L_{\rm 0.1-100}^{\rm intr}$ ($10^{44}$ \eps)&$1.21_{-0.02}^{+0.03}$ & $1.04_{-0.03}^{+0.02}$&  $1.24_{-0.03}^{+0.04}$& $1.10_{-0.03}^{+0.04}$ \\
(20)& $L_{\rm K\alpha}$ (10$^{42}$ \ecs) &$0.99_{-0.07}^{+0.09}$ & $0.95_{-0.07}^{+0.08}$&  $0.97_{-0.12}^{+0.10}$& $0.96_{-0.10}^{+0.08}$ \\
\hline
\end{tabular}
\leftline{(1) Line of sight hydrogen column density (\nhl) in $10^{23}$ \pcm, (2) photon index ($\Gamma$) of the primary emission, (3) cut-off energy ($E_{\rm cut}$) in keV, (4) power-law }.
\leftline{normalization ($N_{\rm PL}$) in 10$^{-2}$\phc, (5) reflection fraction ($R_{\rm refl}$), (6) fraction of scattered primary emission ($f_{\rm Scat}$), (7) Fe K$\alpha$ line energy in keV, }
\leftline{(8) equivalent width of the Fe K$\alpha$ line in eV, (9) normalization of the Fe K$\alpha$ line in $10^{-4}$ \phc, (10) Fe K$\beta$ line energy in keV, (11) equivalent width }
\leftline{of the  Fe K$\beta$ line in eV, (12) normalization of the Fe K$\beta$ line in $10^{-6}$ \phc, (13) Ni K$\alpha$ line energy in keV, (14) equivalent width of the Ni K$\alpha$ line in}
\leftline{eV, (15) normalization of the Ni K$\alpha$ line in $10^{-5}$ \phc, (16) $\chi^2$ value for degrees of freedom (dof), (17) $2-10$~keV observed flux $F_{\rm 2-10}^{\rm obs}$ in 10$^{-11}$}
\leftline{\ecs, (18) $2-10$~keV intrinsic luminosity $L_{\rm 2-10}^{\rm intr}$ in $10^{43}$ \eps, (19) $0.1-100$~keV intrinsic luminosity ($L_{\rm 0.1-100}^{\rm intr}$) in $10^{44}$ \eps, (20) Fe K$\alpha$ line }
\leftline{luminosity ($L_{\rm K\alpha}$) in $2-10$~keV energy ranges in 10$^{42}$ \ecs.}
\end{table*}

\label{sec:MYTC}
\subsection{MYTORUS}
The \textsc{pexrav} model considers reflection from a semi-infinite slab. Hence, it might not provide an accurate representation of the reprocessed radiation in obscured AGN. Thus, to probe the complex absorber, one should consider a more physical torus model. For our spectral analysis, we used the physically-motivated torus model \textsc{mytorus}\footnote{\url{http://www.mytorus.com/}} \citep{MY2009,Yaqoob2012}. This model consists of an absorbing torus, surrounding the X-ray source, with a fixed opening angle of 60$^{\circ{}}$ (i.e., a covering factor of 0.5). \textsc{mytorus} has three spectral components: the zeroth ordered component (\mytz), a scattered/reprocessed component (\myts), and a line component (\mytl). The \mytz~ component describes the absorbed transmitted continuum emission in the line-of-sight. The \myts~ component describes the reprocessed emission from the surrounding torus. The relative normalization ($A_{\rm S}$) of the \myts~ component is estimated using a `\textsc{constant}' in {\tt XSPEC}. The \mytl~ component describes the Fe K$\alpha$ and Fe K$\beta$ line emission. The relative normalization of this component (A$_{\rm L}$) is set to be the same as the relative normalization (A$_{\rm S}$) of the \myts~ component. Any deviation of $A_{\rm S}$ from unity could indicate a time-delay between \mytz~ and \myts~ components, or indicate different geometries of the material with different \nh, or a torus covering factor different than 0.5 \citep{Yaqoob2012}. The \textsc{mytorus} model can be used in two configurations: coupled (\mytc) and decoupled (\mytd). The coupled configuration describes an uniform torus, while the decoupled configuration could be used to describe a clumpy torus \citep{Yaqoob2012}.

\subsubsection{Coupled configuration}
We started our analysis with \textsc{mytorus} model using the coupled configuration. In {\tt XSPEC} the model reads as:

\textsc{phabs * (zpowerlaw1*MYTZ + const1*MYTS + const2*gsmooth*MYTL + const3*zpowerlaw2 + zgauss)}.

In this model, the photon index ($\Gamma$), equatorial hydrogen column density (\nheq), inclination angle ($i$) and normalization of \myts~ and \mytl~ components are tied to those of the \mytz~ component. As recommended, the relative normalization of \myts~ and \mytl~ components are tied, i.e. $A_{\rm S} = A_{\rm L}$. The photon index and normalization of the scattered component (\textsc{zpowerlaw2}) are tied to those of the primary continuum (\textsc{zpowerlaw1}). We also added a Gaussian component for Ni K$\alpha$ line emission, and used a Gaussian convolution model \textsc{gsmooth} to convolve the \mytl~ component.

We fitted all the four spectra simultaneously, setting the inclination angle ($\theta_{\rm incl}$) to be the same. We found that $i = 61.8^{+3.7}_{-0.9}$ degrees, in agreement with the Seyfert~2 classification of the source \citep[e.g.,][]{Beckmann2012}. The equatorial column density was obtained to be \nheq $= 2.3^{+0.6}_{-0.9}$, $2.0^{+0.7}_{-0.8}$, $2.5^{+0.7}_{-0.7}$, and $2.0^{+0.6}_{-0.8} \times 10^{24}$ \pcm, for N1, N2, N3, and N4 observations, respectively. Here, we must note that the variation of equatorial column density (\nheq) is not plausible on timescales of months, and one may need to include an extra varying absorber to address this \citep[e.g.,][]{Ricci2016}. The line-of-sight column density is defined as \citep{MY2009}, 

\begin{equation}
N_{\rm H}^{\rm los} = N_{\rm H}^{\rm eq} [1 - 4 \cos ~i]^{1/2}.
\end{equation}

The line-of-sight column density (\nhl) is found to be $7.5^{+4.2}_{-2.2}$, $6.5^{+2.7}_{-1.4}$, $8.1^{+2.9}_{-1.9}$, and $6.5^{+2.4}_{-1.4} \times 10^{23}$ \pcm~ for the N1, N2, N3, and N4 observation, respectively. The photon index was obtained to be roughly constant ($\Gamma \sim 1.7$). The relative normalization ($A_{\rm S}$) was observed to deviate from unity in all four observations, indicating delayed reprocessed emission or different geometries of the absorbing material with different \nh, or a different geometry from that assumed by {\tt mytorus}. The parameters obtained by using this model are reported in Table~\ref{tab:mytc}. Figure~\ref{fig:mytc-spec} shows the best-fit spectrum obtained with the \mytc~ model. The confidence contour between \nheq~ and $\Gamma$ are shown in Figure~\ref{fig:cntr-mytc}.

\label{sec:MYTD}
\subsubsection{Decoupled configuration}
\mytc~ considers an uniform obscuring torus, however, the torus is likely to be clumpy, and the \mytc~ model does not have the flexibility to accurately capture the variations in the line-of-sight column density. Thus, we used the decoupled configuration of \textsc{mytorus} (\mytd), which mimics a clumpy torus. This can be achieved by decoupling the column density of the \myts~ component from that of the \mytz~ component. In this configuration, we set the inclination angle of \myts~ and \mytl~ components to 0\textdegree ~to mimic the backside reflection from the far side material. Thus, the column density represents the global averaged column density of the obscured material (\nht). The inclination angle of the \mytz~ component is instead set to 90\textdegree. \mytz~ represents the direct component, and its column density is the line of sight column density (\nhl) \citep{Yaqoob2012}. The model reads in {\tt XSPEC} as,

\textsc{phabs*( zpowerlaw*MYTZ(90) + const1*MYTS(0) + const2*gsmooth*MYTL(0) + const3*zpowerlaw + zgauss)}.

The spectral analysis, performed with the \mytd~ model, provided a good fit for all observations. We found, however, that the averaged column density of the obscured materials (\nht) varied in the range of $\sim 1.9-2.4 \times 10^{23}$ \pcm. Within the observation period of $\sim 4$ months, the global obscuration properties is unlikely to change. Thus, we fitted all four spectra simultaneously with \nht tied together, and obtained \nht $=2.2^{+0.6}_{-0.5} \times 10^{23}$ \pcm~. The line of sight column density varied in the range of $7.0\pm 1.4-8.2\pm 1.3 \times 10^{23}$ \pcm. The detailed spectral analysis result using \mytd~ model is given in Table~\ref{tab:mytd}. \mytd~ model fitted unfolded spectra are shown in Figure~\ref{fig:mytd-spec}. Figure~\ref{fig:cntr-mytd} shows the confidence contour between \nht~ and \nhl for all observations, fitted with \mytd.

\label{sec:borus}
\subsection{Borus02}
The \borus\footnote{\url{https://sites.astro.caltech.edu/~mislavb/download/}}~ model consists of a spherical homogeneous torus with two polar cutouts in conical shape \citep{Balokovic2018}. Unlike the \textsc{mytorus} model, the opening angle of the torus is a free parameter in \borus. The \borus~ model also allows to separate the line of sight column density (\nhl) from the torus/obscuring material column density (\nht). The model setup with \borus~ model reads in {\tt XSPEC} as,

\textsc{phabs1*(zphabs2*cabs*zcutoffpl1 + const1*borus02 + const2*cutoffpl2)}.

\textsc{zphabs2*cabs*zcutoffpl1} represents the absorbed direct primary emission. `\textsc{const1}' represents the relative normalization ($A_{\rm S}$) of the reprocessed component. \textsc{const2*cutoffpl} represents the scattered primary emission while \textsc{const2} is the scattering fraction ($f_{\rm Scat}$). The photon index ($\Gamma$), cutoff energy ($E_{\rm cut}$), normalization of \textsc{cutoffpl1}, \textsc{zcutoffpl2}, and \borus~ model are linked together. The column densities of \textsc{cabs} and \textsc{zphabs2} models are tied together, and represents the line of sight column density obscuration.

We simultaneously fitted all four spectra with \borus~ model with \nht~, inclination angle ($i$), iron abundance ($A_{\rm Fe}$) and torus covering factor ($C_{\rm tor}$) tied together. First, we fitted the spectra with a fixed cutoff energy at $E_{\rm cut} = 400$~keV and iron abundance at Solar value ($A_{\rm Fe}=1$). We allowed the cutoff energy to vary, and the fit improved by $\Delta \chi^2 = 8$ for 1 dof. The fit improved significantly ($\Delta \chi^2 = 34$ for 1 dof) when we allowed the Fe abundance ($A_{\rm Fe}$) to vary, which resulted in a sub-Solar value in all four observations, with $A_{\rm Fe} \sim 0.47 \pm 0.07$. We found a column density of the obscuring material of \nht $=2.6^{+0.7}_{-0.6} \times 10^{23}$ \pcm. The line of sight column density was found to be $9.4^{+0.6}_{-0.6} \times 10^{23}$, $8.3^{+0.7}_{-0.6} \times 10^{23}$, $9.8^{+0.6}_{-0.6} \times 10^{23}$, and $8.4^{+0.5}_{-0.5} \times 10^{23}$ \pcm, for N1, N2, N3 and N4, respectively. We obtained the inclination angle to be $i = 64.5^{+5.2}_{-6.3}$\textdegree. The torus covering factor is found to be $C_{\rm tor}=0.58 \pm 0.10$ with the torus opening angle to be obtained in the range of $\sim 47$\textdegree--$60$\textdegree. The results obtained with this fit are reported in Table~\ref{tab:borus}. \borus~ model fitted unfolded spectra are shown in Figure~\ref{fig:borus-spec}. The confidence contour between \nht~ and \nhl~ is shown in Figure~\ref{fig:cntr-borus} for all observations, fitted with \borus.

\label{sec:xclumpy}
\subsection{XCLUMPY}
Next, we used the \xclumpy\footnote{\url{https://github.com/AtsushiTanimoto/XClumpy}}~ model \citep{Tanimoto2019,Tanimoto2020} to fit the {\it NuSTAR} spectra of \source. The model geometry is based on the CLUMPY infrared model developed by \citet{Nenkova2008a,Nenkova2008b}. This model assumes a power-law distribution of clumps in the radial direction between inner and outer radii, and a normal distribution in the elevation direction. The free parameters of this model are the equatorial column density (\nheq), the torus angular width ($\sigma_{\rm tor}$), and the inclination angle ($i$). From the equatorial column density, one can easily calculate the line-of-sight column density using the following equation \citep{Tanimoto2019},

\begin{equation}
\label{eqn:xclumpy}
N_{\rm H}^{\rm los} = N_{\rm H}^{\rm Eq} [\exp ( - \frac{(i - \pi/2)^2}{\sigma_{\rm Tor}^2})]. 
\end{equation}

The \xclumpy~ model has two components, \textsc{xclumpy$\_$R} and \textsc{xclumpy$\_$L}, representing the reprocessed and line emission, respectively. The full model reads in {\tt XSPEC} as,

\textsc{phabs1*(zphabs2*cabs*zcutoffpl1 + xclumpy$\_$R + const1*xclumpy$\_$L + const2*zcutoffpl2)}.

The first term is the direct primary emission. The second and third terms represent the reprocessed and line emission, respectively. The last term represents the scattered primary emission. The parameters of \textsc{xclumpy$\_$R} and \textsc{xclumpy$\_$L} models are linked. The photon index ($\Gamma$), cutoff energy ($E_{\rm cut}$) and normalization of \textsc{xclumpy$\_$R} model are linked with the \textsc{cutoffpl1}. The \textsc{const1} and \textsc{const2} represent the relative normalization of the line emission ($A_{\rm L}$) and scattering fraction ($f_{\rm Scat}$).

We fitted all the four spectra simultaneously using the \xclumpy~ model, with \nheq~ and $i$ tied together. We fitted the spectra with cutoff energy fixed at $E_{\rm cut} = 370$~keV, since the \textsc{xclumpy} table consider a fixed cut-off energy at 370~keV. We found that the inclination angle is $i = 64.1^{+7.4}_{-4.9}$\textdegree, while the equatorial column density is \nheq = $2.1^{+0.6}_{-0.9} \times 10^{24}$ \pcm. The line of sight density was obtained to be, \nhl = $8.1^{+2.4}_{-1.5}\times 10^{23}$, $7.0^{+1.5}_{-1.2} \times 10^{23}$, $8.2^{+1.5}_{-2.2} \times10^{23}$ and $7.1^{+1.5}_{-2.1} \times 10^{23}$, for N1, N2, N3, and N4, respectively. The torus angular width we obtained is roughly constant with $\sigma_{\rm tor} \sim$ 25--26\textdegree. The results of this spectral analysis are reported in Table~\ref{tab:xclumpy}. Figure~\ref{fig:xc-spec} shows the \xclumpy~ model fitted unfolded spectra.  Figure~\ref{fig:cntr-xclumpy} shows the confidence contour between the \nheq and $\sigma_{\rm tor}$, fitted with the \xclumpy~ model.

\label{sec:rxtorus}
\subsection{RXTORUS}
Next, we used \textsc{rxtorus}\footnote{\url{https://www.astro.unige.ch/reflex/}} model \citep{Paltani2017} for the spectral analysis. \textsc{rxtorus} is based on the ray-tracing code for X-ray reprocessing code \textsc{REFLEX}. This model includes absorption and reflection from the torus with varying torus covering factor. The covering factor is defined as the ratio of minor to major axis of the torus (r/R). The \textsc{rxtorus} model has three components : absorbed primary emission (\textsc{rxtorus$\_$cont}), scattered emission (\textsc{rxtorus$\_$scat}) and line emission (\textsc{rxtorus$\_$fluo}). The scattered emission and line emission are merged into a reprocessed component (\textsc{rxtorus$\_$rprc}). The model reads in {\tt XSPEC} as,

\textsc{phabs(rxtorus$\_$cont*cutoffpl + const1*RXTorus$\_$rprc + const2*cutoffpl)}.

Here, \textsc{const1} represents the relative normalization. `\textsc{const*cutoffpl}' is the scattered primary emission. The line of sight column density is give by,

\begin{equation}
\label{eqn:rx}
N_{\rm H}^{\rm los} = N_{\rm H}^{\rm Eq} ~\sqrt{1-(\frac{\cos~i}{r/R})^2}.    
\end{equation}

We fitted all four spectra simultaneously with the inclination angle and equatorial column density tied together. The normalization of the primary emission and \textsc{rxtorus$\_$rprc} components are tied together. The equatorial column density is found to be $N_{\rm H}^{\rm Eq}=2.1^{+0.6}_{-0.5} \times 10^{24}$ \pcm, while the line of sight column density varied between $\sim 6.0^{+0.9}_{-1.1} - 7.0^{+1.6}_{-1.8} \times 10^{23}$ \pcm. The covering factor (r/R) was obtained to be $\sim 0.41-0.42$. The results of the spectral analysis are presented in Table~\ref{tab:rxtorus}. Figure~\ref{fig:rx-spec} shows the \rxtorus~ model fitted unfolded spectra. The confidence contour between the \nheq and $\sigma_{\rm tor}$ is shown in Figure~\ref{fig:rx-cntr}.

\section{Discussion}
We presented the detailed spectral analysis result of \source~ using \nustar~ observations in 3--60 keV energy range. We carried out the spectral analysis with the slab model, \textsc{mytorus}, \borus, \xclumpy~ and \rxtorus~ models.

\begin{table}
\centering
\caption{Variation of line of sight column density (\nhl), obtained from different models}
\begin{tabular}{ccccccc}
\hline
ID & Slab & \mytc & \mytd & \borus & \xclumpy & \rxtorus \\
\hline
N1&$7.6^{+0.5}_{-0.6}$& $7.5^{+4.2}_{-2.2}$&$8.2^{+1.3}_{-1.3}$&$9.4^{+0.6}_{-0.6}$&$8.1^{+2.4}_{-1.5}$&$6.9^{+2.4}_{-1.6}$\\
N2&$6.3^{+0.5}_{-0.5}$& $6.5^{+2.7}_{-1.4}$&$7.0^{+1.4}_{-1.3}$&$8.3^{+0.7}_{-0.6}$&$7.0^{+1.6}_{-1.1}$&$6.0^{+0.9}_{-1.1}$\\
N3&$7.4^{+0.5}_{-0.4}$& $8.1^{+2.9}_{-1.9}$&$8.1^{+1.5}_{-1.5}$&$9.8^{+0.6}_{-0.6}$&$8.2^{+1.6}_{-2.3}$&$7.0^{+1.6}_{-1.8}$\\
N4&$6.7^{+0.3}_{-0.5}$& $6.5^{+2.4}_{-1.4}$&$7.1^{+1.3}_{-1.4}$&$8.4^{+0.5}_{-0.5}$&$7.1^{+1.5}_{-2.2}$&$6.1^{+1.5}_{-2.1}$\\
\hline
\end{tabular}
\leftline{\nhl~ is in unit of 10$^{23}$ \pcm.}
\label{tab:nh-model}
\end{table}

\subsection{Comparison among different Spectral Models}
\label{sec:comparison}
Using the results obtained by our X-ray spectral analysis, we explore the nuclear and obscuration properties of \source~. The main difference among different spectral models is how they treat the reprocessed emission. While \textsc{pexrav} assumed reflection from a cold, semi-infinite slab, physically motivated torus-based model assumed different torus structures and geometries. Therefore, the fits performed using the different models return slightly different results. As the absorber is likely not uniform, given the column density variability found here and in previous works (e.g., \citealp{Braito2013}), we do not discuss the results obtained considering the \mytc~ model. 

The spectral analysis carried out using different models resulted in different values of $\Gamma$. The slab model, \mytd~, \xclumpy~ and \rxtorus~ models returned with $\Gamma \sim 1.6-1.7$, while the \borus~ model indicated slightly flatter spectra with $\Gamma \sim 1.5.$ For all models, the photon index was roughly constant across the different observations.

The variation of \nhl~ was observed to be similar as obtained from different spectral models. The slab model showed that \nhl~ varied in the range of $\sim 6-8\times 10^{23}$ \pcm, while \mytd~ and \xclumpy~ showed that \nhl~ varied in the range of $7-9 \times 10^{23}$ \pcm. The \rxtorus, \borus~ and \xclumpy~ models also returned with similar value of \nhl, varying in the range of $\sim 6-9 \times 10^{23}$ \pcm. The \nhl~ variation, obtained from different spectral model, is listed in Table~\ref{tab:nh-model}.

\subsection{Nuclear Properties}
Our spectral analysis indicated very little variation of the photon index during the observations, and the parameter can be considered constant within the uncertainties. The cutoff energy is found to be in the range of $E_{\rm cut} \sim 121^{+107}_{-41}~{\rm keV}-135^{+58}_{-73}$~keV from the slab model and $75^{+29}_{-15}-97^{+46}_{-18}$~keV from \borus~ model. This is consistent with typical values found for nearby AGN (e.g., \citealp{Ricci2018,Balokovic2020}). Both models indicate a constant cutoff energy within our observations. The intrinsic luminosity in the 2--10 keV energy band is found to be in the range of $L_{\rm 2-10}^{\rm intr} \sim (3.0\pm 0.2 - 3.6\pm 0.3) \times 10^{43}$ \eps. \citet{Vasedevan2009} estimated the bolometric correction factor $\kappa_{\rm bol,2-10~keV} \approx 15-30$ for $\lambda_{\rm Edd} > 0.1$, and $\kappa_{\rm bol,2-10~keV} \approx 10$ for $\lambda_{\rm Edd} \leq 0.1$. Considering the bolometric correction factor $\kappa_{\rm bol,2-10~keV}=20$, we obtained, the bolometric luminosity in the range of $L_{\rm bol}= (5.9\pm 0.4 - 7.2\pm 0.5) \times 10^{44}$ \eps. The Eddington ratio would be $\lambda_{\rm Edd} = L_{\rm bol}/L_{\rm Edd} \sim 0.1$, considering the black hole mass, $M_{\rm BH} = 4.5 \times 10^7$ $M_{\odot}$ \citep{Marinucci2012}. However, \citet{Winter2009} reported of a higher mass ($M_{\rm BH} \sim 10^{8.4}$ $M_{\odot}$), which would lead to $\lambda_{\rm Edd} \sim 10^{-1.7}$. Even, if we consider $\kappa_{\rm bol} \sim 10$, $\lambda_{\rm Edd}$ would be $10^{-2}$. Regardless the assumption of mass or bolometric factor, the Eddington ratios of the source is consistent with that of nearby Seyfert galaxies \citep{Wu2004,Koss2017}.

\subsection{Obscuration properties}
From the spectral analysis, we obtained several torus parameters, such as the line of sight column density (\nhl; from the slab model, \mytc, \mytd, \borus~ \& \xclumpy), the averaged column density of the obscuring materials (\nht; from \mytd~ \& \borus), the equatorial column density (\nheq; from \mytc, \xclumpy~ \& \rxtorus), and the torus covering factor ($CF_{\rm tor}$ or $\theta_{\rm tor}$; from \borus, \xclumpy~ \& \rxtorus). We obtained similar variations of \nhl~ from different spectral models (see Section~\ref{sec:comparison}). The equatorial column density (\nheq) was found to be $\sim 2.1 \pm 0.6 \times 10^{24}$ \pcm, whereas the line of sight column density was found to be $\sim 6-8 \times 10^{23}$ \pcm.

The averaged column density of the obscuring material was obtained to be \nht $=2.2^{+0.6}_{-0.5}$ and $2.6^{+0.7}_{-0.6} \times 10^{23}$ \pcm~ from the \mytd~ and \borus~ model, respectively. \citet{Braito2013} applied the \mytd~ model to the \xmm, {\it Suzaku}, and {\it BeppoSAX} observations obtained between 1997 and 2007, and obtained \nht~ in the range of $\sim (2.4-3.5) \times 10^{23}$ \pcm, which is consistent with our findings. In our analysis, the angular dispersion of the torus is obtained as $\sigma_{\rm tor} \sim 24^{\circ{}}-26$\textdegree. One can easily estimate the torus covering factor ($C_{\rm tor}$) using Equation~\ref{eqn:xclumpy} \citep{Yamada2020}. Using $\theta=\pi/2 - i$, as the elevation angle, we can write Equation~\ref{eqn:xclumpy} as,

\begin{equation}
\label{eqn:covering}
N_{\rm H}^{\rm los} = N_{\rm H}^{\rm Eq} [ \exp ( - \frac{\theta}{\sigma_{\rm tor}})^2]. 
\end{equation}

For \nhl = $10^{22}$ \pcm, we obtained corresponding elevation angle, $\theta =$ 56\textdegree-- 61\textdegree. This transforms to the torus covering factor $C_{\rm tor} = \sin \theta \sim 0.83-87$. Considering Compton-thick obscuration, i.e. setting \nhl = $10^{24}$ \pcm~ in Equation~\ref{eqn:covering}, we obtained a covering factor $C_{\rm tor}^{\rm T} \sim 0.34-0.37$. Our X-ray spectral analysis with the \borus~ and \rxtorus~ models returned $C_{\rm tor} \sim 0.6$ and $\sim 0.4$.

\citet{Ricci2017} showed that the radiation pressure from the AGN could efficiently expel dusty gas when $\lambda_{\rm Edd} \gtrsim 10^{-1.5} $. They showed for the Compton thin material ($N_{\rm H} = 10^{22-24}$ \pcm), a high covering factor is observed ($C_{\rm tor}\sim 0.85$) for $10^{-4} \leq \lambda_{\rm Edd} \leq 10^{-1.5}$, while a much lower covering factor ($C_{\rm tor} \sim 0.4$) is observed for $\lambda_{\rm Edd} \geq 10^{-1.5}$. On the other hand, for the Compton-thick material (\nh~$>10^{24}$ \pcm), the covering factor is predicted to be $C_{\rm tor}^{\rm T} \sim 0.2-0.3$ \citep[see also,][]{Ricci2015}. We obtained an Eddington ratio of $\lambda_{\rm Edd} \sim 0.1$ for $M_{\rm BH} \sim 10^{7.65}$ $M_{\odot}$ \citep{Marinucci2012}. Our estimated torus covering factor is $C_{\rm tor} \sim 0.85$ which is higher than the predicted value for $\lambda_{\rm Edd} \sim 0.1$. However, if we consider a higher mass, $M_{\rm BH} = 10^{8.4}$~$M_{\odot}$ \citep{Winter2009}, the Eddington ratio is $\lambda_{\rm Edd} \sim 10^{-2}$. For this Eddington ratio, the covering factor predicted by \citet{Ricci2017} ($C_{\rm tor} \sim 0.85$) is consistent with the results we obtained from the fit with the \xclumpy~model. This seems to suggest a higher mass of the SMBH. However, the study of \citet{Ricci2017} was conducted using a large sample, that show a variation of covering factor and/or$\lambda_{\rm Edd}$ in a wide range. Hence, it will not be correct to prefer the higher mass for the SMBH solely from this. On the other hand, the covering factors for the Compton-thick material is obtained to be $C_{\rm tor}^{\rm T} \sim 0.35-0.37$, which is slightly higher than the value predicted by \citet{Ricci2017}.

The AGN absorber is complex and there may exist multiple absorbers along the line of sight. In a simpler scenario with two absorbers, one absorber can be considered as varying while the other absorber is non-varying. In IC~751, \citet{Ricci2016} were able to disentangle the column density of the varying clouds from the non-varying absorbers. In NGC~4507, \nhl~ changed about $\sim 15-20$\% in $\sim 35$~days timescale. This implies that $\sim 80-85$\% \nhl~ did not change and there may exist a non-varying absorber. We tried to disentangle the non-varying absorber from the varying one by adding an additional absorber during the spectral analysis. We allowed one absorber to vary and kept the other absorber tied across the observations. However, we were not able to disentangle two absorbers as we could not constraint the column density of two absorbers.

\begin{figure*}
\centering
\includegraphics[width=17cm]{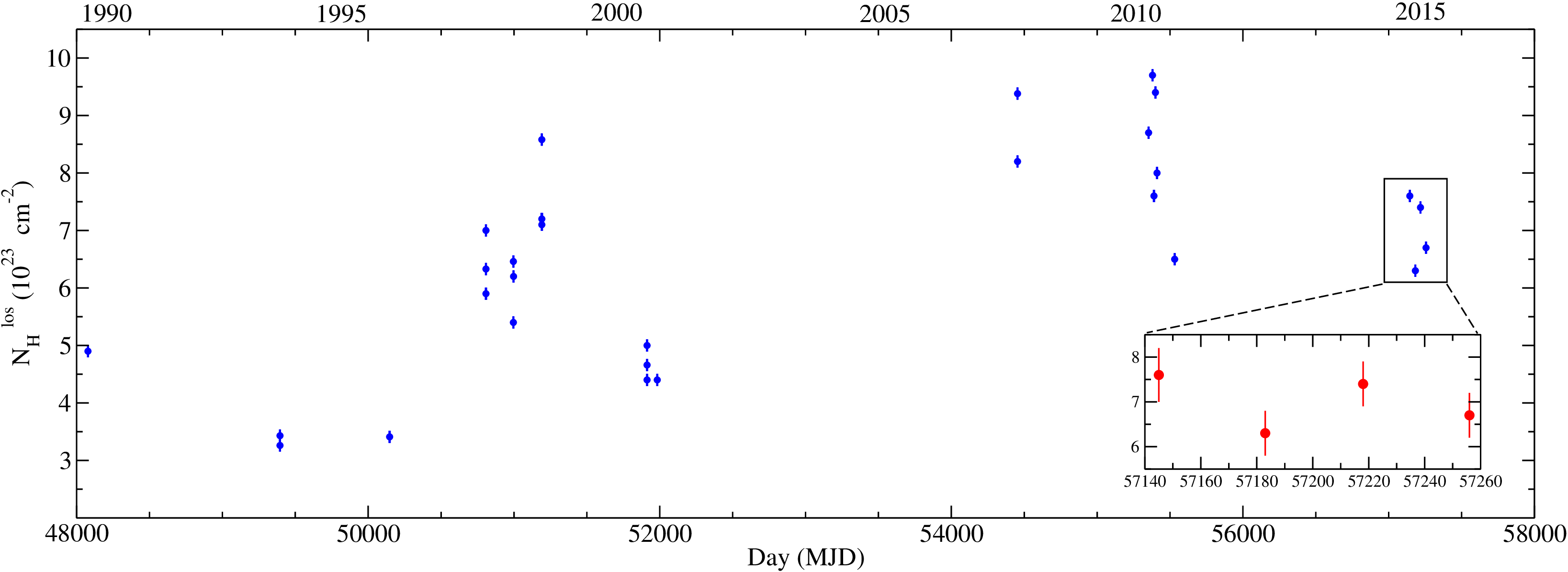}
\caption{Evolution of the line of sight column density (\nhl) in $10^{23}$ \pcm~ between 1990 and 2015. The blue points are taken from the work of \citet{Awaki1991,Comastri1998,Risaliti2002a,Matt2004,Marinucci2013,Braito2013}. The inset figure shows the variation of the spectral parameters for this work.}
\label{fig:nh}
\end{figure*}

\begin{figure}
\centering
\includegraphics[width=8.5cm]{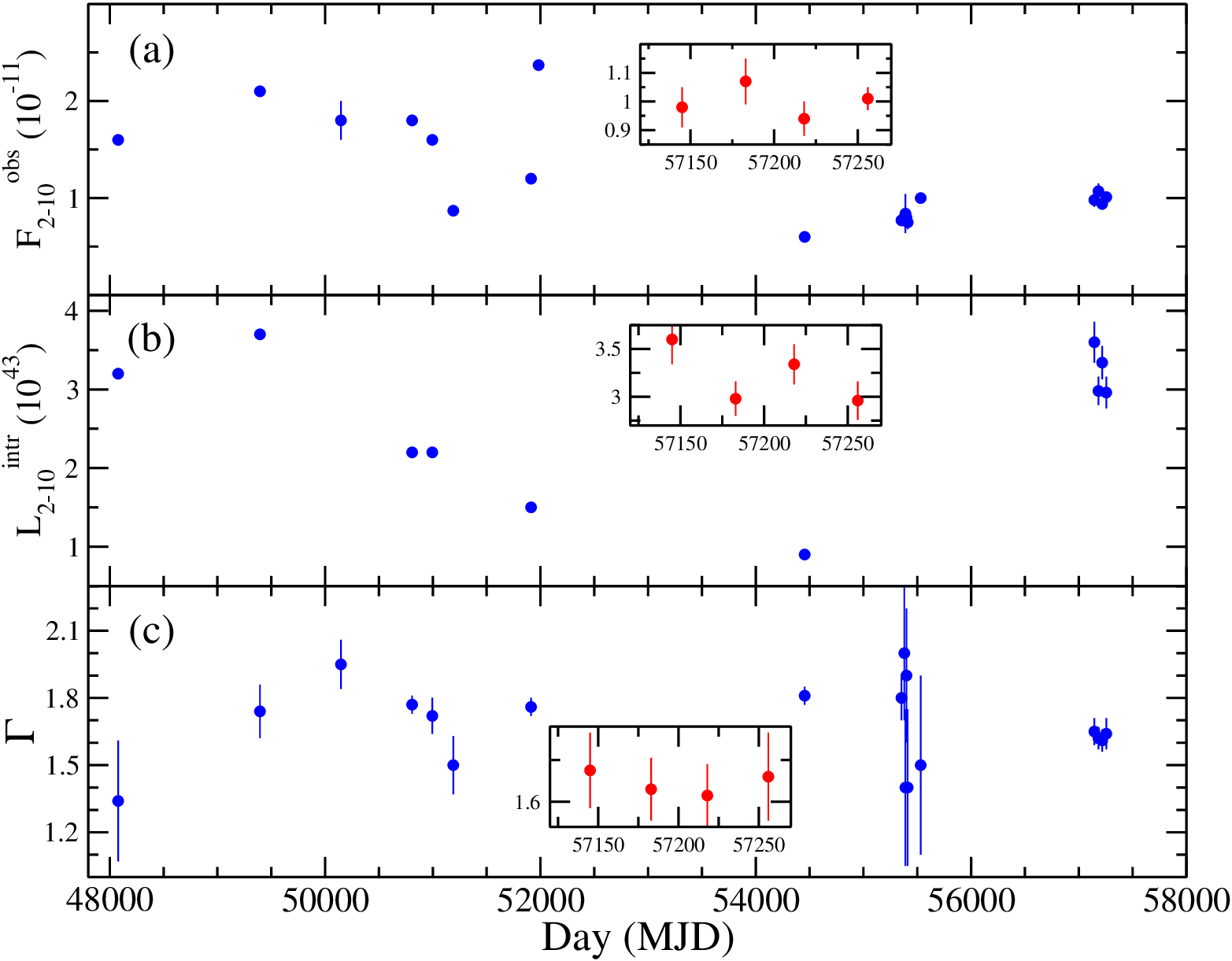}
\caption{Evolution of (a) the line of sight column density (\nhl) in $10^{23}$ \pcm, (b) the $2-10$~keV observed flux ($F^{\rm obs}_{\rm 2-10}$) in $10^{-11}$ \ecs, (c) the $2-10$~keV intrinsic luminosity ($L^{\rm intr}_{\rm 2-10}$) in $10^{43}$ \eps, (d) the photon index ($\Gamma$) between 1990 and 2015. The blue points are taken from the literature. The inset figures in each panel shows the variation of the spectral parameters from this work.}
\label{fig:evo}
\end{figure}

\subsection{Location of the Reprocessing Clouds}
One can easily estimate the distance of the absorbing cloud from the central SMBH from the \nh~ variability \citep[for details, see][]{Risaliti2007,Marinucci2013,Ricci2016}. Assuming that the X-ray source size ($D_{\rm S}$) and cloud size ($D_{\rm C}$) are similar ($D_{\rm S} \simeq D_{\rm C}$), one can easily calculate the transverse velocity of the cloud as $V_{\rm K} = D_{\rm S}/T_{\rm cr}$, where $T_{\rm cr}$ is light-crossing time. Considering Keplerian motion of the cloud, the distance of the absorbing cloud from the central SMBH is given by,

\begin{equation}
R_{\rm C} = \frac{GM_{\rm BH} T_{\rm cr}^2}{D_{\rm S}}.
\end{equation}

The source size could be set to $D_{\rm S} \simeq 10~R_{\rm g}$ \citep{Marinucci2013}, where $R_{\rm g} = GM_{\rm BH}/c^2$ is gravitational radius. We, therefore, obtain

\begin{equation}
R_{\rm C}~\simeq~0.07~R_{\rm 10}^{-2}~M_{\rm 7}~T_{\rm 10}^2~{\rm pc.}
\end{equation}
Here, $R_{\rm 10} = D_{\rm S} /10 R_{\rm g}$, $M_{\rm 7}$ is in unit of $10^7~M_{\odot}$~ and $T_{\rm 10}$ is the crossing time in unit of 10 light-days. The total time span of each \nustar~ observation was $\sim 70$~ks, and no significant variability is observed on that timescale. Thus, the location of the obscuring material must be, $R_{\rm min} > 0.002$~ pc ($\sim 2.4$ light-days). \citet{Marinucci2013} did not observe variability in \nh~ in timescale of $\sim 1.5$~months, although they observed variability on timescales of $\sim 4$~months. From that, they concluded the location of the absorbing material to be $\sim 7-40$~ pc, i.e. farther from the SMBH than the BLR. In this work, we observed $N_{\rm H}$ variability on timescale of $\sim 35$ days which implies that the location of the absorbing clouds is $R_{\rm max} \leq 4$~pc, considering $M_{\rm BH}=10^{7.65}$ $M_{\odot}$. The different result for the location of the obscuring material could be due to different absorbers. Considering a mass of $M_{\rm BH} = 10^{8.4}$ $M_{\odot}$ \citep{Winter2009}, the absorbing cloud location would be $R_{\rm C} \sim 0.01-21$~pc.

Optical reverberation mapping studies have shown that the radius of the BLR scales with the squared root of X-ray luminosity. Considering the H$\beta$ lag, \citet{Kaspi2005} obtained

\begin{equation}
\label{eqn:blr}
\frac{R_{\rm BLR}}{\rm 10 ~lt-days} = 0.86 \times (\frac{L_{\rm 2-10}}{\rm 10^{43} ergs/s})^{0.53}.
\end{equation}

Considering the average intrinsic luminosity of the source as $L_{\rm X} \sim 3.2 \times 10^{43}$ \eps, we obtained, $R_{\rm BLR} \simeq 16$~lt-days ($\sim 0.013$~pc). This indicates that the absorbing cloud is located beyond the BLR. From infrared studies, the near and mid-IR emitting regions are found to scale as the squared root of the X-ray luminosity. From the reverberation mapping in the K-band, the inner edge of the torus ($R_{\rm NIR}$) is estimated to be \citep{Tristram2011}

\begin{equation}
\log(\frac{R_{\rm NIR}}{\rm 1 pc}) = -23.10 + 0.5 \log (L_{\rm 14-195}),
\end{equation}

where $L_{\rm 14-195}$ is the intrinsic X-ray luminosity in the 14--195~keV energy band. One can also estimate the radius of the mid-IR emitting region ($R_{\rm MIR}$; \citealp{Tristram2011}):

\begin{equation}
\log(\frac{R_{\rm MIR}}{\rm 1 pc}) = -21.62 + 0.5 \log (L_{\rm 14-195}).
\end{equation}

The 70-months averaged luminosity of \source~ is $\log (L_{\rm 14-195}) = 43.96$ \citep{Ricci2017apjs}. Using this, we obtained, $R_{\rm NIR} \simeq 0.08~{\rm pc} \simeq 95$~light-days, and $R_{\rm MIR} \simeq 2.32~{\rm pc} \simeq 2760$~light-days. 

From the above calculations, we estimated that $R_{\rm BLR} \sim 0.013$~pc, $R_{\rm NIR} \sim 0.08$~pc and $R_{\rm MIR} \sim 2.3$~pc. The location of the reporecessed material is estimated to be $R_{\rm C} \sim 0.01-21$~pc (for $M_{\rm BH} \sim 10^{8.4}$ \M) or $R_{\rm C} \sim 0.002-4$~pc (for $M_{\rm BH} \sim 10^{7.65}$ \M). These results indicate that the material responsible for the \nh~ variability could be associated either to the BLR or to the torus.

\subsection{Comparison with Previous X-ray Observation}
\source~ was extensively observed in the X-ray wavebands over the years. Figure~\ref{fig:nh} shows the evolution of the line of sight column density (\nhl). The blue points are taken from the literature. The red points in the inset figures are from the current work. Figure~\ref{fig:evo} shows the variation of (a) the $2-10$~keV observed flux, (b) the $2-10$~keV intrinsic luminosity and (c) photon index ($\Gamma$) between 1990 and 2015. The blue points represent the observations taken between 1990 and 2015. 

The line of sight column density was observed to vary in the range of $\sim (3-10) \times 10^{23}$ \pcm~ in the past years. In 1990, a {\it Ginga} observation revealed a Compton-thin nucleus with \nh~ $\sim 5 \times 10^{23}$ \pcm ~\citep{Awaki1991}. An {\it ASCA} observation, carried out in 1994, also showed a Compton-thin nucleus (\nh~ $\sim 3\times 10^{23}$ \pcm). \citet{Risaliti2002b} reported of a Compton-thin absorber (\nh~ $\sim 5-6 \times 10^{23}$ \pcm) by studying three {\it BeppoSAX} observations performed between 1997 and 1999. A column density of \nh~ $\sim 4 \times 10^{23}$ \pcm~ was found by \xmm~ and \chandra~ observations in 2001 \citep{Matt2004}. {\it Suzaku} observed the source in 2007, finding a higher obscuration (\nh~ $\sim 9 \times 10^{23}$ \pcm; \citealp{Braito2013}). \xmm~ and \chandra~ observation campaign in 2010 reported a variable \nh~ in the range of $\sim 6.5-9.7 \times 10^{23}$ \pcm~ \citep{Marinucci2013}. 

In most of the previous studies, the X-ray spectral analysis was performed with a \textsc{power-law} or with the \textsc{pexrav} model. We also performed the spectral analysis using such a phenomenological model, and found that the column density varied in the range $\sim (6-8) \times 10^{23}$ \pcm. \citet{Braito2013} obtained a line-of-sight column density of \nhl $\sim (6-9) \times 10^{23}$ \pcm, after applying the \mytd~ model to the \xmm, {\it Suzaku}, and {\it BeppoSAX} observations performed between 1997 and 2007. They also estimated the \nht~ $\sim (2.4-3.5) \times 10^{23}$ \pcm. In the present work, we found that the global averaged column density of the obscuring material is \nht$ \sim 2.2 \pm 0.6 \times 10^{23}$ \pcm~ from the 2015 observations. From this, we can conclude that the global averaged column density of the absorber was nearly constant over decades, which is expected, as the global properties of the circumnuclear material are unlikely to change on the timescale of months. We tabulated the variation of \nhl~ over the years in Table~\ref{tab:nh}.

In the past 30 years, the source luminosity was observed to be in the range of $L_{\rm 2-10}^{\rm intr} \sim (1-4) \times 10^{43}$ \eps. In our analysis, we found that the source luminosity is in the range of $L_{\rm 2-10}^{\rm intr} \sim (3-3.6) \times 10^{43}$ \eps. This implies a rather steady accretion onto the SMBH at the centre of \source~. The rms fractional variability of $<3\%$ in 35 days timescale also support the steady accretion. At similar luminosity, the rms variability is observed to be in the range of $\sim 0.1-10$\% \citep{Fiore1998}. The rms variability is, thus, consistent with other AGNs in similar timescale \citep{Fiore1998}.

\section{Summary}
We studied \source~ using \nustar~ observations obtained between May 2015 and August 2015. Using the phenomenological slab model and several physically motivated torus-based models, we studied the properties of the X-ray emission and of the obscuring gas. We also estimated various properties of the obscuring material, e.g., the line-of-sight column density, the average density of the obscuring material, and its covering factor. From the variability of the absorption, we also provided some refined constraints on the location of the obscuring materials. Our key findings are given below.

\begin{enumerate}

\item We found that the equatorial column density of the torus is Compton-thick (\nheq $\sim 2\times 10^{24}$ \pcm). 

\item During the period of the observations analyzed here, the line of sight column density (\nhl) was found to vary in the range of \nhl $\sim 6-9 \times 10^{23}$ \pcm. The variability of \nhl~ is observed on timescales of $\leq 35$ days.

\item No variability is observed in the primary emission during the observation period.

\item The source was found to be buried in the obscuring medium, with the torus having a high covering factor $C_{\rm tor} \sim 0.83-0.85$. For the Compton-thick material, the torus covering factor was found to be $C_{\rm tor}^{\rm T} \sim 0.34-0.37$ which is in good agreement with average covering factors of the obscuring materials for nearby AGN \citep[e.g., ][]{Ricci2015}.

\item The reprocessing material is found to be located either at the BLR or in the torus.

\end{enumerate}

\section*{Acknowledgements}
We thank the anonymous reviewer for his/her suggestions and comments that helped us to improve the quality of this manuscript. Research at Physical Research Laboratory is supported by the Department of Space, Government of India for this work. AJ and HK acknowledge the support of the grant from the Ministry of Science and Technology of Taiwan with the grand number MOST 110-2811-M-007-500 and  MOST 111-2811-M-007-002. HK acknowledge the support of the grant from the Ministry of Science and Technology of Taiwan with the grand number MOST 110-2112-M-007-020. CR acknowledges support from the Fondecyt Iniciacion grant 11190831 and ANID BASAL project FB210003. AC and SSH are supported by the Canadian Space Agency and the Natural Sciences and Engineering Research Council of Canada. This research has made use of data and/or software provided by the High Energy Astrophysics Science Archive Research Center (HEASARC), which is a service of the Astrophysics Science Division at NASA/GSFC and the High Energy Astrophysics Division of the Smithsonian Astrophysical Observatory. This work has made use of data obtained from the {\it NuSTAR} mission, a projects led by Caltech, funded by NASA and managed by NASA/JPL, and has utilised the NuSTARDAS software package, jointly developed by the ASDC, Italy and Caltech, USA. 

\section*{Data Availability}
We used archival data of \nustar~ observatories for this work. All the models used in this work, are publicly available. Appropriate links are given in the text.



\bibliographystyle{mnras}
\bibliography{ref-AGN} 

\begin{thebibliography}{}
\makeatletter
\relax
\def\mn@urlcharsother{\let\do\@makeother \do\$\do\&\do\#\do\^\do\_\do\%\do\~}
\def\mn@doi{\begingroup\mn@urlcharsother \@ifnextchar [ {\mn@doi@}
  {\mn@doi@[]}}
\def\mn@doi@[#1]#2{\def\@tempa{#1}\ifx\@tempa\@empty \href
  {http://dx.doi.org/#2} {doi:#2}\else \href {http://dx.doi.org/#2} {#1}\fi
  \endgroup}
\def\mn@eprint#1#2{\mn@eprint@#1:#2::\@nil}
\def\mn@eprint@arXiv#1{\href {http://arxiv.org/abs/#1} {{\tt arXiv:#1}}}
\def\mn@eprint@dblp#1{\href {http://dblp.uni-trier.de/rec/bibtex/#1.xml}
  {dblp:#1}}
\def\mn@eprint@#1:#2:#3:#4\@nil{\def\@tempa {#1}\def\@tempb {#2}\def\@tempc
  {#3}\ifx \@tempc \@empty \let \@tempc \@tempb \let \@tempb \@tempa \fi \ifx
  \@tempb \@empty \def\@tempb {arXiv}\fi \@ifundefined
  {mn@eprint@\@tempb}{\@tempb:\@tempc}{\expandafter \expandafter \csname
  mn@eprint@\@tempb\endcsname \expandafter{\@tempc}}}

\bibitem[\protect\citeauthoryear{{Anders} \& {Grevesse}}{{Anders} \&
  {Grevesse}}{1989}]{Anders1989}
{Anders} E.,  {Grevesse} N.,  1989, \mn@doi [\gca]
  {10.1016/0016-7037(89)90286-X}, \href
  {https://ui.adsabs.harvard.edu/abs/1989GeCoA..53..197A} {53, 197}

\bibitem[\protect\citeauthoryear{{Antonucci}}{{Antonucci}}{1993}]{Antonucci1993}
{Antonucci} R.,  1993, \mn@doi [\araa] {10.1146/annurev.aa.31.090193.002353},
  \href {https://ui.adsabs.harvard.edu/abs/1993ARA&A..31..473A} {31, 473}

\bibitem[\protect\citeauthoryear{{Arnaud}}{{Arnaud}}{1996}]{Arnaud1996}
{Arnaud} K.~A.,  1996, in {Jacoby} G.~H.,  {Barnes} J.,  eds,  Astronomical
  Society of the Pacific Conference Series Vol. 101, Astronomical Data Analysis
  Software and Systems V. p.~17

\bibitem[\protect\citeauthoryear{{Awaki}, {Kunieda}, {Tawara}  \&
  {Koyama}}{{Awaki} et~al.}{1991}]{Awaki1991}
{Awaki} H.,  {Kunieda} H.,  {Tawara} Y.,   {Koyama} K.,  1991, \pasj, \href
  {https://ui.adsabs.harvard.edu/abs/1991PASJ...43L..37A} {43, L37}

\bibitem[\protect\citeauthoryear{{Balokovi{\'c}} et~al.,}{{Balokovi{\'c}}
  et~al.}{2018}]{Balokovic2018}
{Balokovi{\'c}} M.,  et~al., 2018, \mn@doi [\apj] {10.3847/1538-4357/aaa7eb},
  \href {https://ui.adsabs.harvard.edu/abs/2018ApJ...854...42B} {854, 42}

\bibitem[\protect\citeauthoryear{{Balokovi{\'c}} et~al.,}{{Balokovi{\'c}}
  et~al.}{2020}]{Balokovic2020}
{Balokovi{\'c}} M.,  et~al., 2020, \mn@doi [\apj] {10.3847/1538-4357/abc342},
  \href {https://ui.adsabs.harvard.edu/abs/2020ApJ...905...41B} {905, 41}

\bibitem[\protect\citeauthoryear{{Bassani}, {Malaguti}, {Jourdain}, {Roques}
  \& {Johnson}}{{Bassani} et~al.}{1995}]{Bassani1995}
{Bassani} L.,  {Malaguti} G.,  {Jourdain} E.,  {Roques} J.~P.,   {Johnson}
  W.~N.,  1995, \mn@doi [\apjl] {10.1086/187863}, \href
  {https://ui.adsabs.harvard.edu/abs/1995ApJ...444L..73B} {444, L73}

\bibitem[\protect\citeauthoryear{{Beckmann} \& {Shrader}}{{Beckmann} \&
  {Shrader}}{2012}]{Beckmann2012}
{Beckmann} V.,  {Shrader} C.~R.,  2012, {Active Galactic Nuclei}

\bibitem[\protect\citeauthoryear{{Bennett} et~al.,}{{Bennett}
  et~al.}{2003}]{Bennett2003}
{Bennett} C.~L.,  et~al., 2003, \mn@doi [\apjs] {10.1086/377253}, \href
  {https://ui.adsabs.harvard.edu/abs/2003ApJS..148....1B} {148, 1}

\bibitem[\protect\citeauthoryear{{Braito}, {Ballo}, {Reeves}, {Risaliti},
  {Ptak}  \& {Turner}}{{Braito} et~al.}{2013}]{Braito2013}
{Braito} V.,  {Ballo} L.,  {Reeves} J.~N.,  {Risaliti} G.,  {Ptak} A.,
  {Turner} T.~J.,  2013, \mn@doi [\mnras] {10.1093/mnras/sts226}, \href
  {https://ui.adsabs.harvard.edu/abs/2013MNRAS.428.2516B} {428, 2516}

\bibitem[\protect\citeauthoryear{{Comastri}, {Vignali}, {Cappi}, {Matt},
  {Audano}, {Awaki}  \& {Ueno}}{{Comastri} et~al.}{1998}]{Comastri1998}
{Comastri} A.,  {Vignali} C.,  {Cappi} M.,  {Matt} G.,  {Audano} R.,  {Awaki}
  H.,   {Ueno} S.,  1998, \mn@doi [\mnras] {10.1046/j.1365-8711.1998.01302.x},
  \href {https://ui.adsabs.harvard.edu/abs/1998MNRAS.295..443C} {295, 443}

\bibitem[\protect\citeauthoryear{{Edelson}, {Turner}, {Pounds}, {Vaughan},
  {Markowitz}, {Marshall}, {Dobbie}  \& {Warwick}}{{Edelson}
  et~al.}{2002}]{Edelson2002}
{Edelson} R.,  {Turner} T.~J.,  {Pounds} K.,  {Vaughan} S.,  {Markowitz} A.,
  {Marshall} H.,  {Dobbie} P.,   {Warwick} R.,  2002, \mn@doi [\apj]
  {10.1086/323779}, \href
  {https://ui.adsabs.harvard.edu/abs/2002ApJ...568..610E} {568, 610}

\bibitem[\protect\citeauthoryear{{Elitzur}}{{Elitzur}}{2012}]{Elitzur2012}
{Elitzur} M.,  2012, \mn@doi [\apjl] {10.1088/2041-8205/747/2/L33}, \href
  {https://ui.adsabs.harvard.edu/abs/2012ApJ...747L..33E} {747, L33}

\bibitem[\protect\citeauthoryear{{Fiore}, {Laor}, {Elvis}, {Nicastro}  \&
  {Giallongo}}{{Fiore} et~al.}{1998}]{Fiore1998}
{Fiore} F.,  {Laor} A.,  {Elvis} M.,  {Nicastro} F.,   {Giallongo} E.,  1998,
  \mn@doi [\apj] {10.1086/306031}, \href
  {https://ui.adsabs.harvard.edu/abs/1998ApJ...503..607F} {503, 607}

\bibitem[\protect\citeauthoryear{{Guainazzi}}{{Guainazzi}}{2002}]{Guainazzi2002}
{Guainazzi} M.,  2002, \mn@doi [\mnras] {10.1046/j.1365-8711.2002.05132.x},
  \href {https://ui.adsabs.harvard.edu/abs/2002MNRAS.329L..13G} {329, L13}

\bibitem[\protect\citeauthoryear{{Guainazzi}, {Matt}, {Piro}  \&
  {Robba}}{{Guainazzi} et~al.}{1997}]{Guainazzi1997}
{Guainazzi} M.,  {Matt} G.,  {Piro} L.,   {Robba} N.~R.,  1997, \memsai, \href
  {https://ui.adsabs.harvard.edu/abs/1997MmSAI..68..131G} {68, 131}

\bibitem[\protect\citeauthoryear{{Gupta} et~al.,}{{Gupta}
  et~al.}{2021}]{Gupta2021}
{Gupta} K.~K.,  et~al., 2021, \mn@doi [\mnras] {10.1093/mnras/stab839}, \href
  {https://ui.adsabs.harvard.edu/abs/2021MNRAS.504..428G} {504, 428}

\bibitem[\protect\citeauthoryear{{HI4PI Collaboration} et~al.,}{{HI4PI
  Collaboration} et~al.}{2016}]{HI4PI2016}
{HI4PI Collaboration} et~al., 2016, \mn@doi [\aap]
  {10.1051/0004-6361/201629178}, \href
  {https://ui.adsabs.harvard.edu/abs/2016A&A...594A.116H} {594, A116}

\bibitem[\protect\citeauthoryear{{Harrison} et~al.,}{{Harrison}
  et~al.}{2013}]{Harrison2013}
{Harrison} F.~A.,  et~al., 2013, \mn@doi [\apj] {10.1088/0004-637X/770/2/103},
  \href {https://ui.adsabs.harvard.edu/abs/2013ApJ...770..103H} {770, 103}

\bibitem[\protect\citeauthoryear{{Jana}, {Chatterjee}, {Kumari}, {Nandi },
  {Naik}  \& {Patra}}{{Jana} et~al.}{2020}]{AJ2020}
{Jana} A.,  {Chatterjee} A.,  {Kumari} N.,  {Nandi } P.,  {Naik} S.,   {Patra}
  D.,  2020, \mn@doi [\mnras] {10.1093/mnras/staa2552}, \href
  {https://ui.adsabs.harvard.edu/abs/2020MNRAS.499.5396J} {499, 5396}

\bibitem[\protect\citeauthoryear{{Kaspi}, {Maoz}, {Netzer}, {Peterson},
  {Vestergaard}  \& {Jannuzi}}{{Kaspi} et~al.}{2005}]{Kaspi2005}
{Kaspi} S.,  {Maoz} D.,  {Netzer} H.,  {Peterson} B.~M.,  {Vestergaard} M.,
  {Jannuzi} B.~T.,  2005, \mn@doi [\apj] {10.1086/431275}, \href
  {https://ui.adsabs.harvard.edu/abs/2005ApJ...629...61K} {629, 61}

\bibitem[\protect\citeauthoryear{{Koss} et~al.,}{{Koss}
  et~al.}{2017}]{Koss2017}
{Koss} M.,  et~al., 2017, \mn@doi [\apj] {10.3847/1538-4357/aa8ec9}, \href
  {https://ui.adsabs.harvard.edu/abs/2017ApJ...850...74K} {850, 74}

\bibitem[\protect\citeauthoryear{{Magdziarz}, {Blaes}, {Zdziarski}, {Johnson}
  \& {Smith}}{{Magdziarz} et~al.}{1998}]{MZ98}
{Magdziarz} P.,  {Blaes} O.~M.,  {Zdziarski} A.~A.,  {Johnson} W.~N.,   {Smith}
  D.~A.,  1998, \mn@doi [\mnras] {10.1046/j.1365-8711.1998.02015.x}, \href
  {https://ui.adsabs.harvard.edu/abs/1998MNRAS.301..179M} {301, 179}

\bibitem[\protect\citeauthoryear{{Marinucci}, {Bianchi}, {Nicastro}, {Matt}  \&
  {Goulding}}{{Marinucci} et~al.}{2012}]{Marinucci2012}
{Marinucci} A.,  {Bianchi} S.,  {Nicastro} F.,  {Matt} G.,   {Goulding} A.~D.,
  2012, \mn@doi [\apj] {10.1088/0004-637X/748/2/130}, \href
  {https://ui.adsabs.harvard.edu/abs/2012ApJ...748..130M} {748, 130}

\bibitem[\protect\citeauthoryear{{Marinucci}, {Risaliti}, {Wang}, {Bianchi},
  {Elvis}, {Matt}, {Nardini}  \& {Braito}}{{Marinucci}
  et~al.}{2013}]{Marinucci2013}
{Marinucci} A.,  {Risaliti} G.,  {Wang} J.,  {Bianchi} S.,  {Elvis} M.,  {Matt}
  G.,  {Nardini} E.,   {Braito} V.,  2013, \mn@doi [\mnras]
  {10.1093/mnras/sts534}, \href
  {https://ui.adsabs.harvard.edu/abs/2013MNRAS.429.2581M} {429, 2581}

\bibitem[\protect\citeauthoryear{{Markowitz}, {Krumpe}  \&
  {Nikutta}}{{Markowitz} et~al.}{2014}]{Markowitz2014}
{Markowitz} A.~G.,  {Krumpe} M.,   {Nikutta} R.,  2014, \mn@doi [\mnras]
  {10.1093/mnras/stt2492}, \href
  {https://ui.adsabs.harvard.edu/abs/2014MNRAS.439.1403M} {439, 1403}

\bibitem[\protect\citeauthoryear{{Matt}, {Guainazzi}  \& {Maiolino}}{{Matt}
  et~al.}{2003}]{Matt2003}
{Matt} G.,  {Guainazzi} M.,   {Maiolino} R.,  2003, \mn@doi [\mnras]
  {10.1046/j.1365-8711.2003.06539.x}, \href
  {https://ui.adsabs.harvard.edu/abs/2003MNRAS.342..422M} {342, 422}

\bibitem[\protect\citeauthoryear{{Matt}, {Bianchi}, {D'Ammando}  \&
  {Martocchia}}{{Matt} et~al.}{2004}]{Matt2004}
{Matt} G.,  {Bianchi} S.,  {D'Ammando} F.,   {Martocchia} A.,  2004, \mn@doi
  [\aap] {10.1051/0004-6361:20047045}, \href
  {https://ui.adsabs.harvard.edu/abs/2004A&A...421..473M} {421, 473}

\bibitem[\protect\citeauthoryear{{Murphy} \& {Yaqoob}}{{Murphy} \&
  {Yaqoob}}{2009}]{MY2009}
{Murphy} K.~D.,  {Yaqoob} T.,  2009, \mn@doi [\mnras]
  {10.1111/j.1365-2966.2009.15025.x}, \href
  {https://ui.adsabs.harvard.edu/abs/2009MNRAS.397.1549M} {397, 1549}

\bibitem[\protect\citeauthoryear{{Murphy}, {Yaqoob}  \& {Terashima}}{{Murphy}
  et~al.}{2007}]{Murphy2007}
{Murphy} K.~D.,  {Yaqoob} T.,   {Terashima} Y.,  2007, \mn@doi [\apj]
  {10.1086/520039}, \href
  {https://ui.adsabs.harvard.edu/abs/2007ApJ...666...96M} {666, 96}

\bibitem[\protect\citeauthoryear{{Nandra}, {George}, {Mushotzky}, {Turner}  \&
  {Yaqoob}}{{Nandra} et~al.}{1997}]{Nandra1997}
{Nandra} K.,  {George} I.~M.,  {Mushotzky} R.~F.,  {Turner} T.~J.,   {Yaqoob}
  T.,  1997, \mn@doi [\apj] {10.1086/303600}, \href
  {https://ui.adsabs.harvard.edu/abs/1997ApJ...476...70N} {476, 70}

\bibitem[\protect\citeauthoryear{{Nenkova}, {Sirocky}, {Ivezi{\'c}}  \&
  {Elitzur}}{{Nenkova} et~al.}{2008a}]{Nenkova2008a}
{Nenkova} M.,  {Sirocky} M.~M.,  {Ivezi{\'c}} {\v{Z}}.,   {Elitzur} M.,  2008a,
  \mn@doi [\apj] {10.1086/590482}, \href
  {https://ui.adsabs.harvard.edu/abs/2008ApJ...685..147N} {685, 147}

\bibitem[\protect\citeauthoryear{{Nenkova}, {Sirocky}, {Nikutta}, {Ivezi{\'c}}
  \& {Elitzur}}{{Nenkova} et~al.}{2008b}]{Nenkova2008b}
{Nenkova} M.,  {Sirocky} M.~M.,  {Nikutta} R.,  {Ivezi{\'c}} {\v{Z}}.,
  {Elitzur} M.,  2008b, \mn@doi [\apj] {10.1086/590483}, \href
  {https://ui.adsabs.harvard.edu/abs/2008ApJ...685..160N} {685, 160}

\bibitem[\protect\citeauthoryear{{Paltani} \& {Ricci}}{{Paltani} \&
  {Ricci}}{2017}]{Paltani2017}
{Paltani} S.,  {Ricci} C.,  2017, \mn@doi [\aap] {10.1051/0004-6361/201629623},
  \href {https://ui.adsabs.harvard.edu/abs/2017A&A...607A..31P} {607, A31}

\bibitem[\protect\citeauthoryear{{Puccetti}, {Fiore}, {Risaliti}, {Capalbi},
  {Elvis}  \& {Nicastro}}{{Puccetti} et~al.}{2007}]{Puccetti2007}
{Puccetti} S.,  {Fiore} F.,  {Risaliti} G.,  {Capalbi} M.,  {Elvis} M.,
  {Nicastro} F.,  2007, \mn@doi [\mnras] {10.1111/j.1365-2966.2007.11634.x},
  \href {https://ui.adsabs.harvard.edu/abs/2007MNRAS.377..607P} {377, 607}

\bibitem[\protect\citeauthoryear{{Ricci}, {Ueda}, {Koss}, {Trakhtenbrot},
  {Bauer}  \& {Gandhi}}{{Ricci} et~al.}{2015}]{Ricci2015}
{Ricci} C.,  {Ueda} Y.,  {Koss} M.~J.,  {Trakhtenbrot} B.,  {Bauer} F.~E.,
  {Gandhi} P.,  2015, \mn@doi [\apjl] {10.1088/2041-8205/815/1/L13}, \href
  {https://ui.adsabs.harvard.edu/abs/2015ApJ...815L..13R} {815, L13}

\bibitem[\protect\citeauthoryear{{Ricci} et~al.,}{{Ricci}
  et~al.}{2016}]{Ricci2016}
{Ricci} C.,  et~al., 2016, \mn@doi [\apj] {10.3847/0004-637X/820/1/5}, \href
  {https://ui.adsabs.harvard.edu/abs/2016ApJ...820....5R} {820, 5}

\bibitem[\protect\citeauthoryear{{Ricci} et~al.,}{{Ricci}
  et~al.}{2017a}]{Ricci2017apjs}
{Ricci} C.,  et~al., 2017a, \mn@doi [\apjs] {10.3847/1538-4365/aa96ad}, \href
  {https://ui.adsabs.harvard.edu/abs/2017ApJS..233...17R} {233, 17}

\bibitem[\protect\citeauthoryear{{Ricci} et~al.,}{{Ricci}
  et~al.}{2017b}]{Ricci2017}
{Ricci} C.,  et~al., 2017b, \mn@doi [\nat] {10.1038/nature23906}, \href
  {https://ui.adsabs.harvard.edu/abs/2017Natur.549..488R} {549, 488}

\bibitem[\protect\citeauthoryear{{Ricci} et~al.,}{{Ricci}
  et~al.}{2018}]{Ricci2018}
{Ricci} C.,  et~al., 2018, \mn@doi [\mnras] {10.1093/mnras/sty1879}, \href
  {https://ui.adsabs.harvard.edu/abs/2018MNRAS.480.1819R} {480, 1819}

\bibitem[\protect\citeauthoryear{{Risaliti}}{{Risaliti}}{2002}]{Risaliti2002b}
{Risaliti} G.,  2002, \mn@doi [\aap] {10.1051/0004-6361:20020170}, \href
  {https://ui.adsabs.harvard.edu/abs/2002A&A...386..379R} {386, 379}

\bibitem[\protect\citeauthoryear{{Risaliti}, {Elvis}  \& {Nicastro}}{{Risaliti}
  et~al.}{2002}]{Risaliti2002a}
{Risaliti} G.,  {Elvis} M.,   {Nicastro} F.,  2002, \mn@doi [\apj]
  {10.1086/324146}, \href
  {https://ui.adsabs.harvard.edu/abs/2002ApJ...571..234R} {571, 234}

\bibitem[\protect\citeauthoryear{{Risaliti}, {Elvis}, {Fabbiano}, {Baldi},
  {Zezas}  \& {Salvati}}{{Risaliti} et~al.}{2007}]{Risaliti2007}
{Risaliti} G.,  {Elvis} M.,  {Fabbiano} G.,  {Baldi} A.,  {Zezas} A.,
  {Salvati} M.,  2007, \mn@doi [\apjl] {10.1086/517884}, \href
  {https://ui.adsabs.harvard.edu/abs/2007ApJ...659L.111R} {659, L111}

\bibitem[\protect\citeauthoryear{{Risaliti}, {Elvis}, {Bianchi}  \&
  {Matt}}{{Risaliti} et~al.}{2010}]{Risaliti2010}
{Risaliti} G.,  {Elvis} M.,  {Bianchi} S.,   {Matt} G.,  2010, \mn@doi [\mnras]
  {10.1111/j.1745-3933.2010.00873.x}, \href
  {https://ui.adsabs.harvard.edu/abs/2010MNRAS.406L..20R} {406, L20}

\bibitem[\protect\citeauthoryear{{Tanimoto}, {Ueda}, {Odaka}, {Kawaguchi},
  {Fukazawa}  \& {Kawamuro}}{{Tanimoto} et~al.}{2019}]{Tanimoto2019}
{Tanimoto} A.,  {Ueda} Y.,  {Odaka} H.,  {Kawaguchi} T.,  {Fukazawa} Y.,
  {Kawamuro} T.,  2019, \mn@doi [\apj] {10.3847/1538-4357/ab1b20}, \href
  {https://ui.adsabs.harvard.edu/abs/2019ApJ...877...95T} {877, 95}

\bibitem[\protect\citeauthoryear{{Tanimoto}, {Ueda}, {Odaka}, {Ogawa},
  {Yamada}, {Kawaguchi}  \& {Ichikawa}}{{Tanimoto} et~al.}{2020}]{Tanimoto2020}
{Tanimoto} A.,  {Ueda} Y.,  {Odaka} H.,  {Ogawa} S.,  {Yamada} S.,  {Kawaguchi}
  T.,   {Ichikawa} K.,  2020, \mn@doi [\apj] {10.3847/1538-4357/ab96bc}, \href
  {https://ui.adsabs.harvard.edu/abs/2020ApJ...897....2T} {897, 2}

\bibitem[\protect\citeauthoryear{{Tristram} \& {Schartmann}}{{Tristram} \&
  {Schartmann}}{2011}]{Tristram2011}
{Tristram} K.~R.~W.,  {Schartmann} M.,  2011, \mn@doi [\aap]
  {10.1051/0004-6361/201116867}, \href
  {https://ui.adsabs.harvard.edu/abs/2011A&A...531A..99T} {531, A99}

\bibitem[\protect\citeauthoryear{{Tueller}, {Mushotzky}, {Barthelmy},
  {Cannizzo}, {Gehrels}, {Markwardt}, {Skinner}  \& {Winter}}{{Tueller}
  et~al.}{2008}]{Tueller2008}
{Tueller} J.,  {Mushotzky} R.~F.,  {Barthelmy} S.,  {Cannizzo} J.~K.,
  {Gehrels} N.,  {Markwardt} C.~B.,  {Skinner} G.~K.,   {Winter} L.~M.,  2008,
  \mn@doi [\apj] {10.1086/588458}, \href
  {https://ui.adsabs.harvard.edu/abs/2008ApJ...681..113T} {681, 113}

\bibitem[\protect\citeauthoryear{{Turner}, {George}, {Nandra}  \&
  {Mushotzky}}{{Turner} et~al.}{1997}]{Turner1997}
{Turner} T.~J.,  {George} I.~M.,  {Nandra} K.,   {Mushotzky} R.~F.,  1997,
  \mn@doi [\apjs] {10.1086/313053}, \href
  {https://ui.adsabs.harvard.edu/abs/1997ApJS..113...23T} {113, 23}

\bibitem[\protect\citeauthoryear{{Ueda} et~al.,}{{Ueda}
  et~al.}{2007}]{Ueda2007}
{Ueda} Y.,  et~al., 2007, \mn@doi [\apjl] {10.1086/520576}, \href
  {https://ui.adsabs.harvard.edu/abs/2007ApJ...664L..79U} {664, L79}

\bibitem[\protect\citeauthoryear{{Vasudevan} \& {Fabian}}{{Vasudevan} \&
  {Fabian}}{2009}]{Vasedevan2009}
{Vasudevan} R.~V.,  {Fabian} A.~C.,  2009, \mn@doi [\mnras]
  {10.1111/j.1365-2966.2008.14108.x}, \href
  {https://ui.adsabs.harvard.edu/abs/2009MNRAS.392.1124V} {392, 1124}

\bibitem[\protect\citeauthoryear{{Vaughan}, {Edelson}, {Warwick}  \&
  {Uttley}}{{Vaughan} et~al.}{2003}]{Vaughan2003}
{Vaughan} S.,  {Edelson} R.,  {Warwick} R.~S.,   {Uttley} P.,  2003, \mn@doi
  [\mnras] {10.1046/j.1365-2966.2003.07042.x}, \href
  {https://ui.adsabs.harvard.edu/abs/2003MNRAS.345.1271V} {345, 1271}

\bibitem[\protect\citeauthoryear{{Verner}, {Ferland}, {Korista}  \&
  {Yakovlev}}{{Verner} et~al.}{1996}]{Verner1996}
{Verner} D.~A.,  {Ferland} G.~J.,  {Korista} K.~T.,   {Yakovlev} D.~G.,  1996,
  \mn@doi [\apj] {10.1086/177435}, \href
  {https://ui.adsabs.harvard.edu/abs/1996ApJ...465..487V} {465, 487}

\bibitem[\protect\citeauthoryear{{Weaver}, {Nousek}, {Yaqoob}, {Mushotzky},
  {Makino}  \& {Otani}}{{Weaver} et~al.}{1996}]{Weaver1996}
{Weaver} K.~A.,  {Nousek} J.,  {Yaqoob} T.,  {Mushotzky} R.~F.,  {Makino} F.,
  {Otani} C.,  1996, \mn@doi [\apj] {10.1086/176800}, \href
  {https://ui.adsabs.harvard.edu/abs/1996ApJ...458..160W} {458, 160}

\bibitem[\protect\citeauthoryear{{Winter}, {Mushotzky}, {Reynolds}  \&
  {Tueller}}{{Winter} et~al.}{2009}]{Winter2009}
{Winter} L.~M.,  {Mushotzky} R.~F.,  {Reynolds} C.~S.,   {Tueller} J.,  2009,
  \mn@doi [\apj] {10.1088/0004-637X/690/2/1322}, \href
  {https://ui.adsabs.harvard.edu/abs/2009ApJ...690.1322W} {690, 1322}

\bibitem[\protect\citeauthoryear{{Wu} \& {Liu}}{{Wu} \& {Liu}}{2004}]{Wu2004}
{Wu} X.-B.,  {Liu} F.~K.,  2004, \mn@doi [\apj] {10.1086/423446}, \href
  {https://ui.adsabs.harvard.edu/abs/2004ApJ...614...91W} {614, 91}

\bibitem[\protect\citeauthoryear{{Yamada}, {Ueda}, {Tanimoto}, {Oda},
  {Imanishi}, {Toba}  \& {Ricci}}{{Yamada} et~al.}{2020}]{Yamada2020}
{Yamada} S.,  {Ueda} Y.,  {Tanimoto} A.,  {Oda} S.,  {Imanishi} M.,  {Toba} Y.,
    {Ricci} C.,  2020, \mn@doi [\apj] {10.3847/1538-4357/ab94b1}, \href
  {https://ui.adsabs.harvard.edu/abs/2020ApJ...897..107Y} {897, 107}

\bibitem[\protect\citeauthoryear{{Yaqoob}}{{Yaqoob}}{2012}]{Yaqoob2012}
{Yaqoob} T.,  2012, \mn@doi [\mnras] {10.1111/j.1365-2966.2012.21129.x}, \href
  {https://ui.adsabs.harvard.edu/abs/2012MNRAS.423.3360Y} {423, 3360}

\bibitem[\protect\citeauthoryear{{Yaqoob}, {Tatum}, {Scholtes}, {Gottlieb}  \&
  {Turner}}{{Yaqoob} et~al.}{2015}]{Yaqoob2015}
{Yaqoob} T.,  {Tatum} M.~M.,  {Scholtes} A.,  {Gottlieb} A.,   {Turner} T.~J.,
  2015, \mn@doi [\mnras] {10.1093/mnras/stv2021}, \href
  {https://ui.adsabs.harvard.edu/abs/2015MNRAS.454..973Y} {454, 973}

\makeatother
\end{thebibliography}




\appendix
\section{Tables}

\begin{table*}
\centering
\caption{MYTORUS model fitted spectral analysis result for coupled configuration.}
\label{tab:mytc}
\begin{tabular}{lcccccccccccc}
\hline
ID & \nheq & \nhl &  $\Gamma$ & $N_{\rm PL}$ & $i^{\rm t}$ & $A_{\rm S}$ & $f_{\rm Scat}$ & $\chi^2$/dof \\
 & (10$^{24}$ \pcm) & (10$^{23}$ \pcm) & & (10$^{-2}$ \phc) & (degree) & & (10$^{-2}$) & \\
 (1) & (2) & (3) & (4) & (5) & (6) & (7) & (8) & (9) \\
\hline
N1& $2.3^{+0.6}_{-0.9}$&$ 7.5^{+4.2}_{-2.2}$ &$1.75^{+0.06}_{-0.05}$& $3.01^{+0.24}_{-0.17}$& $61.8^{+3.7}_{-0.9}$& $0.79^{+0.11}_{-0.16}$ & $ 1.49^{+0.02}_{-0.03}$ & 664/592 \\
\\
N2& $2.0^{+0.7}_{-0.8}$&$ 6.5^{+2.7}_{-1.4}$ &$1.71^{+0.04}_{-0.05}$& $2.15^{+0.12}_{-0.22}$& " & $0.71^{+0.08}_{-0.15}$ & $ 1.96^{+0.02}_{-0.03}$ &620/627 \\ 
\\
N3& $2.5^{+0.7}_{-0.7}$&$ 8.1^{+2.9}_{-1.9}$ &$1.70^{+0.03}_{-0.05}$& $2.89^{+0.12}_{-0.23}$& " & $0.78^{+0.15}_{-0.24}$ & $ 1.04^{+0.02}_{-0.03}$ &634/611 \\ 
\\
N4& $2.0^{+0.6}_{-0.8}$& $6.5^{+2.4}_{-1.4}$ & $1.75^{+0.06}_{-0.05}$& $2.09^{+0.18}_{-0.13}$& " & $0.73^{+0.12}_{-0.15}$ & $ 1.84^{+0.03}_{-0.08}$ & 584/588 \\
\hline
\end{tabular}
\leftline{(1) ID of the observation, (2) equatorial hydrogen column density (\nheq) in $10^{24}$ \pcm, (3) line of sight hydrogen column density (\nhl) in $10^{23}$ \pcm,}.
\leftline{(4) photon index ($\Gamma$) of the primary emission, (5) power-law normalization ($N_{\rm PL}$) in 10$^{-2}$\phc, (6) inclination angle ($i$) in degree,}
\leftline{(7) relative normalization of the line emission ($A_{\rm S}$), (9) fraction of scattered primary emission ($f_{\rm Scat}$).}
\leftline{$^{\rm t}$ parameter (6) are tied across the observations.}
\end{table*}

\begin{table*}
\centering
\caption{MYTORUS model fitted spectral analysis result for decoupled configuration.}
\label{tab:mytd}
\begin{tabular}{lcccccccccccc}
\hline
 ID  & \nhl & \nht & $\Gamma$ & $N_{\rm PL}$ & $A_{\rm S}$ & $f_{\rm Scat}$ & $\chi^2$/dof \\
 & (10$^{23}$ \pcm) & (10$^{23}$ \pcm) &  & (10$^{-2}$ \phc) & & (10$^{-2}$) & \\
 (1) & (2) & (3) & (4) & (5) & (6) & (7) & (8) \\
\hline
N1 &$8.2^{+1.3}_{-1.3}$&$ 2.2^{+0.6}_{-0.5}$&$ 1.67^{+0.04}_{-0.05}$&$ 3.00^{+0.18}_{-0.31}$&$ 0.71^{+0.03}_{-0.05}$&$ 0.70^{+0.03}_{-0.02}$& 663/592 \\
\\
N2 &$7.0^{+1.4}_{-1.3}$& "  &$ 1.65^{+0.08}_{-0.09}$&$ 2.15^{+0.09}_{-0.14}$&$ 0.70^{+0.02}_{-0.04}$&$ 1.40^{+0.01}_{-0.03}$& 624/625 \\
\\
N3 &$8.1^{+1.5}_{-1.5}$& " &$ 1.66^{+0.06}_{-0.06}$&$ 2.69^{+0.05}_{-0.12}$&$ 0.60^{+0.06}_{-0.08}$&$ 0.96^{+0.06}_{-0.02}$& 633/610 \\
\\
N4 &$7.1^{+1.3}_{-1.4}$& " &$ 1.64^{+0.06}_{-0.09}$&$ 1.98^{+0.06}_{-0.07}$&$ 0.71^{+0.05}_{-0.04}$&$ 1.39^{+0.05}_{-0.03}$ & 589/588 \\
\hline
\end{tabular}
\leftline{(1) ID of the observation, (2) line o of sight hydrogen column density (\nhl) in $10^{23}$ \pcm, (3) global averaged hydrogen column density of the }.
\leftline{obscured materials (\nht) in $10^{23}$ \pcm, (4) photon index ($\Gamma$) of the primary emission, (5) power-law normalization ($N_{\rm PL}$) in 10$^{-2}$\phc, }
\leftline{(6) relative normalization of the line emission ($A_{\rm S}$), (7) fraction of scattered primary emission ($f_{\rm Scat}$).}
\leftline{$^{\rm t}$ parameter (3) are tied across the observations.}
\end{table*}

\begin{table*}
\centering
\hspace*{-0.5cm}
\caption{\borus~ model fitted spectral analysis result.}
\label{tab:borus}
\begin{tabular}{lccccccccccccc}
\hline
ID & \nhl & $^{\rm t}$\nht & $\Gamma$ & $E_{\rm cut}$&$N_{\rm PL}$ & $C_{\rm tor}$ & $i^{\rm t}$ & $^{\rm t}A_{\rm Fe}$ & $f_{\rm Scat}$ & $\chi^2$/dof \\
& (10$^{23}$) & (10$^{23}$) & &(keV) & (10$^{-2}$\phc) & & (degree) & (A$_{\odot}$) & (10$^{-2}$) & \\
(1) & (2) & (3) & (4) & (5) & (6) & (7) & (8) & (9) & (10) & (11) \\
\hline
N1&$ 9.4^{+0.6}_{-0.6}$&$ 2.6^{+0.7}_{-0.6}$&$ 1.48^{+0.06}_{-0.05}$&$75^{+29}_{-15}$&$ 1.57^{+0.08}_{-0.11} $&$ 0.58^{+0.10}_{-0.08}$&$ 64.5^{+5.2}_{-6.3}$& $ 0.47^{+0.07}_{-0.06}$&$ 1.02^{+0.11}_{-0.13}$& 656/592 \\
\\
N2&$ 8.3^{+0.7}_{-0.6}$& " &$ 1.50^{+0.02}_{-0.04}$&$97^{+46}_{-18}$ &$ 1.39^{+0.12}_{-0.10} $& " & " & " &$ 1.74^{+0.04}_{-0.03}$& 620/629 \\
\\
N3&$ 9.8^{+0.6}_{-0.6}$& " &$ 1.47^{+0.04}_{-0.04}$& $91^{+30}_{-23}$ &$ 1.51^{+0.16}_{-0.08} $& " & " & " &$ 1.14^{+0.08}_{-0.12}$& 631/613 \\
\\
N4&$ 8.4^{+0.5}_{-0.5}$& " &$ 1.46^{+0.04}_{-0.04}$&$ 89^{+43}_{-12}$ &$ 1.31^{+0.11}_{-0.13} $& " & "  & " &$ 1.61^{+0.09}_{-0.12}$& 577/592 \\
\hline
\end{tabular}
\leftline{(1) ID of the observation, (2) line o of sight hydrogen column density (\nhl) in $10^{23}$ \pcm, (3) averaged hydrogen column density of the obscured materials}.
\leftline{(\nht) in $10^{23}$ \pcm, (4) cut-off energy ($E_{\rm cut}$) in keV, (5) photon index ($\Gamma$) of the primary emission, (6) power-law normalization ($N_{\rm PL}$) in 10$^{-2}$\phc,}
\leftline{(7) covering factor the obscured materials, (8) inclination angle ($i$) in degree, (9) iron abundances ($A_{\rm Fe}$) in solar value ($A_{\odot}$), (10) fraction of scattered} 
\leftline{primary emission ($f_{\rm Scat}$).}
\leftline{$^{\rm t}$ parameter (3) and (8) are tied across the observations.}
\end{table*}

\begin{table*}
\centering
\hspace*{-0.5cm}
\caption{{\tt XCLUMPY} model fitted spectral analysis result.}
\label{tab:xclumpy}
\begin{tabular}{lccccccccccccc}
\hline
 ID & $^{\rm t}$\nheq & \nhl & $\Gamma$ & $N_{\rm PL}$ & $\sigma_{\rm tor}$ & $i^{\rm t}$ &$A_{\rm L}$ & $f_{\rm Scat}$ & $\chi^2$/dof \\
 & (10$^{24}$) & (10$^{23}$) & & (10$^{-2}$\phc) & (degree) & (degree) & & (10$^{-2}$) & \\
 (1) & (2) & (3) & (4) & (5) & (6) & (7) & (8) & (9) & (10) \\
\hline
N1&$ 2.1^{+0.6}_{-0.3}$&$  8.1^{+2.4}_{-1.5}$&$ 1.67^{+0.06}_{-0.12}$& $ 1.78^{+0.10}_{-0.18}$&$ 26.3^{+8.4}_{-5.3}$&$ 64.1^{+7.4}_{-4.9}$&$ 0.78^{+0.10}_{-0.08}$&$ 1.78^{+0.42}_{-0.65}$& 652/595 \\
\\
N2& " & $ 7.0^{+1.6}_{-1.1}$&$ 1.66^{+0.12}_{-0.07}$& $ 0.88^{+0.14}_{-0.18}$&$ 24.9^{+4.6}_{-7.1}$& " &$ 0.84^{+0.07}_{-0.11}$& $ 4.30^{+0.53}_{-0.86}$& 616/628 \\
\\
N3&"& $ 8.2^{+1.6}_{-2.3}$& $ 1.63^{+0.09}_{-0.12}$ & $ 1.63^{+0.18}_{-0.25}$&$ 25.7^{+5.2}_{-5.8}$& " &$ 0.76^{+0.08}_{-0.11}$& $ 2.12^{+0.34}_{-0.29}$& 629/613 \\
\\
N4&"&$ 7.1^{+1.5}_{-2.2}$&$ 1.59^{+0.12}_{-0.15}$&$ 1.31^{+0.18}_{-0.12}$&$25.2^{+4.6}_{-4.9}$& " &$ 0.88^{+0.10}_{-0.09}$&$ 2.57^{+0.27}_{-0.38}$& 579/590 \\
\hline
\end{tabular}
\leftline{(1) ID of the observation, (2) equatorial hydrogen column density (\nheq) in $10^{24}$ \pcm, (3) line of sight hydrogen column density (\nhl) in $10^{23}$ \pcm,}.
\leftline{(4) photon index ($\Gamma$) of the primary emission, (5) power-law normalization ($N_{\rm PL}$) in 10$^{-2}$\phc, (6) torus angular width ($\sigma_{\rm tor}$) in degrees,}
\leftline{(7) inclination angle ($i$) in degree, (8) relative normalization of the line emission ($A_{\rm L}$), (9) fraction of scattered primary emission ($f_{\rm Scat}$).}
\leftline{$^{\rm t}$ parameter (2) and (7) are tied across the observations.}
\end{table*}

\begin{table*}
\centering
\hspace*{-0.5cm}
\caption{\textsc{rxtorus} model fitted spectral analysis result.}
\label{tab:rxtorus}
\begin{tabular}{lcccccccccccccc}
\hline
 ID & $^{\rm t}$\nheq & \nhl & $\Gamma$ & $i^{\rm t}$ & r/R & $N_{\rm PL}$ & $A_{\rm rpcr}$& $f_{\rm Scat}$ & $\chi^2$/dof \\
 & (10$^{24}$) & (10$^{23}$) & &(degree) & & (10$^{-2}$\phc) & & ($10^{-2}$)  & \\
 (1) & (2) & (3) & (4) & (5) & (6) & (7) & (8) & (9) & (10) \\
\hline
N1&$ 2.1^{+0.6}_{-0.5}$ & $6.9^{+2.4}_{-1.6}$&$ 1.63^{+0.07}_{-0.06}$& $ 66.7^{+4.5}_{-7.2}$&$0.42^{+0.07}_{-0.09}$&$1.75^{+0.08}_{-0.07}$& $1.04^{+0.15}_{-0.11}$ &$ 1.58^{+0.45}_{-0.75}$& 652/596 \\
\\
N2 & " & $6.0^{+0.9}_{-1.1}$ & $1.58^{+0.10}_{-0.08}$ & " & $0.41^{+0.09}_{-0.07}$ & $1.46^{+0.12}_{-0.16}$ & $1.03^{+0.14}_{-0.19} $&$2.14^{+0.09}_{-0.12}$ & 615/629 \\
\\
N3 & " & $7.0^{+1.6}_{-1.8}$ & $1.60^{+0.06}_{-0.07}$ & " & $0.42^{+0.08}_{-0.07}$ & $1.70^{+0.08}_{-0.12}$ & $0.97^{+0.06}_{-0.12} $& $1.74^{+0.08}_{-0.11}$ & 625/615 \\
\\
N4 & " & $6.1^{+1.5}_{-2.1}$ & $1.57^{+0.06}_{-0.07}$ & " & $0.41^{+0.08}_{-0.08}$ & $1.36^{+0.10}_{-0.11}$ &$ 1.06^{+0.16}_{-0.21}$ & $2.11^{+0.18}_{-0.22}$ & 578/592 \\
\hline
\end{tabular}
\leftline{(1) ID of the observation, (2) equatorial hydrogen column density (\nheq) in $10^{24}$ \pcm, (3) line of sight hydrogen column density (\nhl) in $10^{23}$ \pcm,}.
\leftline{(4) photon index ($\Gamma$) of the primary emission, (5) inclination angle ($i$) in degree, (6) torus covering factor (r/R), (7) power-law normalization ($N_{\rm PL}$) in}
\leftline{10$^{-2}$\phc, (8) relative normalization of the reprocessed emission ($A_{\rm rpcr}$), (9) fraction of scattered primary emission ($f_{\rm Scat}$).}
\leftline{$^{\rm t}$ parameter (2) and (7) are tied across the observations.}
\leftline{Line of sight hydrogen column density is calculated using Equation~\ref{eqn:rx}.}
\end{table*}

\begin{table*}
\caption{Variation of line-of-sight column density (\nhl)}
\label{tab:nh}
\centering
\begin{tabular}{ccccc}
\hline
Date & \nhl & $F_{2-10~{\rm keV}}^{\rm obs}$ & Observatories & Ref.  \\
(YYYY-MM-DD) & ($10^{23}$ \pcm) & ($10^{-11}$ \ecs) & & \\
(1) & (2) & (3) & (4) & (5) \\
\hline
1990-07-07&$  4.9  \pm0.7  $ &$1.6^*       $  & {\it Ginga} & \citet{Awaki1991}    \\ 
1994-02-12&$  3.26 \pm0.7  $ &$2.1^*       $  & {\it ASCA} & \citet{Comastri1998}        \\
1996-03-05&$  3.41 \pm0.11 $ &$1.8  \pm0.2   $  & {\it RXTE} & \citet{Guainazzi1997} \\
1997-12-26&$  7.00 \pm0.45 $ &$1.8^*       $  & {\it BeppoSAX} & \citet{Braito2013} \\
1998-07-02&$  6.20 \pm0.50 $ &$1.6^*      $  & {\it BeppoSAX} & \citet{Braito2013}   \\
1999-01-13&$  6.40 \pm0.95 $ &$0.87^*      $  & {\it BeppoSAX} & \citet{Risaliti2002b}  \\
2001-01-04&$  5.0  \pm0.25 $ &$1.2^*      $  & \xmm & \citet{Braito2013}   \\
2001-03-15&$  4.0^*      $ &$2.37^*    $  & \chandra & \citet{Matt2004} \\
2007-12-20&$  8.2  \pm0.6  $ &$0.6^*       $  & {\it Suzaku} & \citet{Braito2013}   \\
2010-06-24&$  8.7  \pm0.7  $ &$7.7 \pm0.3   $  & \xmm & \citet{Marinucci2013} \\ 
2010-07-03&$  9.7  \pm0.9  $ &$8.0  \pm0.3   $  & \xmm & \citet{Marinucci2013} \\
2010-07-13&$  7.6  \pm1.1  $ &$8.4  \pm0.2   $  & \xmm & \citet{Marinucci2013} \\
2010-07-23&$  9.4  \pm1.1  $ &$8.0  \pm0.3   $  & \xmm & \citet{Marinucci2013} \\
2010-08-03&$  8.0  \pm0.7  $ &$7.5  \pm0.7   $  & \xmm & \citet{Marinucci2013} \\
2010-12-02&$  6.5  \pm0.7  $ &$10.0 \pm0.4   $  & \chandra & \citet{Marinucci2013} \\
2015-05-03&$  0.79 \pm0.03 $ &$0.98 \pm0.07  $  &  \nustar &  This work\\
2015-06-10&$  0.69 \pm0.02 $ &$1.07 \pm0.08  $  &  \nustar & This work\\
2015-07-15&$  0.78 \pm0.03 $ &$0.94 \pm0.06  $  &  \nustar & This work\\
2015-08-22&$  0.73 \pm0.02 $ &$1.01 \pm0.04  $  &  \nustar & This work\\
\hline
\end{tabular}
\leftline{ $^*$ no error is quoted.}
\end{table*}

\section{Spectra}

\begin{figure*}
\centering
\includegraphics[width=8.4cm]{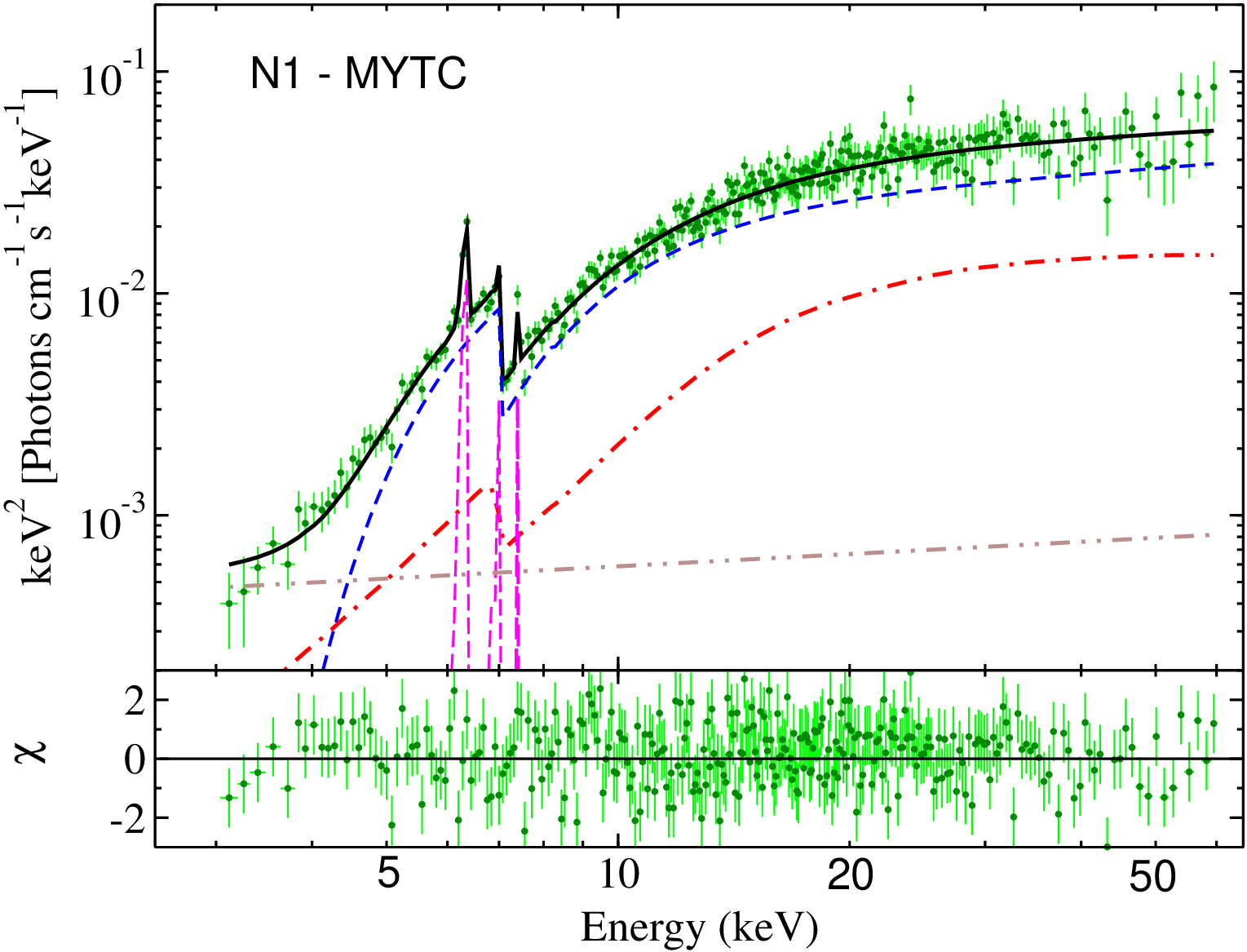}
\includegraphics[width=8.4cm]{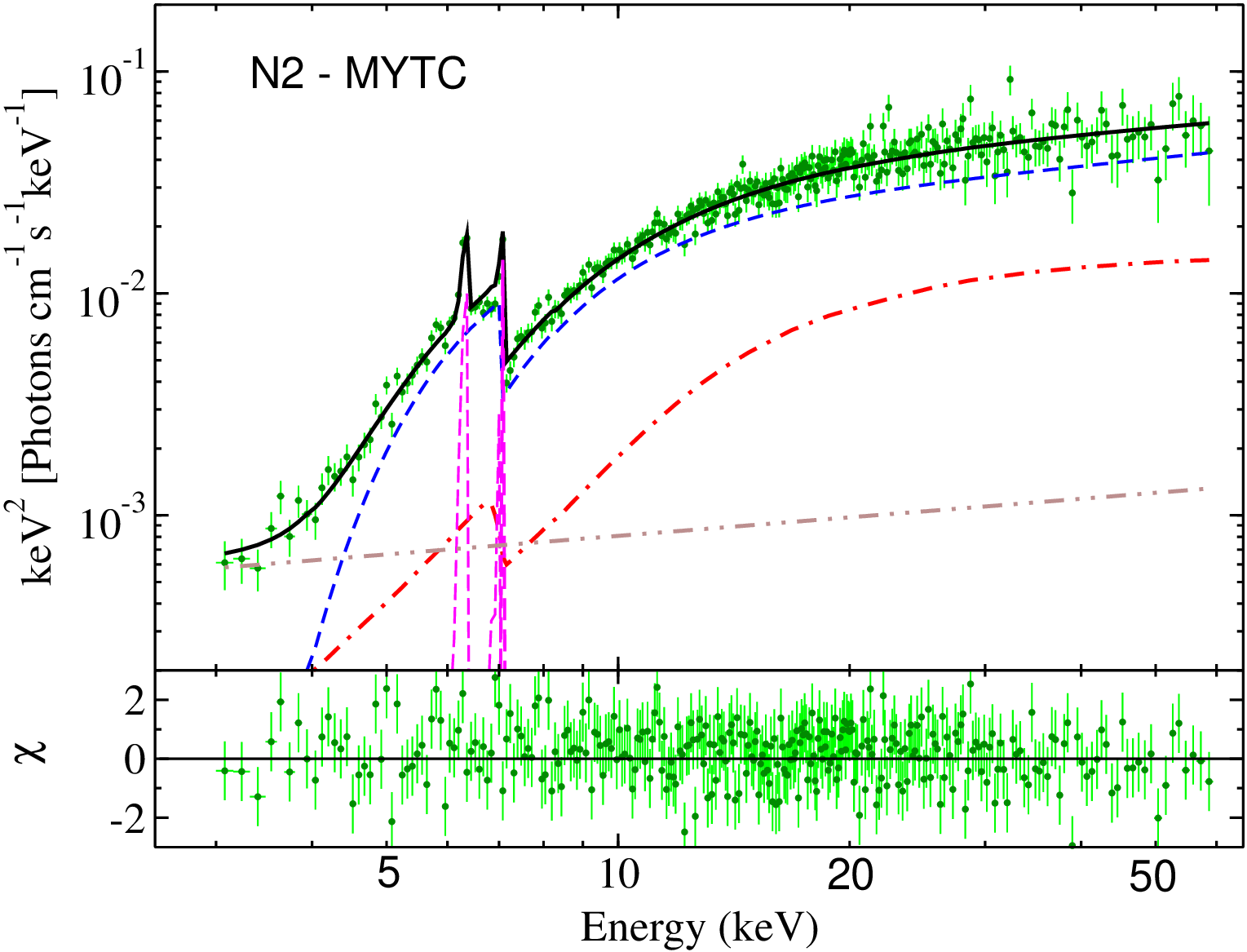}
\includegraphics[width=8.4cm]{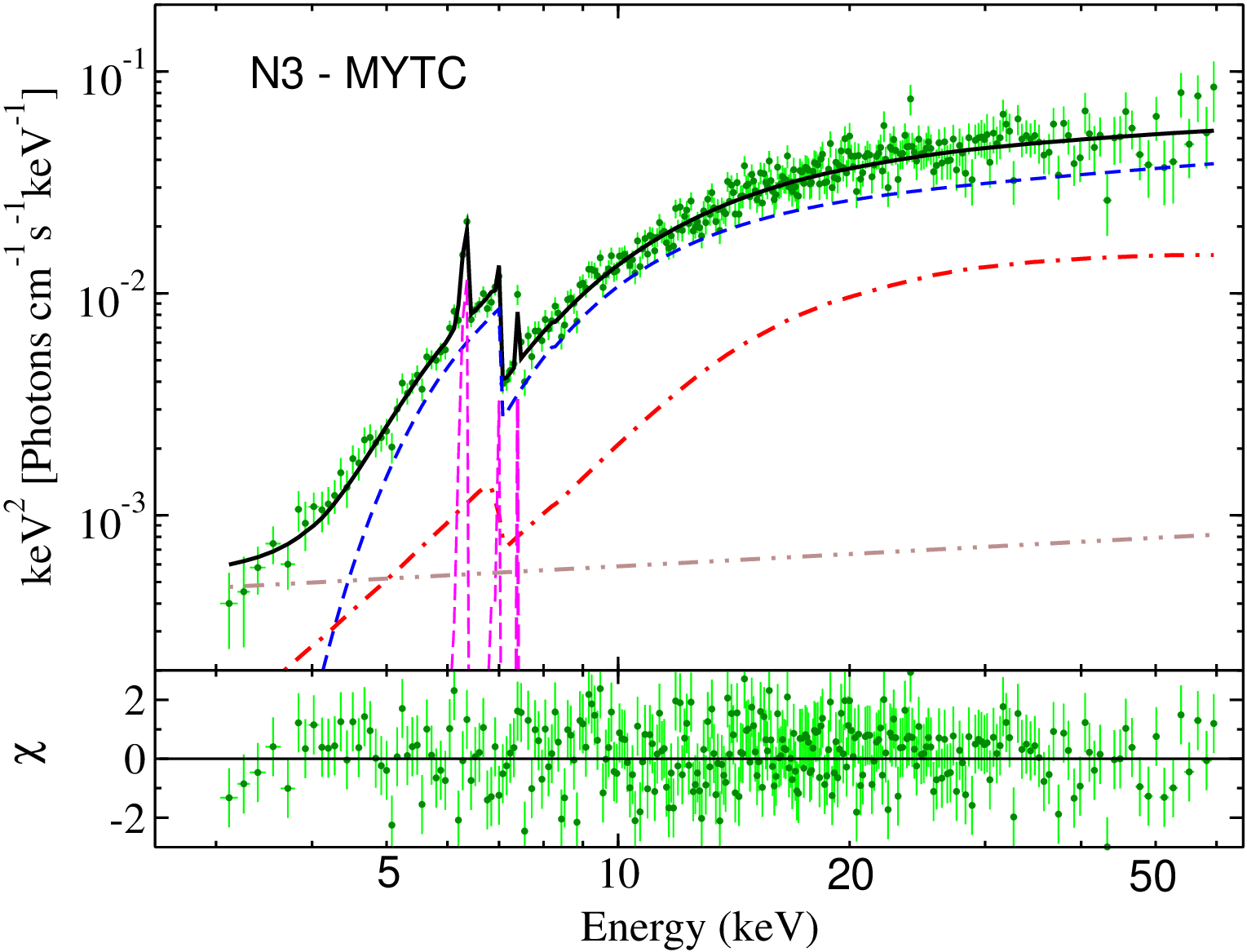}
\includegraphics[width=8.4cm]{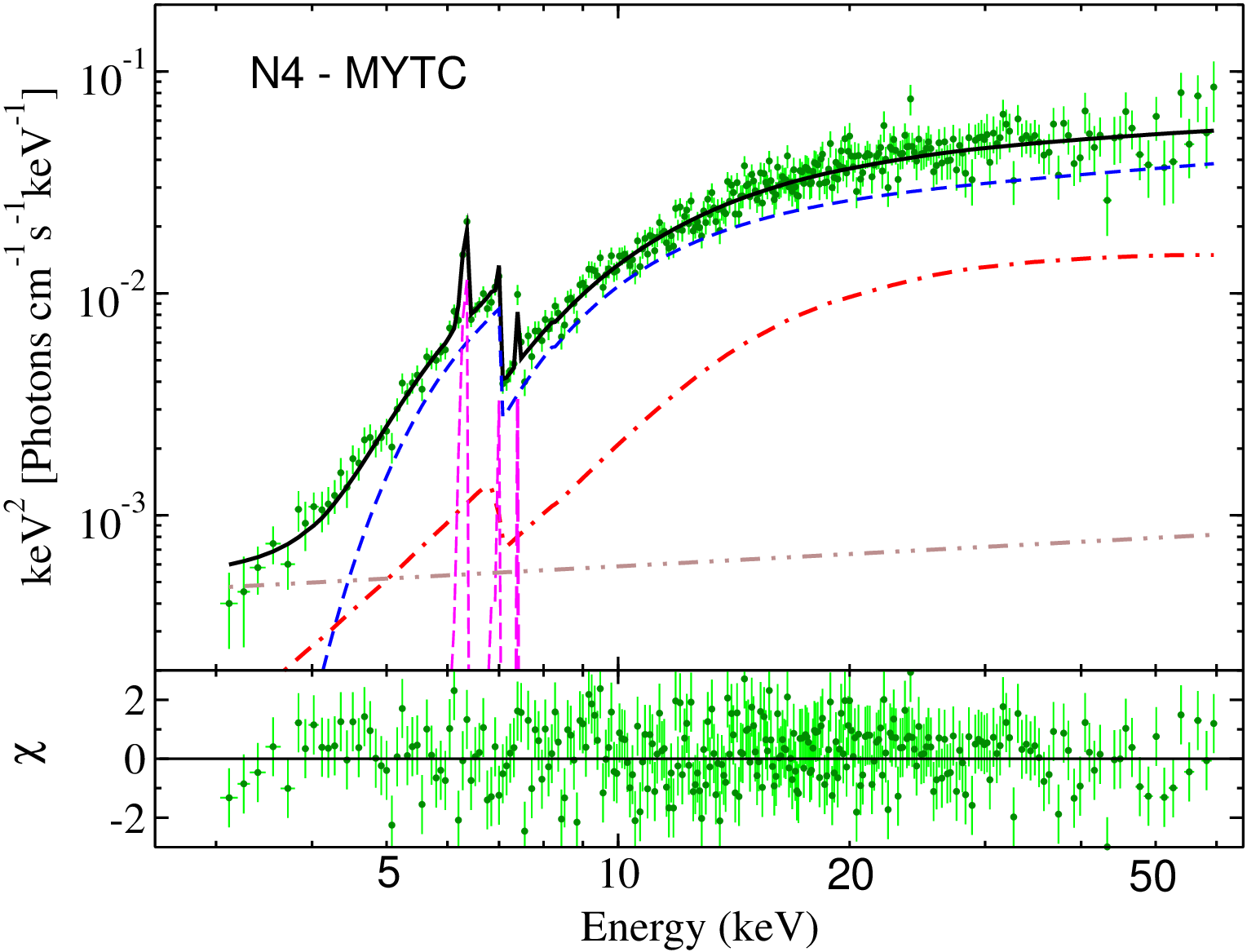}
\caption{Unfolded spectra fitted with \mytc~ model for observation N1 (top left), N2 (top right), N3 (bottom left) and N4 (bottom right). Upper panel : Green points represent the data. The black, blue, red, magenta and brown lines represent the total emission, primary emission, reprocessed emission, line emission (Fe K$\alpha$, Fe K$\beta$ and Ni K $\alpha$), and scattered primary emission. Bottom panel: Corresponding residual.}
\label{fig:mytc-spec}
\end{figure*}

\begin{figure*}
\centering
\includegraphics[width=8.4cm]{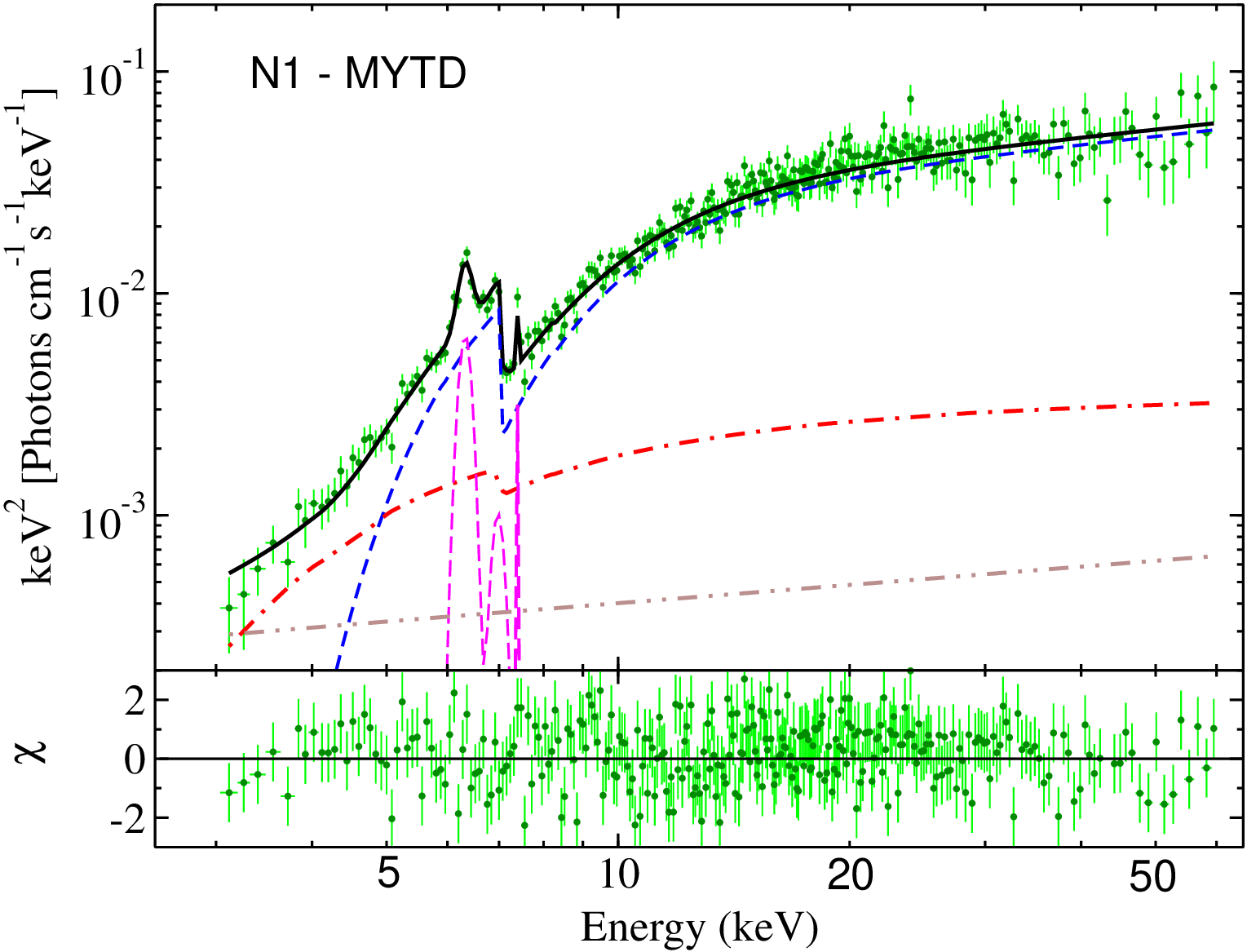}
\includegraphics[width=8.4cm]{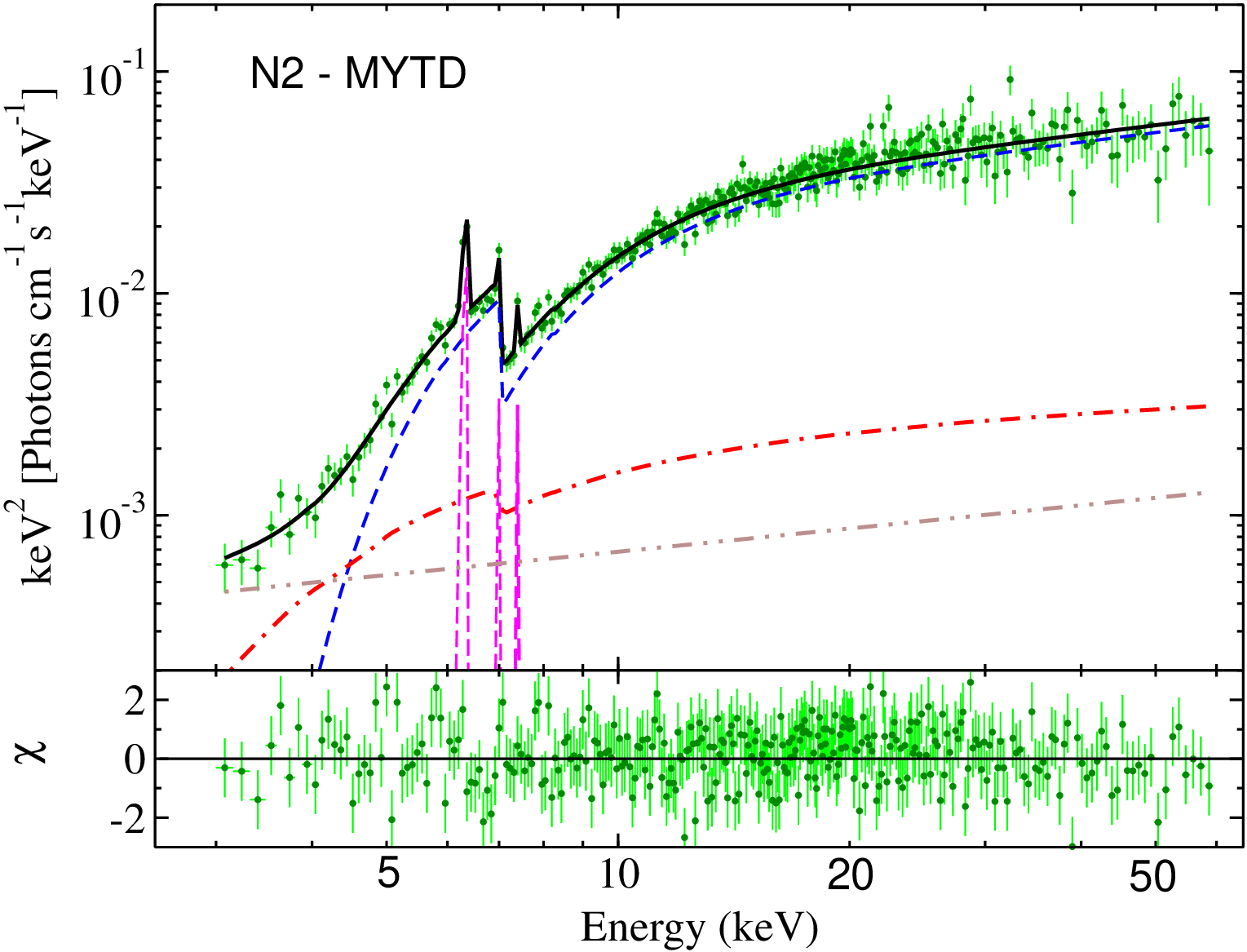}
\includegraphics[width=8.4cm]{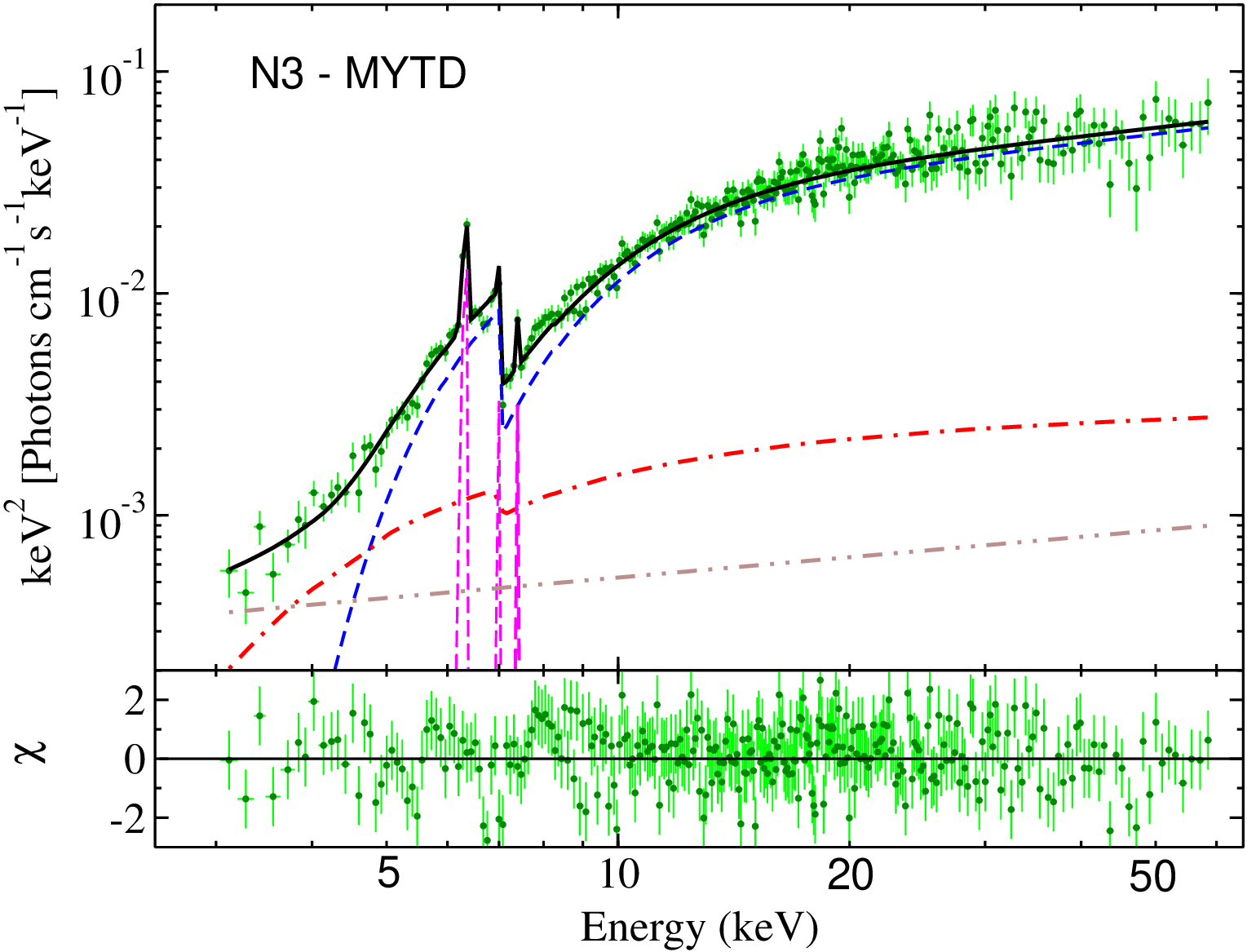}
\includegraphics[width=8.4cm]{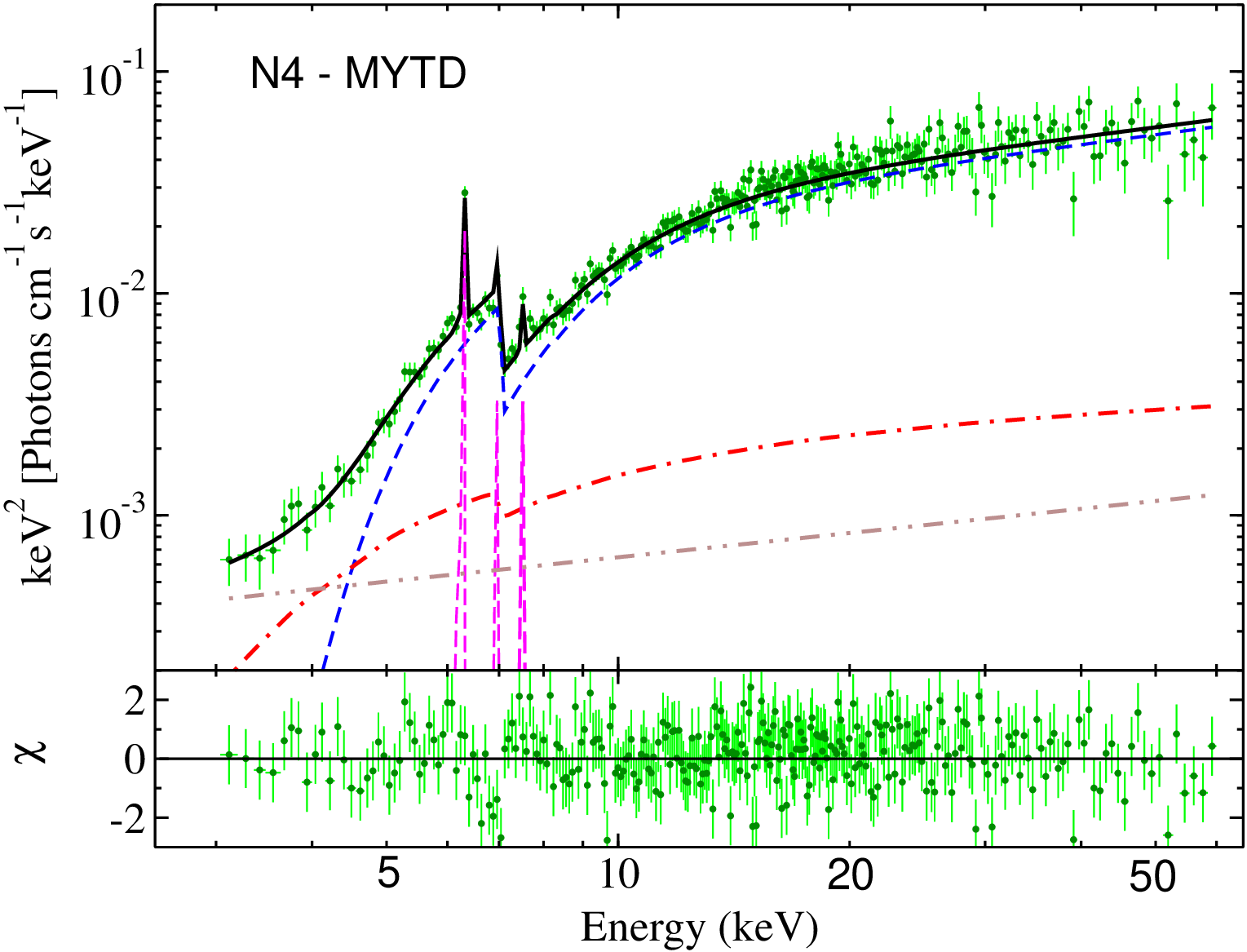}
\caption{Unfolded spectra fitted with \mytd~ model for observation N1 (top left), N2 (top right), N3 (bottom left) and N4 (bottom right). Upper panel : Green points represent the data. The black, blue, red, magenta and brown lines represent the total emission, primary emission, reprocessed emission, line emission (Fe K$\alpha$, Fe K$\beta$ and Ni K $\alpha$), and scattered primary emission. Bottom panel: Corresponding residual.}
\label{fig:mytd-spec}
\end{figure*}

\begin{figure*}
\centering
\includegraphics[width=8.4cm]{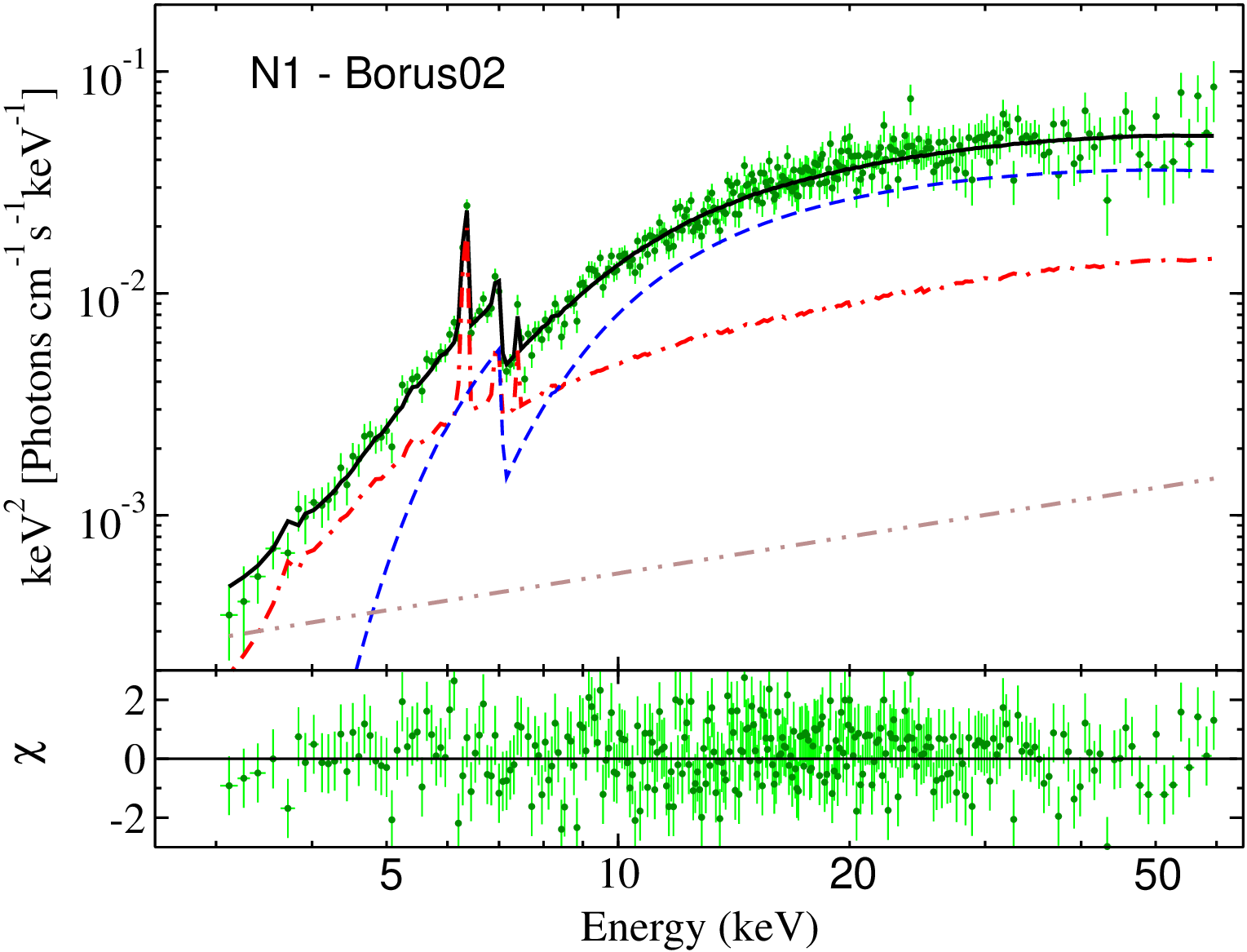}
\includegraphics[width=8.4cm]{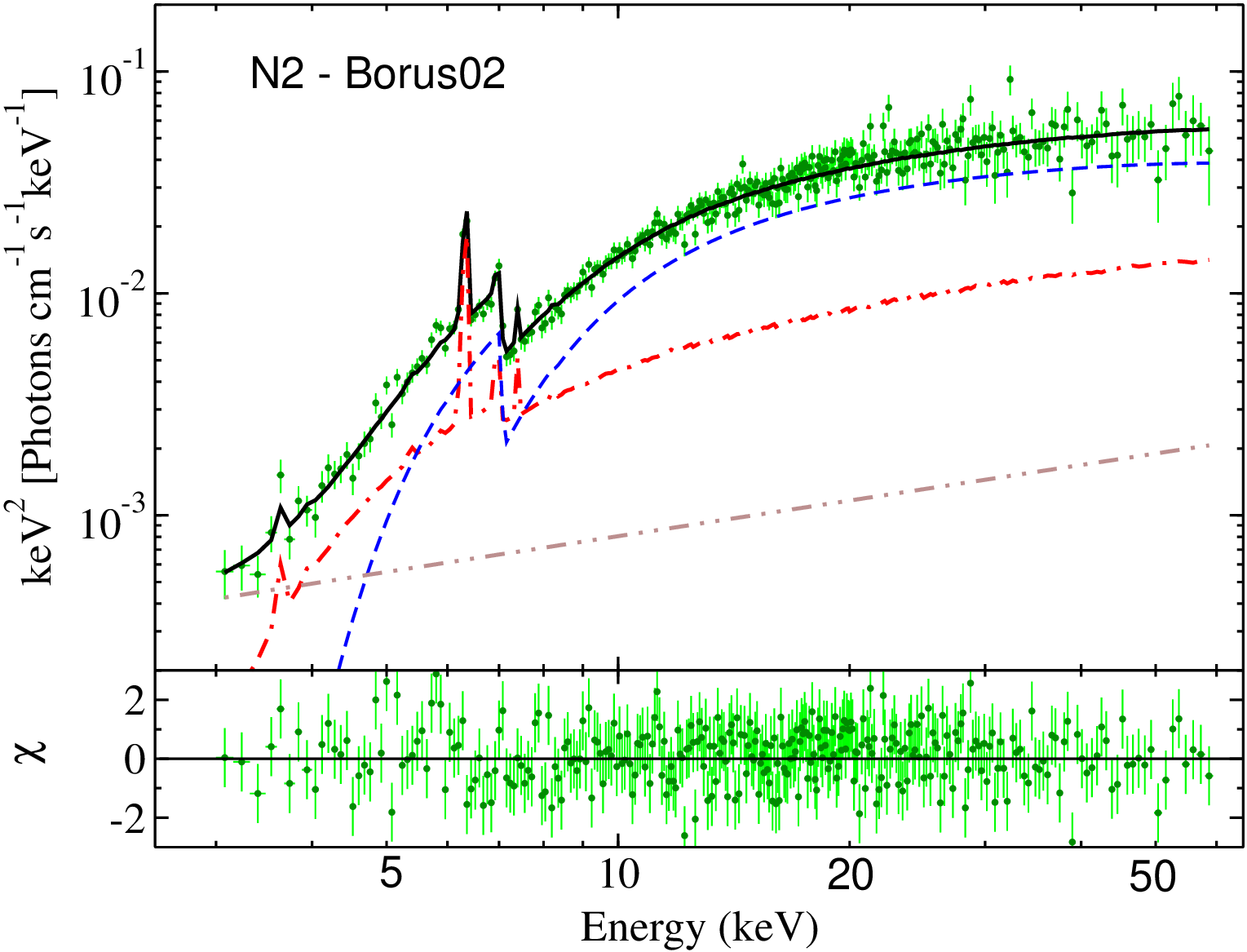}
\includegraphics[width=8.4cm]{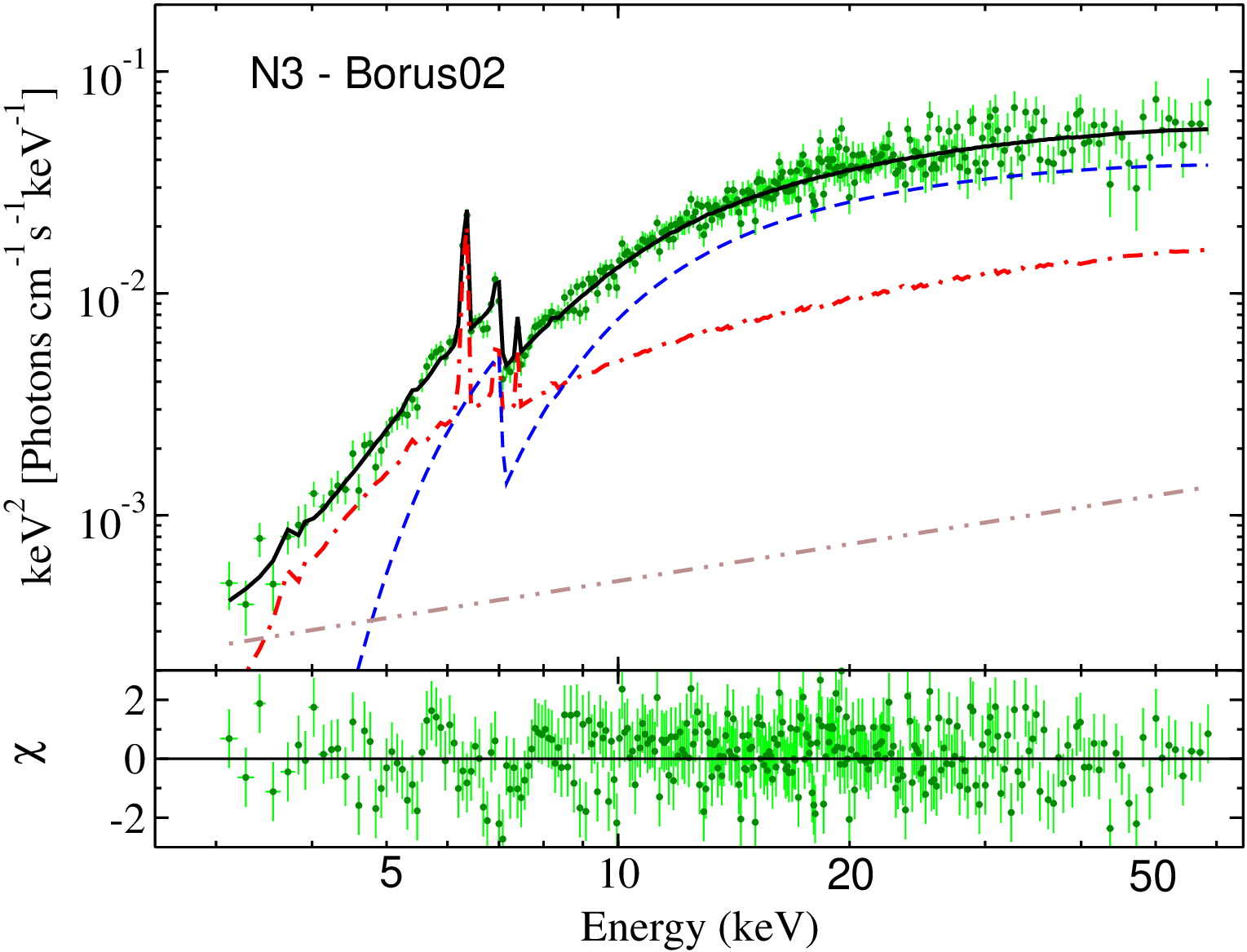}
\includegraphics[width=8.4cm]{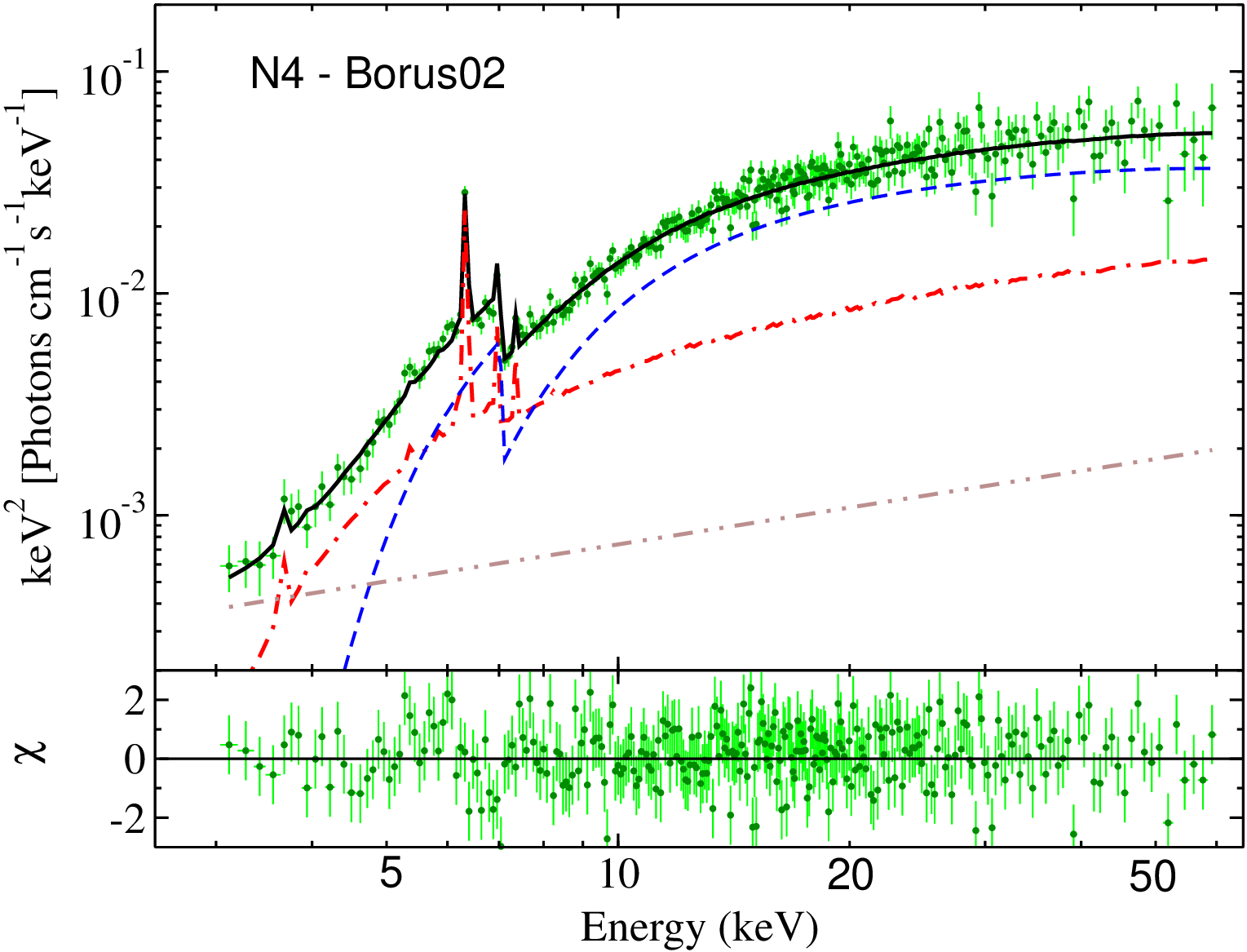}
\caption{Unfolded spectra fitted with \borus~ model for observation N1 (top left), N2 (top right), N3 (bottom left) and N4 (bottom right). Upper panel : Green points represent the data. The black, blue, red, magenta and brown lines represent the total emission, primary emission, reprocessed emission, line emission (Fe K$\alpha$, Fe K$\beta$ and Ni K $\alpha$), and scattered primary emission. Bottom panel: Corresponding residual.}
\label{fig:borus-spec}
\end{figure*}

\begin{figure*}
\centering
\includegraphics[width=8.4cm]{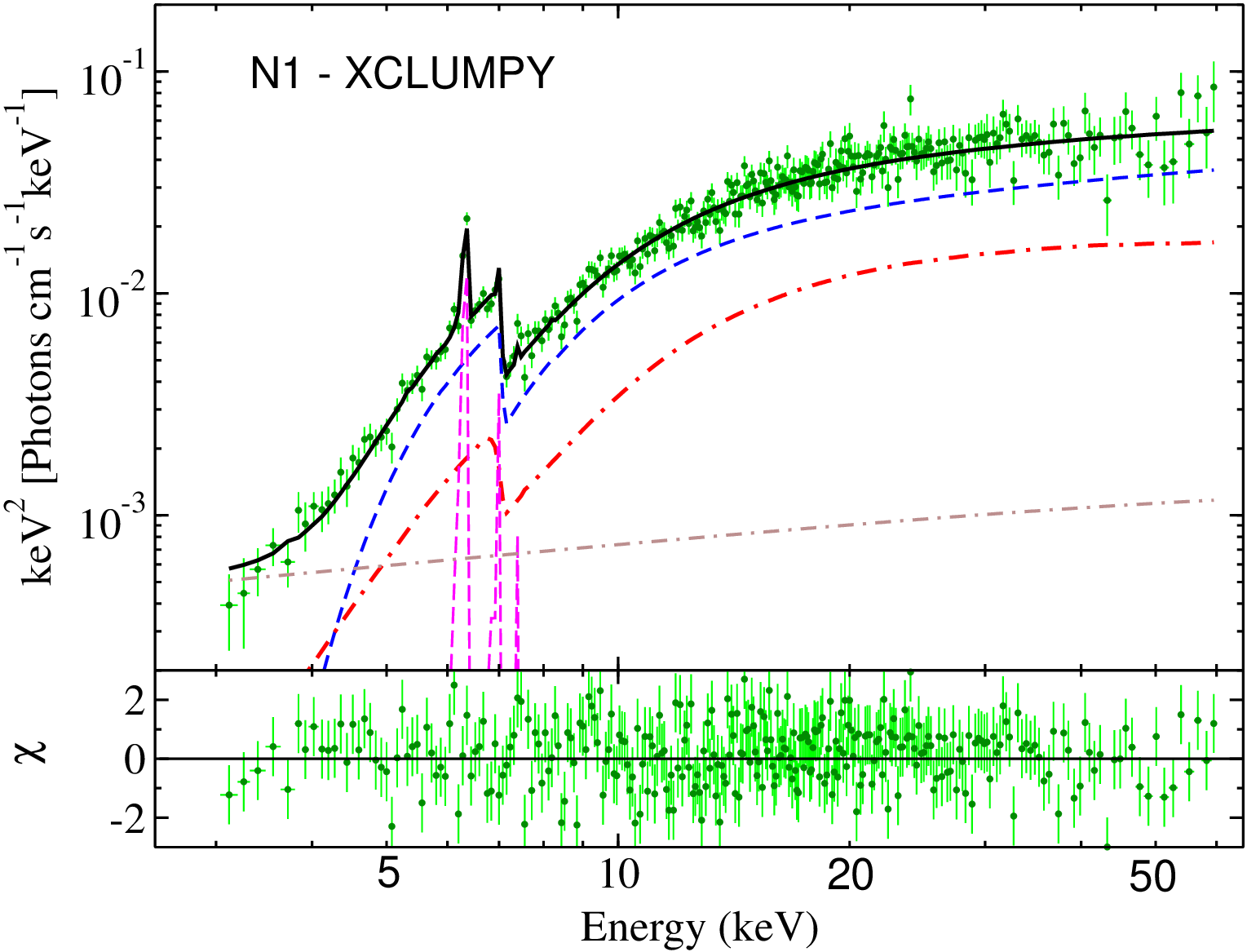}
\includegraphics[width=8.4cm]{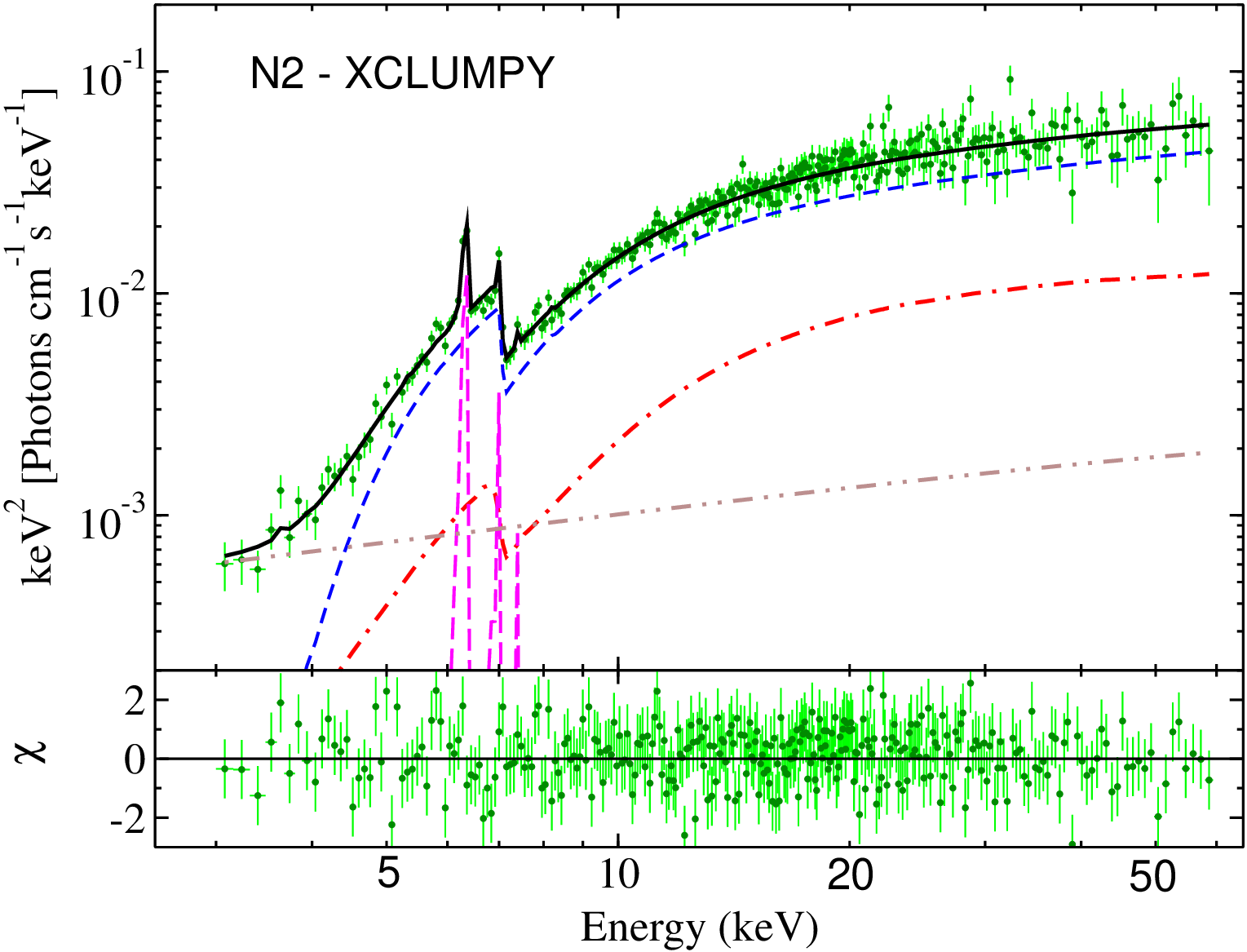}
\includegraphics[width=8.4cm]{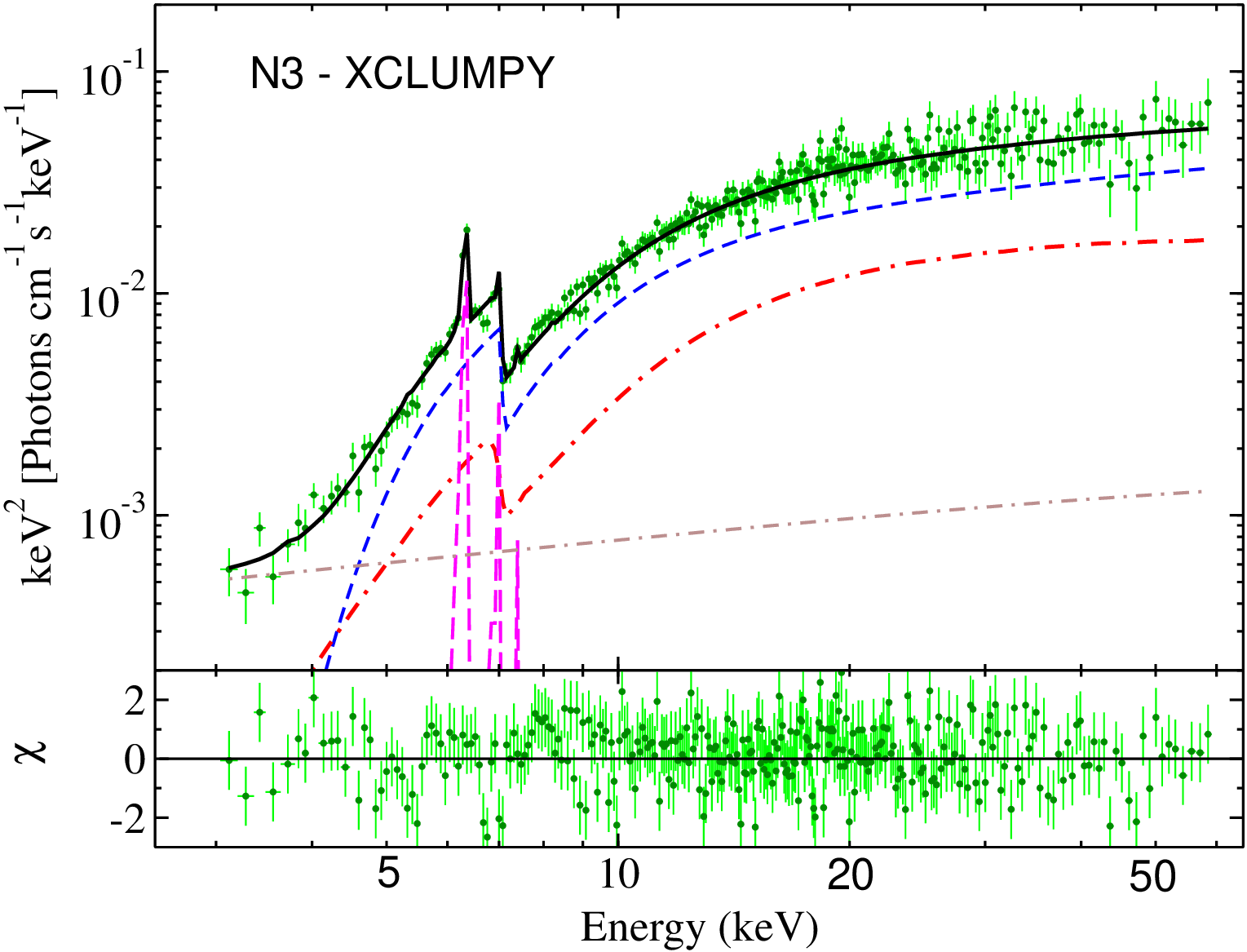}
\includegraphics[width=8.4cm]{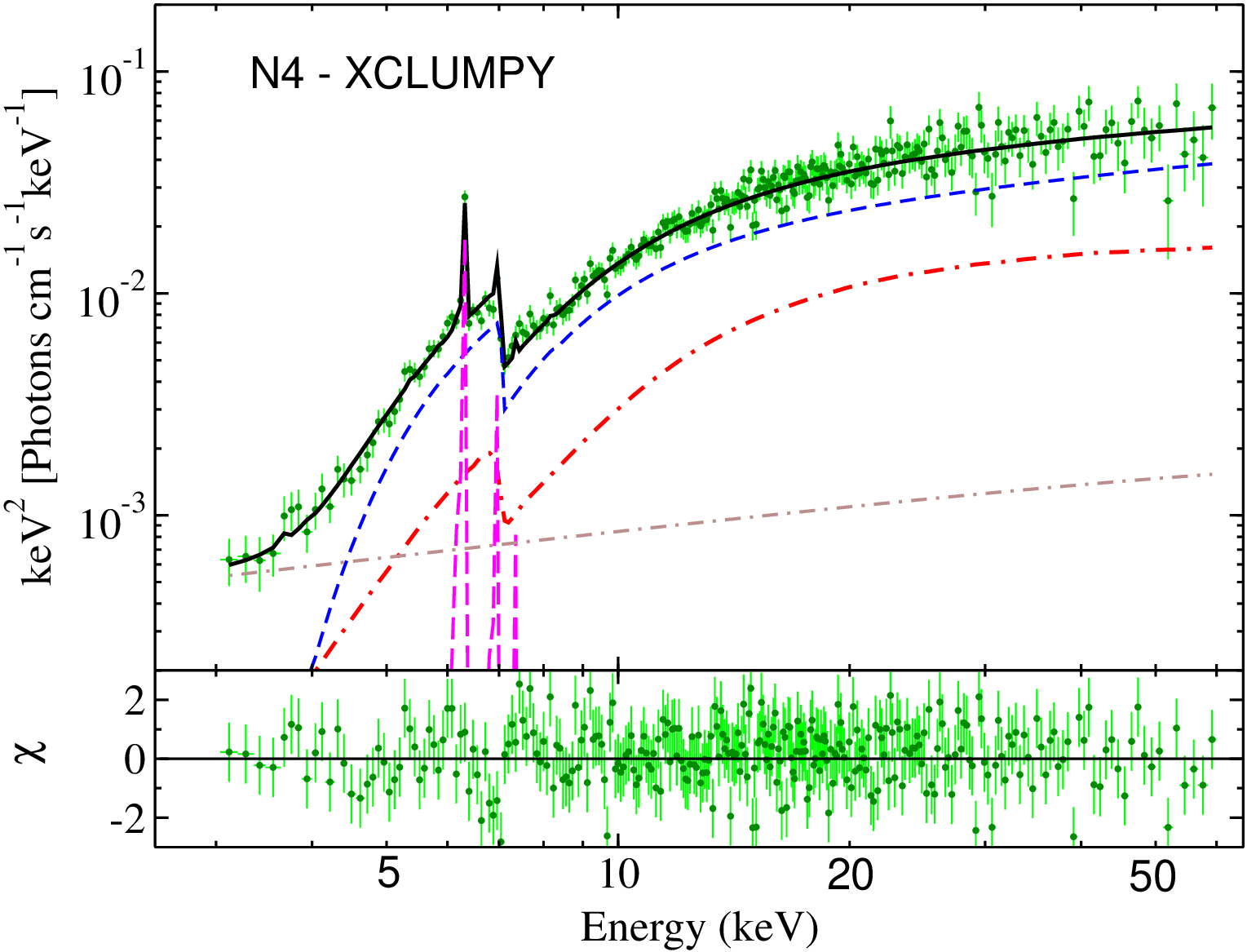}
\caption{Unfolded spectra fitted with \xclumpy~ model for observation N1 (top left), N2 (top right), N3 (bottom left) and N4 (bottom right). Upper panel : Green points represent the data. The black, blue, red, magenta and brown lines represent the total emission, primary emission, reprocessed emission, line emission (Fe K$\alpha$, Fe K$\beta$ and Ni K $\alpha$), and scattered primary emission. Bottom panel: Corresponding residual.}
\label{fig:xc-spec}
\end{figure*}

\begin{figure*}
\centering
\includegraphics[width=8.4cm]{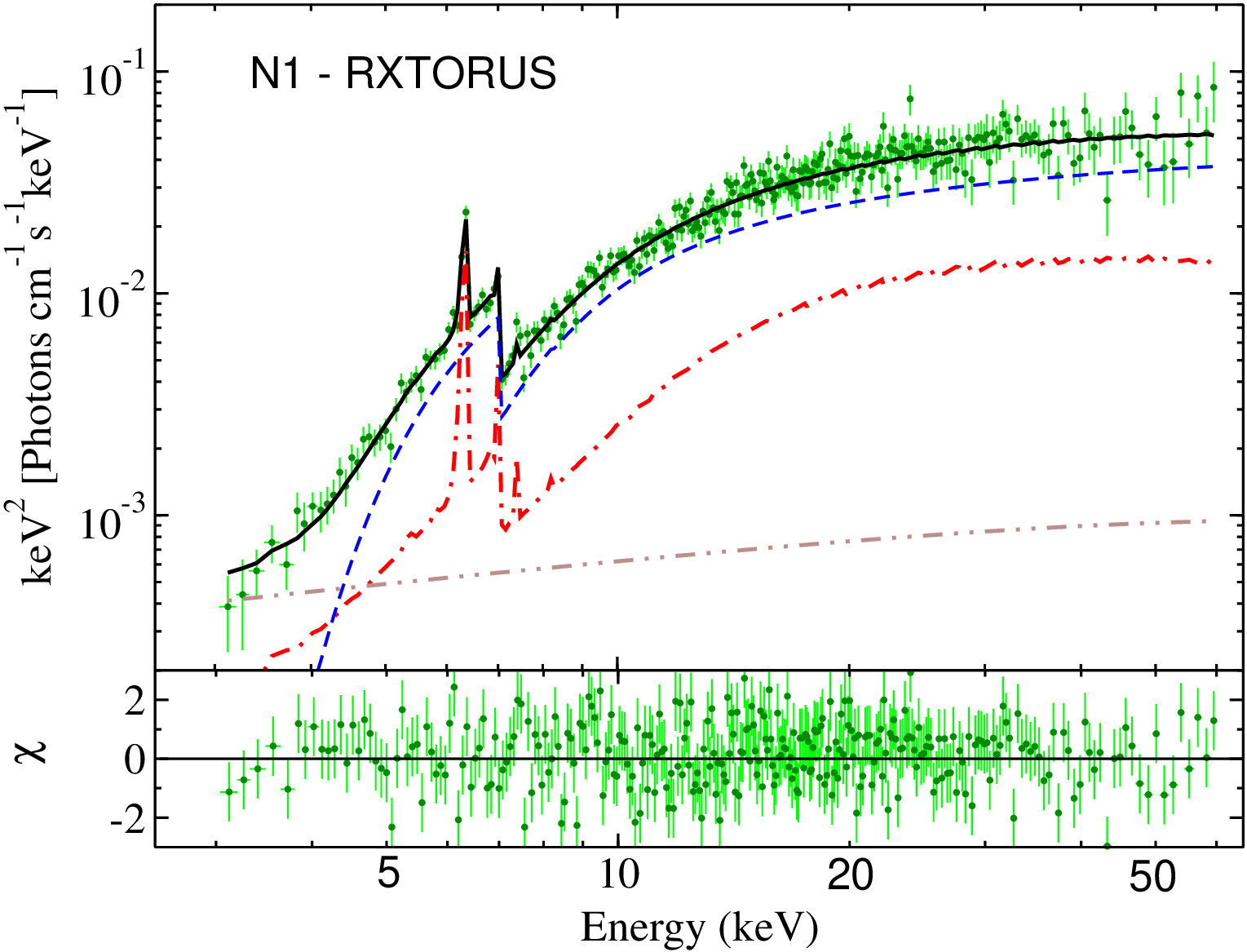}
\includegraphics[width=8.4cm]{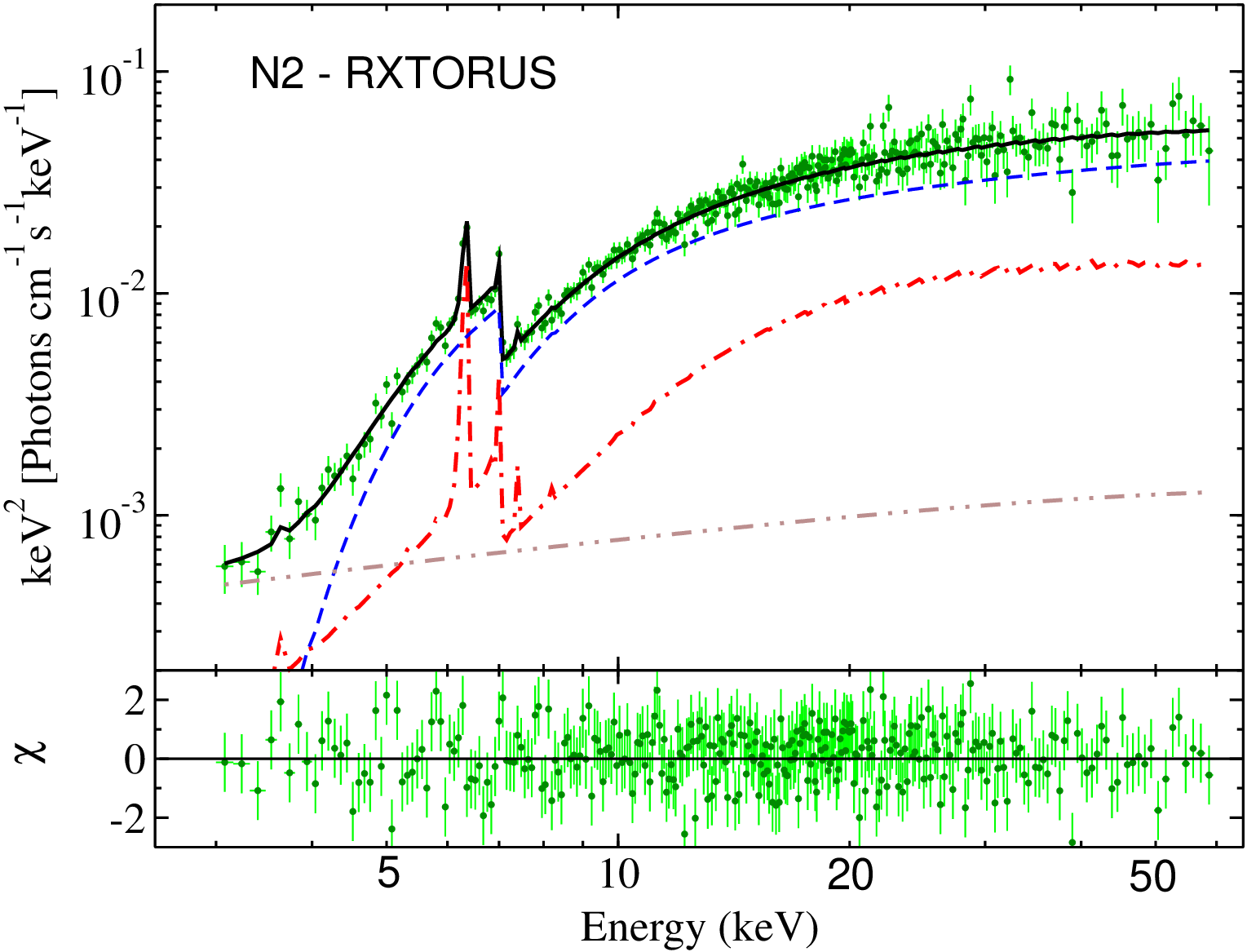}
\includegraphics[width=8.4cm]{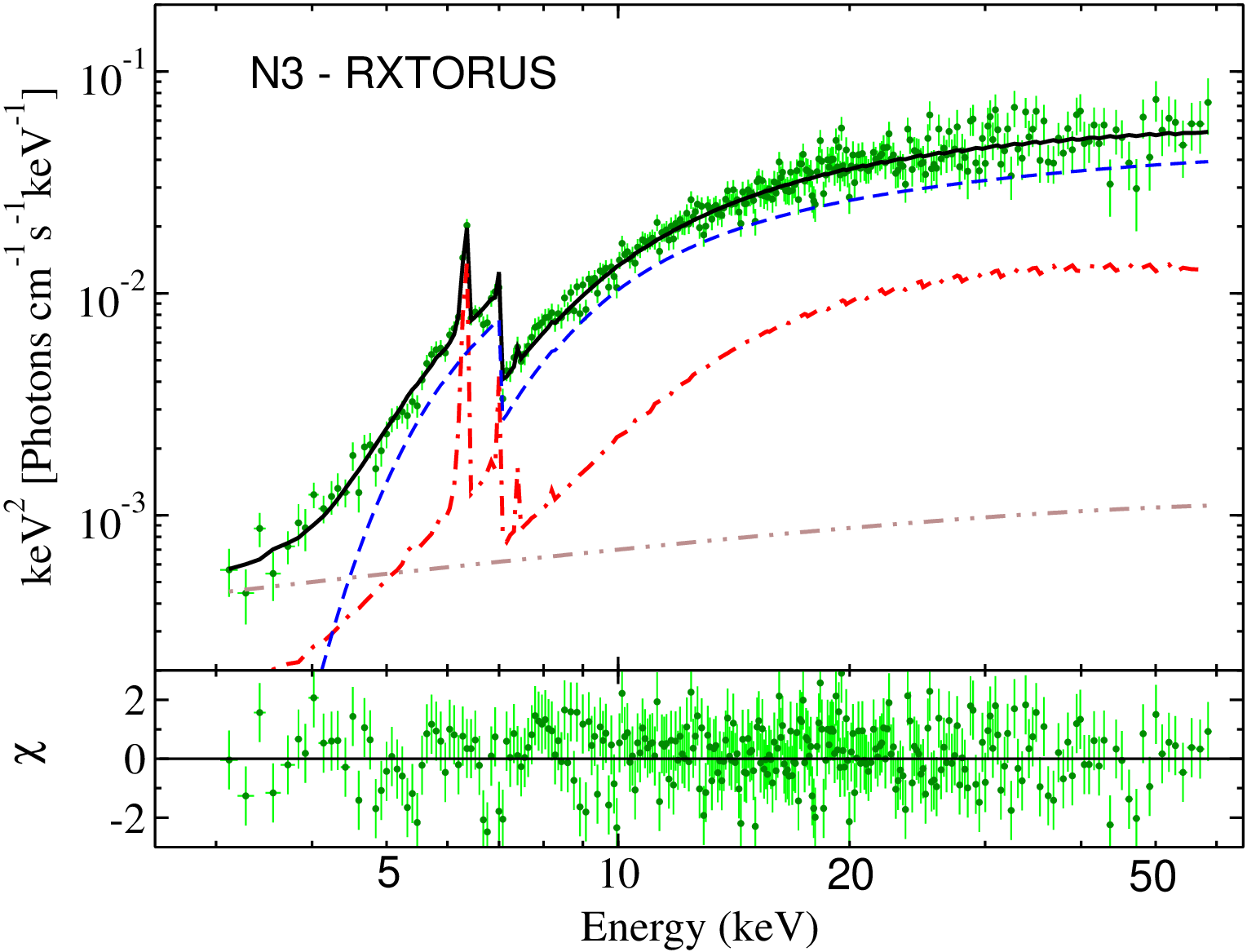}
\includegraphics[width=8.4cm]{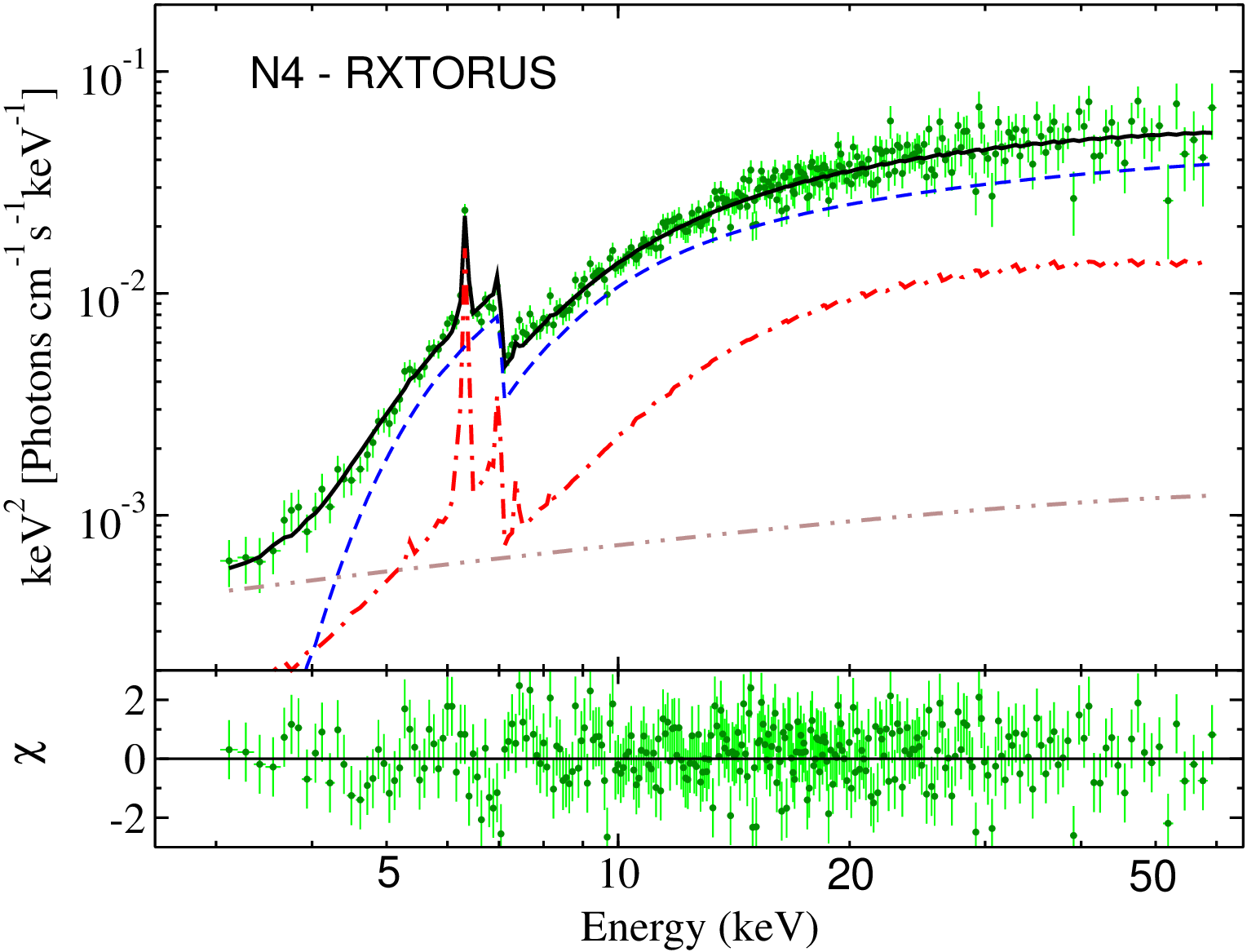}
\caption{Unfolded spectra fitted with \rxtorus~ model for observation N1 (top left), N2 (top right), N3 (bottom left) and N4 (bottom right). Upper panel : Green points represent the data. The black, blue, red, magenta and brown lines represent the total emission, primary emission, reprocessed emission, line emission (Fe K$\alpha$, Fe K$\beta$ and Ni K $\alpha$), and scattered primary emission. Bottom panel: Corresponding residual.}
\label{fig:rx-spec}
\end{figure*}

\newpage

\section{Contour}

\begin{figure}
\centering
\includegraphics[angle=270,width=8.5cm]{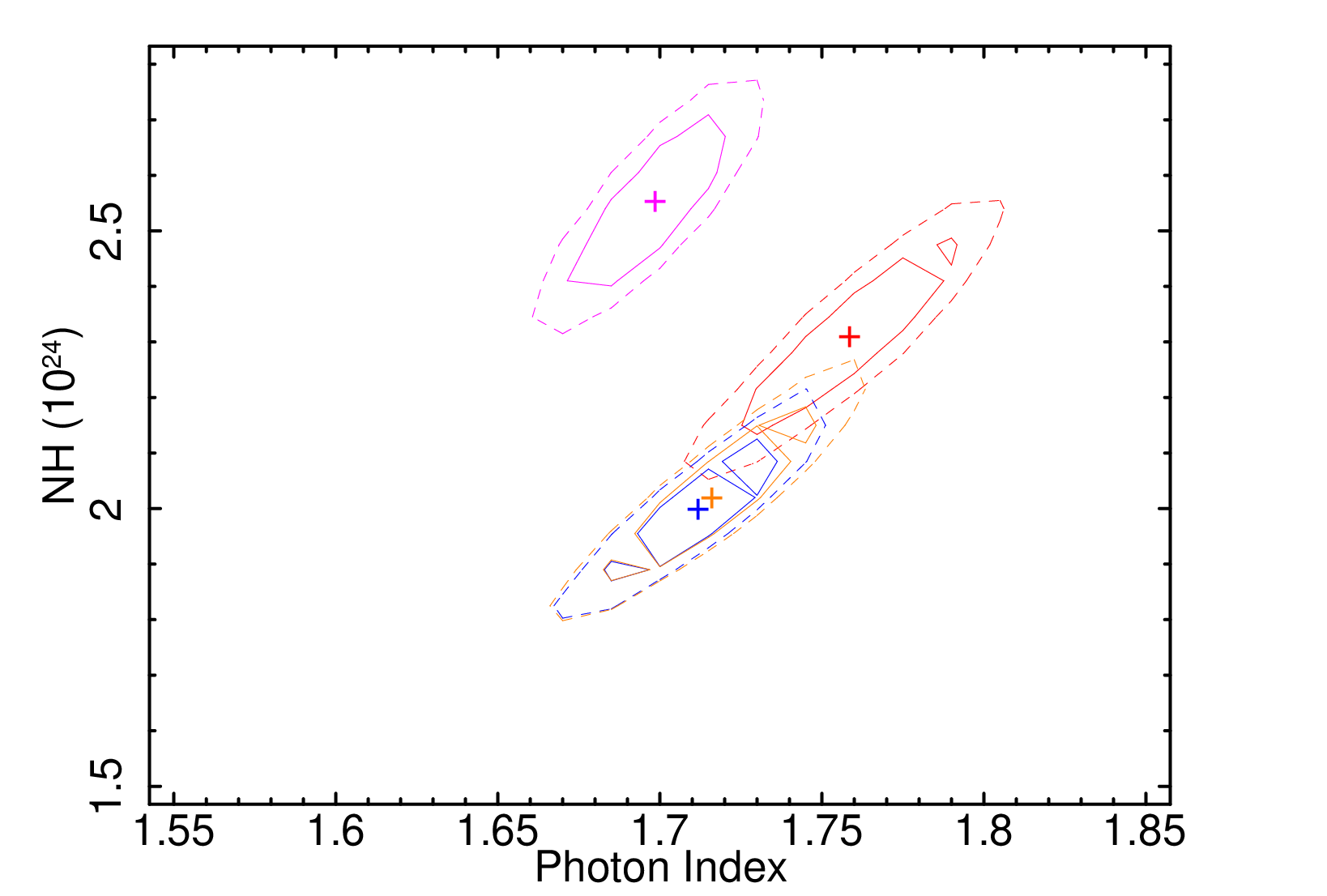}
\caption{Confidence contour between the photon index ($\Gamma$) and equatorial column density (\nh) in $10^{22}$ \pcm, fitted with the \mytc~ model. The red, blue, magenta and orange lines represent the the observation from N1, N2, N3 and N4, respectively. The solid and dashed line represent the contour at $1~\sigma$ and $2~\sigma$ level, respectively.}
\label{fig:cntr-mytc}
\end{figure}

\begin{figure}
\centering
\includegraphics[angle=270,width=8.5cm]{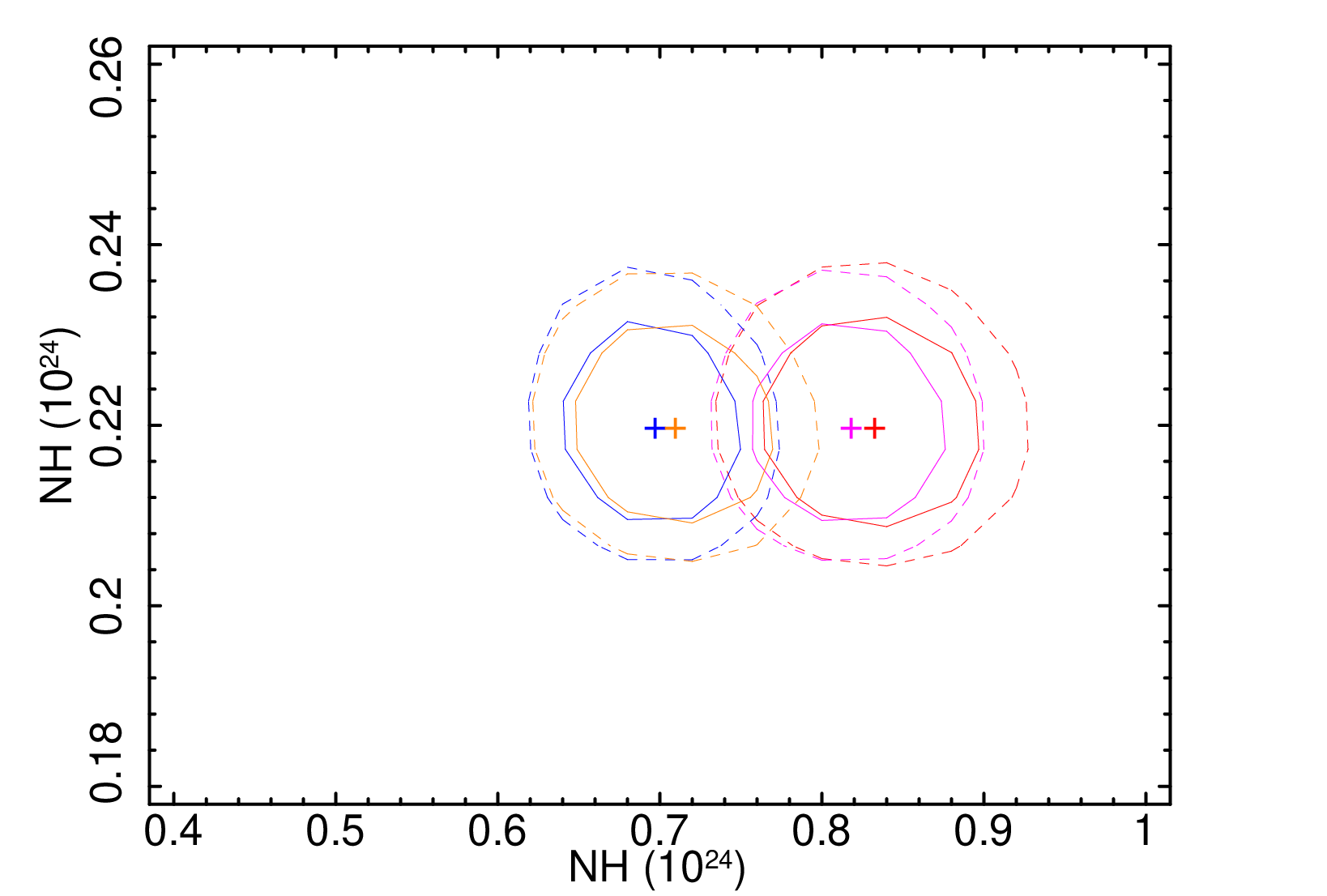}
\caption{Confidence contour between the line of sight column density (\nh) and averaged torus column density in $10^{22}$ \pcm, fitted with the \mytd~ model. The red, blue, magenta and orange lines represent the the observation from N1, N2, N3 and N4, respectively. The solid and dashed line represent the contour at $1~\sigma$ and $2~\sigma$ level, respectively.}
\label{fig:cntr-mytd}
\end{figure}

\begin{figure}
\centering
\includegraphics[angle=270,width=8.5cm]{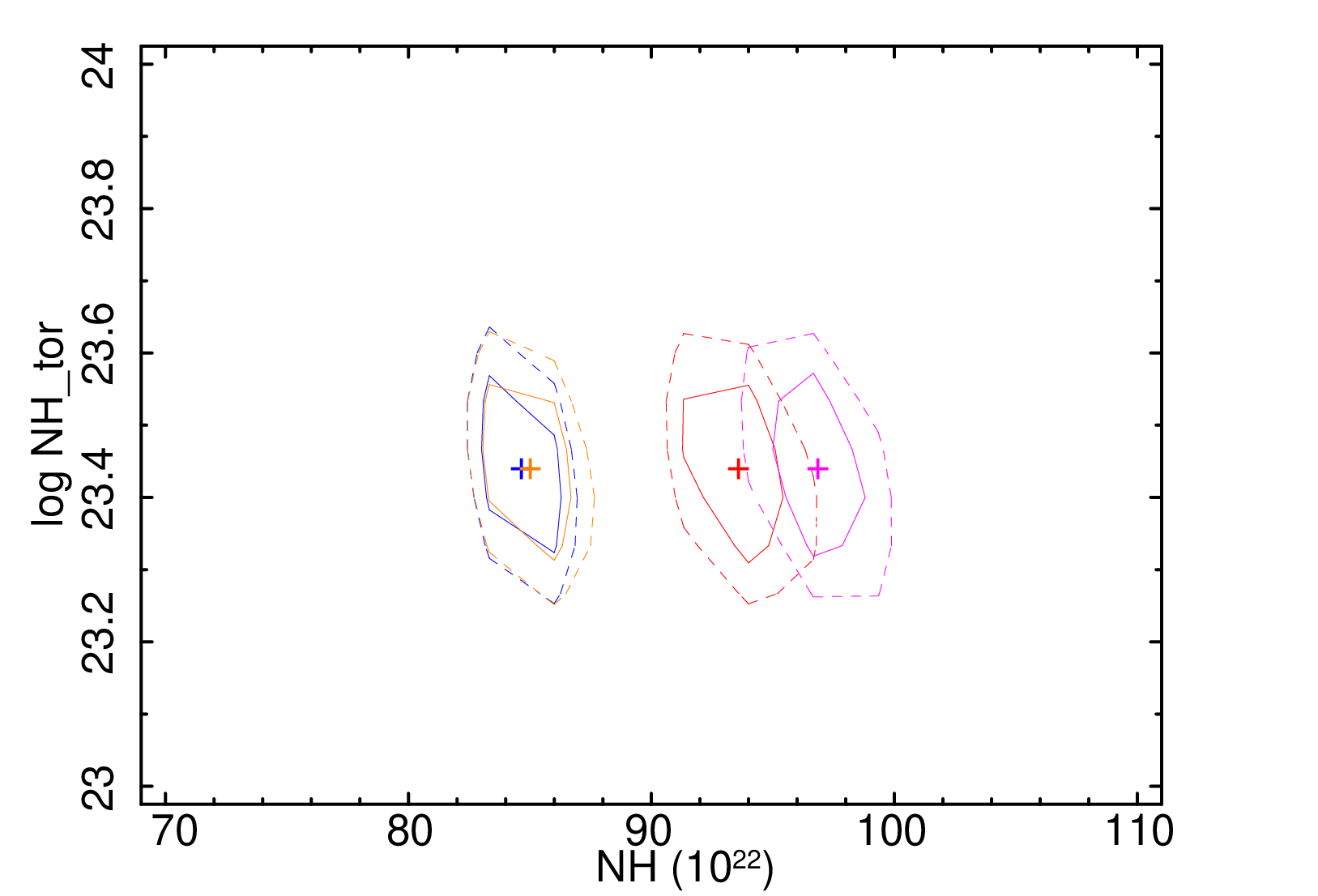}
\caption{Confidence contour between the line of sight column density (\nh) and averaged torus column density in $10^{22}$ \pcm, fitted with the \borus~ model.  The red, blue, magenta and orange lines represent the the observation from N1, N2, N3 and N4, respectively. The solid and dashed line represent the contour at $1~\sigma$ and $2~\sigma$ level, respectively.}
\label{fig:cntr-borus}
\end{figure}

\begin{figure}
\centering
\includegraphics[angle=270,width=8.5cm]{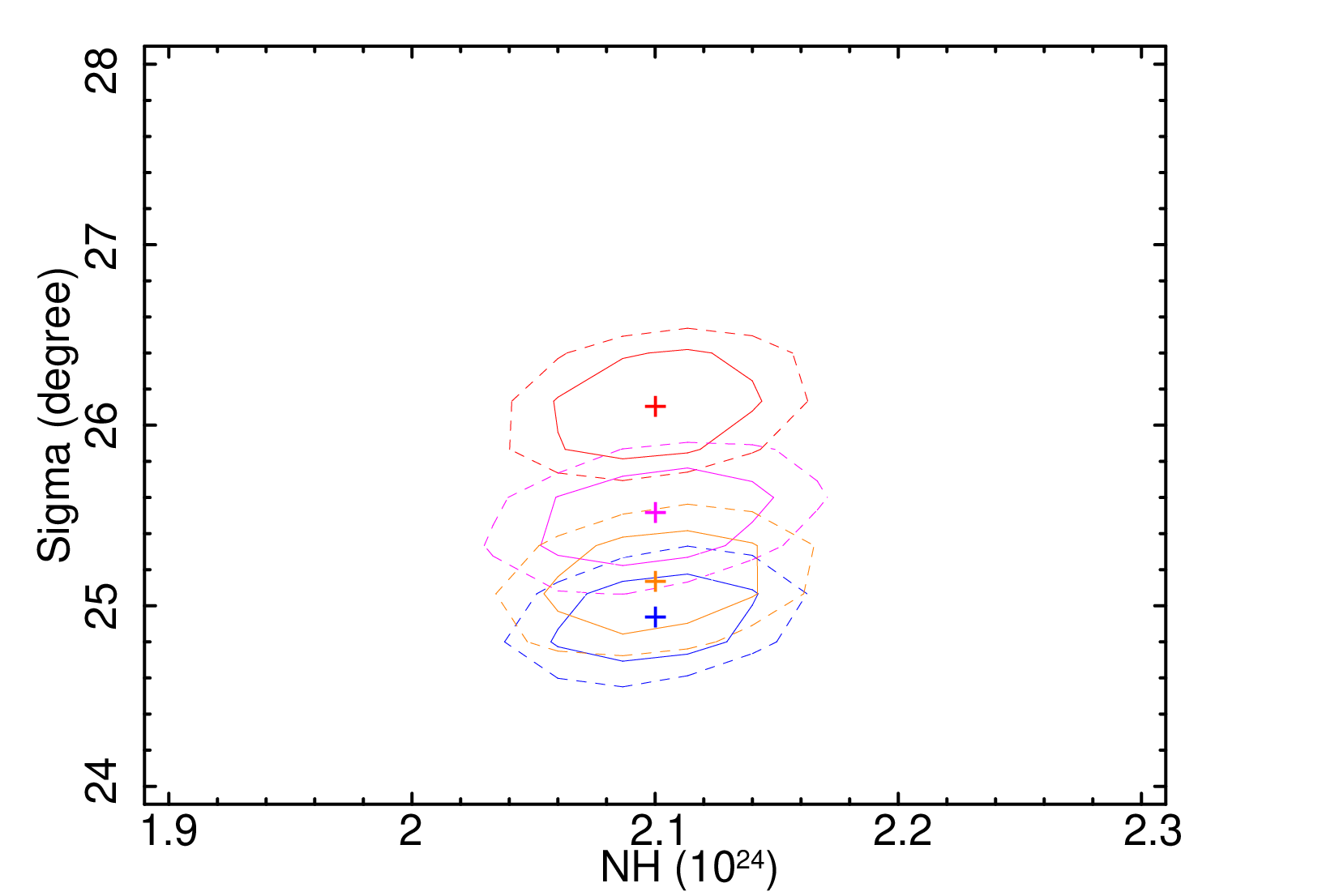}
\caption{Confidence contour between the equatorial column density (\nheq) and torus angular width, fitted with the \xclumpy~ model. The red, blue, magenta and orange lines represent the the observation from N1, N2, N3 and N4, respectively. The solid and dashed line represent the contour at $1~\sigma$ and $2~\sigma$ level, respectively.}
\label{fig:cntr-xclumpy}
\end{figure}

\begin{figure}
\centering
\includegraphics[angle=270,width=8.5cm]{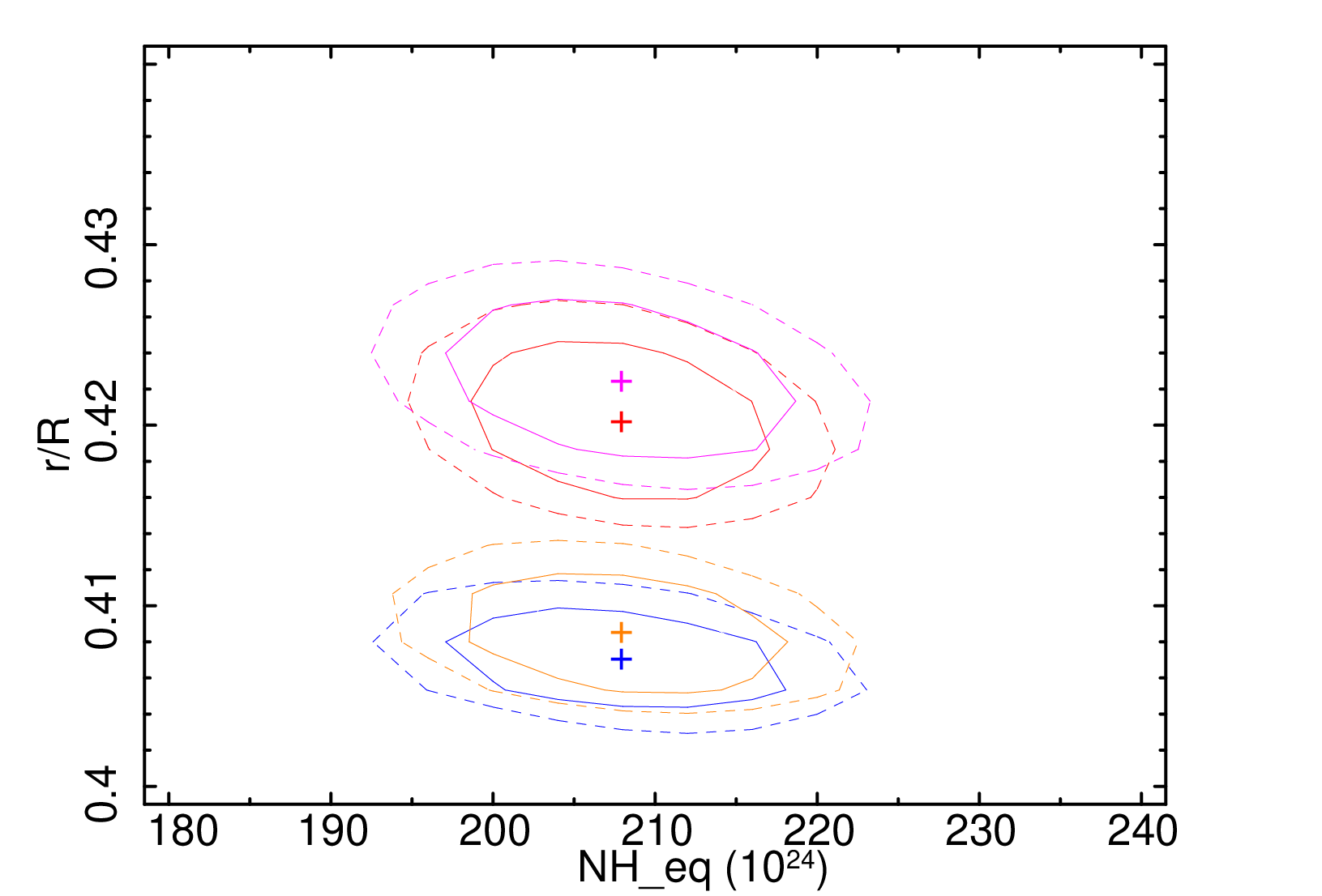}
\caption{Confidence contour between the equatorial column density (\nheq) and covering factor, fitted with the \rxtorus~ model. The red, blue, magenta and orange lines represent the the observation from N1, N2, N3 and N4, respectively. The solid and dashed line represent the contour at $1~\sigma$ and $2~\sigma$ level, respectively.}
\label{fig:rx-cntr}
\end{figure}



\bsp	
\label{lastpage}
\end{document}